\documentclass[paper,toc,nohyper]{JHEP3}
\usepackage{epsfig}
\usepackage{amsbsy,ams} 
\usepackage{subfigure,epsfig}
\usepackage{graphicx}
\usepackage{pifont}
\usepackage{amsmath}

\def\ph#1{\phantom{.}}
\def\JET{J}

\def\bom#1{{\mbox{\boldmath $#1$}}}

\def\doubletilde#1{\widetilde{\vphantom{\raise 1.5pt \hbox{#1}}\smash{\kern -2pt\widetilde{#1}}}}

\allowdisplaybreaks

\preprint{
  IPPP/10/22\\
  \\
  \today}

\title{Antenna subtraction for gluon scattering at NNLO}

\author{E. W. Nigel Glover,
    \ Joao Pires \ 
    	\\
	Department of Physics, University of Durham, Durham, DH1 3LE, UK
	\\
	E-mails: 
        {\tt e.w.n.glover@durham.ac.uk}, 
        	{\tt joao.pires@durham.ac.uk}.
}

\abstract{
We use the antenna subtraction method to isolate the double real radiation infrared singularities present in gluonic scattering amplitudes at next-to-next-to-leading order. The antenna subtraction framework has been successfully applied to the calculation of NNLO corrections to the 3-jet cross section and related event shape distributions in electron-positron annihilation. Here we consider processes 
with two coloured particles in the initial state, and in particular two-jet production at hadron colliders such as the Large Hadron Collider (LHC). We construct a subtraction term that describes the single and double unresolved contributions from the six-gluon tree-level process using antenna functions with initial state partons and show numerically that the subtraction term correctly approximates the matrix elements in the various single and double unresolved configurations.
\\
\today
}

\begin{document}
\tableofcontents
\newpage
\section{Introduction}

In proton-proton collisions, the factorised form of the inclusive cross section is given by,
\begin{equation}
{\rm d}\sigma =\sum_{i,j} \int   
{\rm d}\hat\sigma_{ij}
 f_i(\xi_1) f_j(\xi_2) \frac{d\xi_1}{\xi_1} \frac{d\xi_2}{\xi_2} \nonumber
\end{equation}
where ${\rm d}\hat\sigma_{ij}$ is the parton-level scattering cross section for parton $i$ to scatter off parton $j$ normalised to the hadron-hadron flux\footnote{The partonic cross section normalised to the parton-parton flux is obtained by absorbing the inverse factors of $\xi_1$ and $\xi_2$ into ${\rm d}\hat\sigma_{ij}$.} and the sum runs over the possible parton types $i$ and $j$. The probability of finding a parton of type $i$ in the proton carrying a momentum fraction $\xi$ is described by the parton distribution function $f_i(\xi)d\xi$.  By applying suitable cuts, one can study more exclusive observables such as the transverse momentum distribution or rapidity distributions of the hard objects (jets or vector bosons, higgs bosons or other new particles) produced in the hard scattering.  The leading-order (LO) prediction is a useful guide to the rough size of the cross section, but is usually subject to large uncertainties from the dependence on the unphysical renormalisation and factorisation scales, as well as possible mismatches between the (theoretical) parton-level and the (experimental) hadron-level.  

Frequently, the theoretical prediction may be improved by including 
higher order perturbative predictions which have the effect of (a) reducing the renormalisation/factorisation scale dependence and (b) improving the matching of the parton level event topology with the experimentally observed hadronic final state~\cite{Glover:2002gz}.
The partonic cross section ${\rm d}\hat\sigma_{ij}$ has the perturbative expansion 
\begin{equation}
{\rm d}\hat\sigma_{ij} = {\rm d}\hat\sigma_{ij}^{LO}
+\left(\frac{\alpha_s}{2\pi}\right){\rm d}\hat\sigma_{ij}^{NLO}
+\left(\frac{\alpha_s}{2\pi}\right)^2{\rm d}\hat\sigma_{ij}^{NNLO}
+{\cal O}(\alpha_s^3)
\end{equation}
where the next-to-leading order (NLO) and next-to-next-to-leading order (NNLO) strong corrections are identified. 
Any calculation of these higher-order corrections requires a systematic procedure for extracting the infrared singularities that arise when one or more final state particles become soft and/or collinear.  These singularities are present in the real radiation contribution at next-to-leading order (NLO), and in double real radiation and mixed real-virtual contributions at next-to-next-to-leading order (NNLO). 

For example, let us consider the $m$-jet cross section.  This is obtained by evaluating the tree-level cross section for processes with $m$-partons in the final state (i.e. the processes involving $(m+2)$-partons with two partons in the initial state), and requiring that each final state parton is identified as a jet by some jet algorithm.     For simplicity, we will focus on the contribution involving only gluons and will therefore drop the parton labels.
Suppose we now want to compute the $m$-jet cross section to NLO. For this, we have to consider the real radiation cross section ${\rm d}\hat\sigma^R$ with ($m+1$)-partons in the final state, the one-loop correction ${\rm d}\hat\sigma^V$ with $m$-partons in the final state, and a mass factorisation counterterm ${\rm d}\hat\sigma^{MF}$ to absorb the divergences arising from initial state collinear radiation into the parton densities:
\begin{eqnarray}
{\rm d}\hat\sigma_{NLO}=\int_{{\rm d}\Phi_{m+1}}{\rm d}\hat\sigma^R_{NLO}+\int_{{\rm d}\Phi_{m}}{\rm d}\hat\sigma^V_{NLO}
+\int_{{\rm d}\Phi_{m}}{\rm d}\hat\sigma^{MF}_{NLO}.\label{eq:NLOxsec}
\end{eqnarray}
Here the integration is over the appropriate $N$-parton final state subject to the constraint that precisely $m$-jets are observed by the jet algorithm $J_m^{(N)}$,
\begin{equation}
\int_{{\rm{d}}\Phi_{N}} \equiv \int {\rm{d}}\Phi_{N} J_m^{(N)}.
\end{equation}
The terms on the right hand side of (\ref{eq:NLOxsec}) are separately divergent although their sum is finite. To write a Monte Carlo program to compute those integrals we must first isolate and cancel the singularities of the different pieces and then numerically evaluate the finite remainders to obtain the NLO contribution to the cross section.

Subtraction schemes are a well established solution to this problem. They work by finding a suitable counterterm ${\rm d}\hat\sigma^S_{NLO}$ for ${\rm d}\hat\sigma^R_{NLO}$. It has to satisfy two properties, namely it must have the same singular behaviour in all appropriate limits as ${\rm d}\hat\sigma^R_{NLO}$ and yet be simple enough to be integrated analytically over all singular regions of the $(m+1)$-parton phase space in $d$ dimensions. We proceed by rewriting (\ref{eq:NLOxsec}) in the following form:
\begin{eqnarray}
{\rm d}\hat\sigma_{NLO}=\int_{{\rm d}\Phi_{m+1}}\left({\rm d}\hat\sigma_{NLO}^R-{\rm d}\hat\sigma_{NLO}^S\right)
+\int_{{\rm d}\Phi_m}\left(\int_1{\rm d}\hat\sigma_{NLO}^S+{\rm d}\hat\sigma_{NLO}^V+{\rm d}\hat\sigma_{NLO}^{MF}\right).
\label{eq:NLOxsec2} 
\end{eqnarray}

In its unintegrated form ${\rm d}\hat\sigma_{NLO}^S$ has the same singular behaviour as ${\rm d}\hat\sigma_{NLO}^R$ such that the first integral is finite by definition and can be integrated numerically in four dimensions over the $(m+1)$-parton phase space. The integrated form of the counterterm ${\rm d}\hat\sigma_{NLO}^S$  analytically cancels the explicit singularities of the virtual contribution ${\rm d}\hat\sigma_{NLO}^V$ and the mass factorisation counterterm ${\rm d}\hat\sigma_{NLO}^{MF}$. After checking the cancellation of the pole pieces, we can take the finite remainders of these contributions and perform the last integral on the right hand side of (\ref{eq:NLOxsec2}) numerically over the $m$-parton phase space.

The actual form of the counterterm ${\rm d}\hat\sigma_{NLO}^S$ depends on the subtraction formalism employed because there are many ways of approximating the matrix elements in the neighbourhood of its soft and collinear singularities. Several methods for constructing NLO subtraction terms systematically have been proposed in the literature \cite{Catani:1996vz,Frixione:1995ms,Nagy:1996bz,Frixione:1997np,Somogyi:2006cz} and both the Catani-Seymour \cite{Catani:1996vz} and FKS \cite{Frixione:1995ms} subtraction formalisms have been implemented in an automated way in \cite{Gleisberg:2007md,Seymour:2008mu,Hasegawa:2008ae,Hasegawa:2009tx,Frederix:2008hu,Czakon:2009ss} and \cite{Frederix:2009yq} respectively. These packages aim to automatically generate the subtraction terms and real emission amplitudes for any given process. Used in conjunction with recent automated packages that compute the virtual contribution \cite{Berger:2008sj,Giele:2008bc,Ossola:2007ax,Binoth:2008uq}, there is the exciting possibility of having an automated NLO QCD parton level event generator available in the near future.

Nevertheless, for some processes, particularly the main $2 \to1$ or $2 \to 2$ scattering processes such as Drell-Yan, Higgs production, di-jet production,
vector-boson plus jet, vector-boson pair production or heavy quark pair production,  the NLO corrections   
still have a large theoretical uncertainty and it is necessary to include the
NNLO perturbative corrections.  In addition to reducing the renormalisation and factorisation scale dependence 
there is an improved matching of the parton level theoretical jet algorithm with the hadron level
experimental jet algorithm because the jet structure can be modeled by the
presence of a third parton.  For di-jet production, the resulting theoretical uncertainty
at NNLO is estimated to be at the few per-cent level~\cite{Glover:2002gz}.

At NNLO, there are three distinct contributions due to double real radiation radiation ${\rm{d}}\hat\sigma_{NNLO}^R$, mixed real-virtual radiation ${\rm{d}}\hat\sigma_{NNLO}^{V,1}$ and double virtual radiation ${\rm{d}}\hat\sigma_{NNLO}^{V,2}$, that are given by
\begin{eqnarray}
{\rm d}\hat\sigma_{NNLO}&=&\int_{{\rm{d}}\Phi_{m+2}} {\rm{d}}\hat\sigma_{NNLO}^R 
+\int_{{\rm{d}}\Phi_{m+1}} {\rm{d}}\hat\sigma_{NNLO}^{V,1} 
+\int_{{\rm{d}}\Phi_m}{\rm{d}}\hat\sigma_{NNLO}^{V,2}
\end{eqnarray}
where the integration is over the appropriate $N$-particle final state subject to the constraint that precisely $m$-jets are observed. As usual the individual contributions in the $m$, $(m+1)$ and $(m+2)$-parton final states are all separately infrared divergent although, after renormalisation and factorisation, their sum is finite. 

For processes with two partons in the initial state, the parton level cross sections are related to the interference of $M$-particle $i$-loop and $j$-loop amplitudes $[\langle{\cal M}^{(i)}|{\cal M}^{(j)}\rangle]_M$ by
\begin{eqnarray}
{\rm{d}}\hat\sigma_{NNLO}^R &\sim& \left[\langle{\cal M}^{(0)}|{\cal M}^{(0)}\rangle\right]_{m+4},\nonumber\\
{\rm{d}}\hat\sigma_{NNLO}^{V,1} &\sim& \left[\langle{\cal M}^{(0)}|{\cal M}^{(1)}\rangle+\langle{\cal M}^{(1)}|{\cal M}^{(0)}\rangle\right]_{m+3},\nonumber\\
{\rm{d}}\hat\sigma_{NNLO}^{V,2} &\sim& \left[\langle{\cal M}^{(1)}|{\cal M}^{(1)}\rangle+\langle{\cal M}^{(0)}|{\cal M}^{(2)}\rangle+\langle{\cal M}^{(2)}|{\cal M}^{(0)}\rangle\right]_{m+2}.
\end{eqnarray}

\begin{figure}[t]
\begin{center}
\includegraphics[width=15cm]{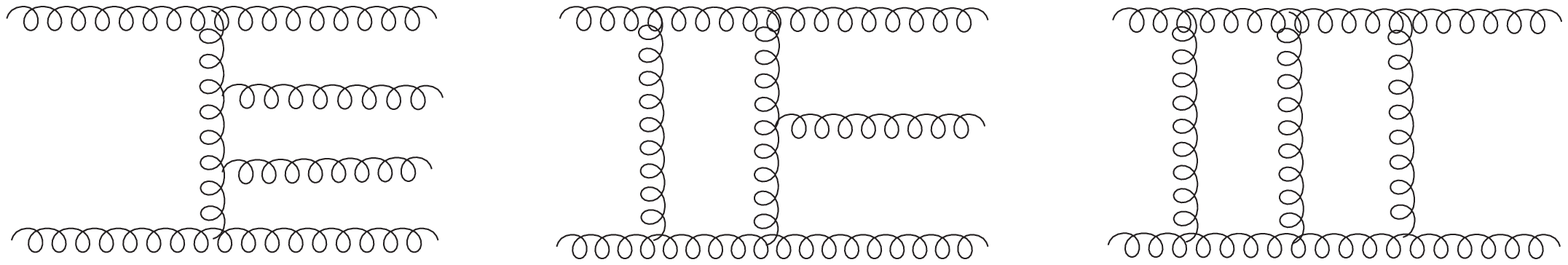}\end{center}\vspace{-0.4cm}
\qquad${\cal M}_6^{(0)}(gg\to gggg)$\qquad\qquad\qquad${\cal M}_5^{(1)}(gg\to ggg)$\qquad\qquad\qquad ${\cal M}_4^{(2)}(gg\to gg)$
\caption
{Sample diagrams contributing to the dijet inclusive rate at NNLO}
\label{fig:jetNNLO}
\end{figure}

In this paper, we specialise to the gluonic contributions to dijet production.
Sample diagrams for each ingredient with $m=2$ in the pure gluon channel is given in figure~\ref{fig:jetNNLO}.
Explicit expressions for the interference of the four-gluon tree-level and two-loop amplitudes are available in Refs.~\cite{Glover:2001af,Bern:2002tk}, while the self interference of the four-gluon one-loop amplitude is given in \cite{Glover:2001rd}. The remaining two-loop matrix elements, for quark-quark and quark-gluon scattering were obtained in \cite{Anastasiou:2000kg,Anastasiou:2000ue,Anastasiou:2001sv,Bern:2003ck,DeFreitas:2004tk}. These two-loop contributions contain explicit infrared divergences coming from the integration over the loop momentum. This singular behaviour is predictable with the Catani formula for the IR pole structure for a general on-shell QCD amplitude at two loop order derived in \cite{Catani:1998bh}.
The one-loop helicity amplitudes for the five gluon amplitude are given in \cite{Bern:1993mq}. This contribution contains explicit infrared divergences coming from integrating over the loop momenta and implicit poles in the regions of the phase space where one of the final state partons becomes unresolved. This corresponds to the soft and collinear regions of the one-loop amplitude that were analyzed in \cite{Bern:1998sc,Bern:1999ry,Catani:2000pi}.
The double real six-gluon matrix elements were derived in \cite{Berends:1987cv,Mangano:1987xk,Mangano:1990by}. Here the singularities occur in the phase space regions corresponding to two gluons becoming simultaneously soft and/or collinear. The ``double'' unresolved behaviour is universal and was discussed in \cite{GehrmannDeRidder:1997gf,Campbell:1997hg, Catani:1998nv,Catani:1999ss}.

Understanding the origin of the singularities of the different contributions is fundamental to constructing a subtraction procedure that can achieve their analytic cancellation.
There have been several approaches to build a general subtraction scheme at NNLO \cite{GehrmannDeRidder:2005cm, Weinzierl:2003fx,Frixione:2004is,Somogyi:2005xz,Somogyi:2006da,Somogyi:2006db,Somogyi:2008fc,Aglietti:2008fe,Somogyi:2009ri,Bolzoni:2009ye}. 
Another subtraction type scheme has been proposed in~\cite{Catani:2007vq}. 
It is not a general NNLO subtraction scheme, but can nevertheless deal with an entire class of processes in hadron-hadron collisions and has been explicitly applied to several observables~\cite{Catani:2007vq,Grazzini:2008tf,Catani:2009sm,Catani:2010en}.
In addition, there is the completely independent sector decomposition
approach which avoids the need for analytical integration and 
which has been developed for virtual~\cite{Binoth:2000ps,Binoth:2003ak,Heinrich:2008si} 
and real radiation~\cite{Heinrich:2002rc,Anastasiou:2003gr, Binoth:2004jv, Heinrich:2006ku} corrections to
NNLO, and applied to several observables already~\cite{Anastasiou:2004qd,Anastasiou:2004xq,Anastasiou:2005qj,Melnikov:2006di}.

We will follow the NNLO antenna subtraction method which was derived in \cite{GehrmannDeRidder:2005cm} for processes involving only (massless) final state partons. This formalism has been applied in the computation of NNLO corrections to three-jet production in electron-positron annihilation \cite{GehrmannDeRidder:2007jk,GehrmannDeRidder:2008ug,Weinzierl:2008iv,Weinzierl:2009nz} and related event shapes \cite{GehrmannDeRidder:2007bj,GehrmannDeRidder:2007hr,GehrmannDeRidder:2009dp,Weinzierl:2009ms,Weinzierl:2009yz}, which were subsequently used in precision determinations of the strong coupling constant \cite{Dissertori:2007xa,Dissertori:2009ik,Dissertori:2009qa,Bethke:2008hf,Gehrmann:2009eh}.  It has also been extended at NNLO to include one hadron in the initial state relevant for electron-proton scattering~\cite{Daleo:2009yj,Daleo:2010tz} while in ref.~\cite{Boughezal:2010ty} the extension of the antenna subtraction method to include two hadrons in the initial state is discussed. 

The general form for the subtraction terms for an $m$-particle final state at NNLO is given by~\cite{GehrmannDeRidder:2005cm}:
\begin{eqnarray}
{\rm d}\hat\sigma_{NNLO}&=&\int_{{\rm{d}}\Phi_{m+2}}\left({\rm{d}}\hat\sigma_{NNLO}^R-{\rm{d}}\hat\sigma_{NNLO}^S\right)
+\int_{{\rm{d}}\Phi_{m+2}}{\rm{d}}\hat\sigma_{NNLO}^S\nonumber\\
&+&\int_{{\rm{d}}\Phi_{m+1}}\left({\rm{d}}\hat\sigma_{NNLO}^{V,1}-{\rm{d}}\hat\sigma_{NNLO}^{V S,1}\right)
+\int_{{\rm{d}}\Phi_{m+1}}{\rm{d}}\hat\sigma_{NNLO}^{V S,1}
+\int_{{\rm{d}}\Phi_{m+1}}{\rm{d}}\hat\sigma_{NNLO}^{MF,1}\nonumber\\
&+&\int_{{\rm{d}}\Phi_m}{\rm{d}}\hat\sigma_{NNLO}^{V,2}
+\int_{{\rm{d}}\Phi_m}{\rm{d}}\hat\sigma_{NNLO}^{MF,2}.
\end{eqnarray}
Here, ${\rm d} \hat\sigma^{S}_{NNLO}$ denotes the subtraction term for the $(m+2)$-parton final state which behaves like the double real radiation contribution
${\rm d} \hat\sigma^{R}_{NNLO}$ in all singular limits. 
Likewise, ${\rm d} \hat\sigma^{VS,1}_{NNLO}$ is the one-loop virtual subtraction term 
coinciding with the one-loop $(m+1)$-final state ${\rm d} \hat\sigma^{V,1}_{NNLO}$ in all singular limits. 
The two-loop correction 
to the $(m+2)$-parton final state is denoted by ${\rm d}\hat\sigma^{V,2}_{NNLO}$.  In addition, when there are partons in the initial state, 
there are two mass factorisation contributions, 
${\rm d}\hat\sigma^{MF,1}_{NNLO}$ and ${\rm d}\hat\sigma^{MF,2}_{NNLO}$, for the $(m+1)$- and $m$-particle final states respectively.

In this paper, we will develop the antenna subtraction method for two partons in the initial state.  As a first example of how to use the formalism, we focus on the double real radiation contribution to dijet production in hadron-hadron collisions and specialise to the case where only gluons participate in the interaction.   At NNLO, dijet production receives contributions from the double real six-gluon process.  This particular contribution has the most complicated infrared structure that we can find at NNLO and we study that in detail.

Our paper is organised as follows.   In section~\ref{sec:gluon}, we review the structure of the multi-gluon amplitudes and the associated multi-gluon cross sections.   In section~\ref{sec:NLOantenna} we review the antenna method for gluons-only processes at NLO, and apply it to di-jet production in sect.~\ref{sec:NLOdijet}.   We develop the antenna subtraction terms for doubly unresolved radiation in section~\ref{sec:NNLOantenna}, and derive the subtraction terms for the six-gluon process in sect.~\ref{sec:DsigmaSNNLO}.   In section~\ref{sec:numerics}, we test the validity of the subtraction term by numerically studying the subtracted matrix elements in all of the singly- and doubly-unresolved limits. In particular, we show that in all cases the ratio of the double real cross section and the subtraction term approaches unity.  Finally, our findings are summarised in section~\ref{sec:conc}.

\section{Gluonic amplitudes}
\label{sec:gluon}
    
The $n$-gluon tree amplitude can be written as
\begin{eqnarray}
{\bf A}^0_n(\{p_i,\lambda_i,a_i\})=2^{n/2} g^{n-2}\sum_{\sigma\in S_n/Z_n}\textrm{Tr}(T^{a_{\sigma(1)}}\cdots T^{a_{\sigma(n)}})
{\cal A}^0_n(\sigma(1),\cdots,\sigma(n))\nonumber\\\label{eq:cdecomp}
\end{eqnarray}
where $g$ is the gauge coupling $({g^2}={4\pi}\alpha_s$) and the  permutation sum, $S_n/Z_n$ is the group of non-cyclic permutations of $n$ symbols.
We will frequently denote the momentum $p_j$ of gluon $j$ by $j$. The helicity information is not relevant to the discussion of the subtraction terms and from now on, we will systematically suppress the helicity labels. The $T^a$ are fundamental representation $SU(N)$ colour matrices, normalised such that ${\rm Tr}(T^aT^b)= \delta^{ab}/2$.

${\cal M}_n(1,\cdots,n)$ is the colour ordered \textit{partial amplitude}~\cite{Berends:1987cv,Kosower:1987ic,Mangano:1987xk,Mangano:1990by}.
It is gauge invariant, as well as being invariant under cyclic permutations of the gluons.

The leading colour contribution to the squared matrix elements summed over colours (and helicities) is given by the sum of squares of partial amplitudes,
\begin{eqnarray}
|{\bf A}^0_n|^2=\left({g^2N}\right)^{n-2}(N^2-1)\sum_{\sigma\in S_n/Z_n}\left\lbrace \ |{\cal A}^0_n(\sigma(1),\cdots,\sigma(n))|^2+{\cal O}\left(\frac{1}{N^2}\right) \right\rbrace.\nonumber\\
\end{eqnarray}

The infrared singularity properties of the tree-level colour ordered matrix elements have been well studied~\cite{Berends:1988zn,Kosower:1987ic,GehrmannDeRidder:1997gf,Campbell:1997hg,Catani:1998nv,Catani:1999ss,DelDuca:1999ha,Birthwright:2005ak,Birthwright:2005vi}.  
At tree-level, the colour-ordered gluon amplitude has a QED-like factorisation in the limit where gluon $b$ is soft, into an eikonal-type singular factor and a colour ordered tree-level squared amplitude where gluon $b$ has been removed.   For the squared amplitude we have,   
\begin{eqnarray}
|{\cal A}^0_{n+1}(1,\ldots,a,b,c,\ldots,n+1)|^2&\stackrel{b_g\to0}{\longrightarrow}&S_{abc}\;|{\cal A}_n(1,\ldots,a,c,\ldots,n+1)|^2,
\label{eq:ssoft}
\end{eqnarray}
with the eikonal factor given by,
\begin{eqnarray}
S_{abc}=\frac{2s_{ac}}{s_{ab}s_{bc}}.\label{eq:seik}
\end{eqnarray}

Similarly in the limit where two gluons become collinear, the colour-ordered amplitudes factorise. If gluons $a$ and $b$ become collinear and form gluon $c$, then adjacent gluons give a singular contribution,
\begin{eqnarray}
|{\cal A}^0_{n+1}(1,\ldots,a,b,\ldots,n+1)|^2&\stackrel{a//b}{\longrightarrow}&\frac{1}{s_{ab}}P_{gg\to g}(z)
|{\cal A}^0_n(1,\ldots,c,\ldots,n+1)|^2\nonumber\\\label{eq:collimit}
\end{eqnarray}
while separated gluons do not,
\begin{eqnarray}
|{\cal A}^0_{n+1}(1,\ldots,a,\ldots,b,\ldots,n+1)|^2&\stackrel{a//b}{\longrightarrow}&\textrm{finite}.
\end{eqnarray}

In equation (\ref{eq:collimit}), $z$ is the fraction of momentum carried by one of the gluons and, after integrating over the azimuthal angle of the plane containing the collinear particles with respect to the hard process, collinear splitting function $P_{gg\to g}$ is given by,
\begin{eqnarray}
P_{gg\to g}(z)=2\left(\frac{z}{1-z}+\frac{1-z}{z}+z(1-z)\right).\label{eq:Pgg}
\end{eqnarray}

In the case of two real unresolved particles there are a variety of different configurations extensively studied in \cite{GehrmannDeRidder:1997gf,Campbell:1997hg,Catani:1998nv,Catani:1999ss}. The expressions for these universal limits are organised according to whether the two unresolved particles are colour connected or not. In the unconnected case, i.e. where the gluons are not adjacent in the colour ordered amplitude, the singular limits are merely obtained by multiplying single unresolved factors. However, when the particles are colour connected, i.e. where the gluons are adjacent in the colour ordered amplitude, the structure is more involved.

At higher numbers of loops, the leading colour contribution to the 
$n$-gluon $m$-loop amplitude can be written as~\cite{Bern:1994zx},
\begin{eqnarray}
{\bf A}^m_n(\{p_i,\lambda_i,a_i\})=2^{n/2} g^{n-2+2m} N^{m} \sum_{\sigma\in S_n/Z_n}\textrm{Tr}(T^{a_{\sigma(1)}}\cdots T^{a_{\sigma(n)}})
{\cal A}^m_n(\sigma(1),\cdots,\sigma(n)).\label{eq:cdecomploop}
\end{eqnarray}
The single unresolved infrared singularity structure of one- and two-loop amplitudes has been studied in Refs.~\cite{Bern:1994zx,Bern:1998sc,Kosower:1999xi,Kosower:1999rx,Bern:1999ry,Catani:2000pi,Catani:2003vu,Kosower:2003cz,Weinzierl:2003ra,Bern:2004cz,Badger:2004uk}.

For convenience, we introduce the additional notation
\begin{equation}
{A}^M_n(\sigma(1),\ldots,\sigma(n)) = \sum_{\rm helicities} \sum_{i=0..M}
{\cal A}_n^{i\dagger}(\sigma(1),\ldots,\sigma(n))
{\cal A}_n^{M-i}(\sigma(1),\ldots,\sigma(n)).
\end{equation}
We will encounter ${A}_5^0$ and ${A}_4^1$ in the example of the antenna subtraction at NLO given in sections~\ref{sec:NLOreal} and \ref{sec:NLOvirtual} respectively, while ${A}^0_6$, ${A}^1_5$ and ${ A}^2_4$ are relevant for the NNLO dijet cross section.  In section~\ref{sec:NNLOfullRR}, we will explicitly derive the NNLO subtraction term for 
${A}^0_6$.

\section{Antenna subtraction at NLO}
\label{sec:NLOantenna}

The leading-order parton-level contribution from the $(m+2)$-parton processes to the $m$-jet cross section at LO in $pp$ collisions is given by  
\begin{eqnarray}
{\rm d}\hat\sigma_{LO}&=&{\cal N}\sum_{\textrm{perms}}{\rm d}\Phi_{m}(p_3,\hdots,p_{m+2};p_1,p_2)\frac{1}{S_{m}}\nonumber\\
&&\times|{\cal M}_{m+2}(1,\ldots,m+2)|^2 J_{m}^{(m)}(p_3,\hdots,p_{m+2})
\label{eq:LOcross}
\end{eqnarray}
where to make the subsequent discussion more general, we denote a generic tree-level colour ordered amplitude by the symbol ${\cal M}(1,\ldots,m+2)$. The sum over ${\rm perms}$ is the appropriate sum over colour ordered amplitudes.  For gluonic amplitudes this is the sum over the group of non-cyclic permutations of $n$ symbols denoted by $S_n/Z_n$ in the previous section.  The symmetry factor $S_{m}$ accounts for identical partons in the final state.
The normalisation factor ${\cal N}$ includes all QCD-independent factors as well as the dependence on the renormalised QCD coupling constant $\alpha_s$.  The initial state momenta are labeled as usual as $p_1$ and $p_2$ whereas the $m$-momenta in the final state are labeled $p_3,\hdots,p_{m+2}$. ${\rm d}\Phi_{m}$ is the $2\to m$ particle phase space:
\begin{eqnarray}
&&{\rm d}\Phi_{m}(p_3,\hdots,p_{m+2};p_1,p_2)=\nonumber\\
&&\frac{{\rm d}^{d-1}p_3}{2E_3(2\pi)^{d-1}}\hdots\frac{{\rm d}^{d-1}p_{m+2}}{2E_{m+2}(2\pi)^{d-1}}
(2\pi)^d\delta^d(p_1+p_2-p_3-\hdots-p_{m+2}).
\end{eqnarray}
The jet function $J_{m}^{(m)}(p_3,\hdots,p_{m+2})$ defines the procedure for building $m$-jets from ($m$)-partons. The main property of $J_{m}^{(m)}$ is that the jet observable defined above is collinear and infrared safe.

We now focus on the antenna subtraction at NLO \cite{Campbell:1998nn,GehrmannDeRidder:2005cm,Daleo:2006xa,GehrmannDeRidder:2009fz}. In this approach, antenna functions describe the colour-ordered radiation of unresolved partons between a pair of hard (radiator) partons. We must distinguish three possible configurations of radiators: final-final when both radiators are final state partons, initial-final when one radiator is an initial state parton and the other radiator is a final state parton and, finally, initial-initial where both radiators are initial state partons.

This means that we will derive subtraction formulae decomposed in these three configurations and label them with a superscript $(ff),(if),(ii)$. In table {\ref{table1}} we distinguish the configurations that are needed according to the scattering process that we are interested in.
\begin{table}
\begin{center}
\begin{tabular}{|c|c|c|c|}
\hline
 & final-final & initial-final & initial-initial \\ \hline
$e^+e^-$ & \checkmark & \ding{55} & \ding{55}\\
\hline
$ep$ & \checkmark & \checkmark & \ding{55}\\
\hline
$pp$ & \checkmark & \checkmark & \checkmark\\
\hline
\end{tabular}
\caption[Antennae configurations]{Antennae configurations needed according to the scattering process}
\label{table1}
\end{center}
\end{table}
Even though we will write down all formulae specifically for $pp$ collisions these formulae can be easily adapted to $ep$ or $e^+e^-$ for the configurations they have in common. To achieve that we should only modify the number of partons that enter the matrix elements in these formulae. This is because we obtain an $m$-jet production at leading order from an $m$-parton matrix element in $e^+e^-$, an ($m+1$)-parton matrix element in $ep$ and an ($m+2$)-parton matrix element in $pp$. 

The $m$-jet real radiation cross section at NLO in $pp$ collisions is given by,
\begin{eqnarray}
{\rm d}\hat\sigma_{NLO}^R&=&{\cal N}\sum_{\textrm{perms}}{\rm d}\Phi_{m+1}(p_3,\hdots,p_{m+3};p_1,p_2)\frac{1}{S_{m+1}}\nonumber\\
&&\times|{\cal M}_{m+3}(1,\ldots,m+3)|^2 J_{m}^{(m+1)}(p_3,\hdots,p_{m+3}).
\label{eq:NLOrealcor}
\end{eqnarray}

\subsection{Final-Final configuration}
\label{sec:NLOff}
\begin{figure}[t]
\begin{center}
\includegraphics[width=0.8\textwidth]{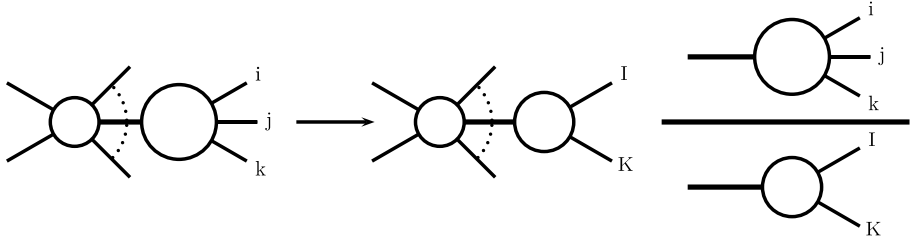}
\caption[Final-final antenna factorisation]
{Illustration of NLO antenna factorisation representing the
factorisation of both the squared matrix elements and the $(m+1)$-particle 
phase
space when the unresolved particle $j$ is colour connected between two final state radiators $i$ and $k$.}
\label{fig:final-final}
\end{center}
\end{figure}

When summing over all colour orderings in equation (\ref{eq:NLOrealcor}) we find terms of this type:
\begin{eqnarray}
|{\cal M}_{m+3}(\hdots,i,j,k,\hdots)|^2
\end{eqnarray}
where $i,j,k$ are colour connected final state partons. This configuration contains a singularity when $j$ is unresolved between $i$ and $k$ which can be approximated by,
\begin{eqnarray}
X_{ijk}|{\cal M}_{m+2}(\hdots,I,K,\hdots)|^2,
\end{eqnarray}
where $X_{ijk}$ is a final-final antenna function that describes all singular configurations (for this colour-ordered amplitude) where parton $j$ is unresolved. The momenta for the new partons $I$ and $J$ are linear combinations of $p_i$, $p_j$, $p_k$ obtained with a final-final mapping that we describe in section \ref{sec:ff}. Both radiator partons $i$ and $k$ are in the final state and we call this situation a final-final antenna, depicted in figure \ref{fig:final-final}. We still have to sum over all possible unresolved partons in this colour ordered amplitude. After that we sum over all colour orderings to obtain the full subtraction term, for the final-final configuration, to use with (\ref{eq:NLOrealcor}).
The subtraction term for this configuration reads:
\begin{eqnarray}
&&{\rm d}\hat\sigma^{S,(ff)}={\cal N}\sum_{\textrm{perms}}{\rm d}\Phi_{m+1}(p_3,\hdots,p_{m+3};p_1,p_2)\frac{1}{S_{m+1}}\nonumber\\
&&\times\sum_j X_{ijk}^0 |{\cal M}_{m+2}(\hdots,I,K,\hdots)|^2
J_m^{(m)}(p_3,\hdots,p_I,p_K,\hdots,p_{m+3}).
\label{eq:sub1}
\end{eqnarray}

The subtraction term involves an $(m+2)$-parton amplitude that does not depend on $p_i$, $p_j$, $p_k$ but on the redefined on-shell momenta $p_I$ and $p_K$, where $p_I$ and $p_K$ are linear combinations of $p_{i}$, $p_{j}$, ${p}_{k}$, while the tree antenna function $X^0_{ijk}$ depends only on $p_i$, $p_j$ and $p_k$. 

The jet function $J^{(m)}_m$ in (\ref{eq:sub1}) does not depend on the individual momenta ${p}_{i}$, $p_j$ and ${p}_{k}$, but only on $p_I$ and $p_K$. One can therefore carry out the integration over the unresolved dipole phase space appropriate for ${p}_{i}$, $p_j$ and ${p}_{k}$ analytically, exploiting the factorisation of the phase space,
\begin{eqnarray}
\label{eq:psx3}
\lefteqn{{\rm d} \Phi_{m+1}(p_{3},\ldots,p_{m+3};p_1,p_2)  = }
\nonumber \\ &&
{\rm d} \Phi_{m}(p_{3},\ldots,p_I,p_K,\ldots,p_{m+3};p_1,p_2)
\cdot 
{\rm d} \Phi_{X_{ijk}} (p_i,p_j,p_k;p_I+p_K).
\end{eqnarray}
The NLO antenna phase space ${\rm d} \Phi_{X_{ijk}}$ is proportional to the three-particle phase space relevant to a $1\to 3$ 
decay.

For the analytic integration, we can use (\ref{eq:psx3}) to rewrite each of the subtraction terms in (\ref{eq:sub1}) in the form,
\begin{displaymath}
|{\cal M}_{m+2}|^2\,
J_{m}^{(m)}\; 
{\rm d}\Phi_{m}
\int {\rm d} \Phi_{X_{ijk}}\;X^0_{ijk}.
\end{displaymath}
The analytic integral of the subtraction term is therefore defined as the antenna function integrated over the fully inclusive antenna phase space, normalised appropriately,
\begin{equation}
\label{eq:x3int}
{\cal X}^0_{ijk}(s_{ijk}) = \frac{1}{C(\epsilon)} 
\int {\rm d} \Phi_{X_{ijk}}\;X^0_{ijk},
\end{equation}
where
\begin{equation}
C(\epsilon)=(4\pi)^{\epsilon}\frac{e^{-\epsilon\gamma}}{8\pi^2}.\label{eq:ceps}
\end{equation}
The integration is performed analytically in $d$ dimensions to make the infrared singularities explicit and then
added directly to the one-loop $m$-particle contribution ${\rm d}\hat\sigma^V_{NLO}$. 

\subsection{Initial-Final configuration}
\label{sec:NLOif}
\begin{figure}[t]
\begin{center}
\includegraphics[width=0.8\textwidth]{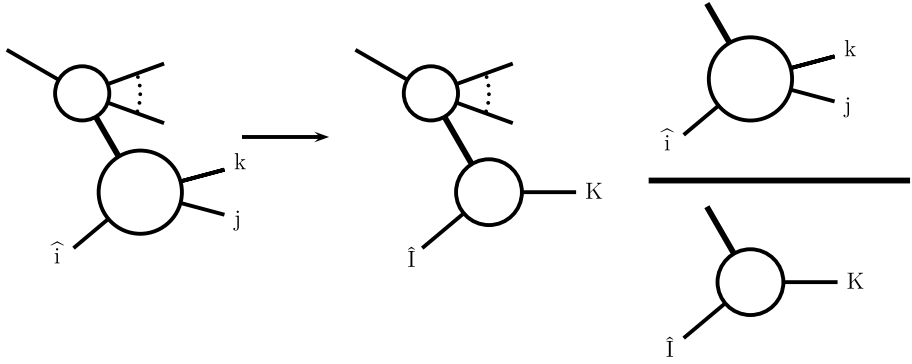}
\caption[Initial-final antenna factorisation]
{Illustration of NLO antenna factorisation representing the
factorisation of both the squared matrix elements and the $(m+1)$-particle 
phase
space when the unresolved particle $j$ is colour connected between an initial state radiator $\hat{i}$ and a final state radiator $k$.}
\label{fig:initial-final}
\end{center}
\end{figure}

In the presence of hadrons in the initial state, matrix elements exhibit singularities that are not accounted by the subtraction terms discussed in the previous section. These singularities are due to soft or collinear radiation within an antenna where one or the two hard partons are in the initial state \cite{Daleo:2006xa}.
This occurs in equation (\ref{eq:NLOrealcor}) when in the ordered amplitude parton $j$ is unresolved between an initial state parton $\hat i$ and a final state parton $k$,
\begin{equation}
|{\cal M}_{m+3}(\hdots,\hat{i},j,k,\hdots)|^2.
\end{equation}
Here and subsequently, a hat denotes an initial state parton. 
The infrared singularities can be approximated by,
\begin{equation}
X_{i,jk}|{\cal M}_{m+2}(\hdots,\hat I,K,\hdots)|^2
\end{equation}
where $X_{i,jk}$ is an initial-final antenna function that describes configurations (for this colour-ordered amplitude) where parton $j$ is unresolved. This configuration is depicted in figure \ref{fig:initial-final}. The mapping that generates the new momenta $p_K$ and $p_{\hat I}$ (which is simply a rescaling of incoming momentum $p_i$ by an amount $x$ such that
$
p_{\hat I} \equiv \bar{p}_i = x_i p_i,
$)
will be discussed in \ref{sec:if}.  The full subtraction term for the the initial-final configuration reads,
\begin{eqnarray}\label{eq:subif}
&&{\rm d}\hat\sigma^{S,(if)}={\cal N}\sum_{\textrm{perms}}{\rm d}\Phi_{m+1}(p_3,\dots,p_{m+3};p_1,p_2)
  \,\frac{1}{S_{m+1}}\nonumber\\
&&\times\sum_{ i=1,2} \sum_{j} X^{0}_{i,jk}
  \left|{\cal M}_{m+2}(\hdots,\hat I,K,\hdots)\right|^2\,
  J^{(m)}_{m}(p_3,\hdots,p_K,\hdots,p_{m+3}). 
\end{eqnarray}

The terms necessary to subtract singularities associated with coloured particles in the initial state can then be simply obtained 
by crossing the corresponding antennae for final state singularities ($X_{ijk}\to X_{i,jk}$). Due to the different kinematics involved, the factorisation of the phase space is slightly more involved and the corresponding phase space mappings are 
different. To cancel explicit infrared poles in virtual contributions and in terms arising from  parton distribution mass factorisation, the crossed antennae must be integrated, analytically, over the corresponding phase space.

As in the final-final case, the tree antenna, $X^{0}_{i,jk}$, depends only on the original momenta $p_i$, $p_j$ and $p_k$, and contains all the configurations in which parton $j$ becomes unresolved. Similarly, the $(m+2)$-parton amplitude depends only on the redefined on-shell momenta $p_{\hat I}$ (and therefore on the momentum fraction $x$) and $p_K$ and not on $p_i$, $p_j$ and $p_k$. The jet function, $J^{(m)}_{m}$ depends on the momenta $p_i$, $p_j$ and $p_k$ only through $p_K$. Thus, provided a suitable factorisation of the phase space, one can perform the integration of the antennae analytically. 

The phase space can be factorised as an $m$-parton phase space convoluted with a two particle phase space \cite{Daleo:2006xa} such as the special case $i=2$,
\begin{eqnarray}
{\rm d}\Phi_{m+1}(p_3,\dots,p_{m+3};p_1,p_2)&=&{\rm d}\Phi_{m}(p_3,\dots,p_K,\dots,p_{m+3};p_1,x_2p_2)\nonumber\\
&\times&\frac{Q^2}{2\pi}{\rm d}\Phi_{2}(p_j,p_k;p_2,q)\frac{{\rm d} x_2}{x_2}\,\label{eq:IFPSfact}
\end{eqnarray}
with $Q^2=-q^2$ and $q=p_j+p_k-p_2$.
Replacing the phase space in (\ref{eq:subif}), we can explicitly carry out the integration of the antenna factors over the two particle phase space. When combining the integrated subtraction terms with virtual contributions and mass factorisation terms, it turns out to be convenient to normalise the integrated antennae as follows 
\begin{equation}\label{eq:aint}
{\cal X}_{i,jk}(x_i)=\frac{1}{C(\epsilon)}\int {\rm d}\Phi_2 \frac{Q^2}{2\pi} X_{i,jk}\,,
\end{equation}
where $C(\epsilon)$ is given in (\ref{eq:ceps}).

\subsection{Initial-Initial configuration}
\label{sec:NLOii}
\begin{figure}[t]
\begin{center}
\includegraphics[width=0.8\textwidth]{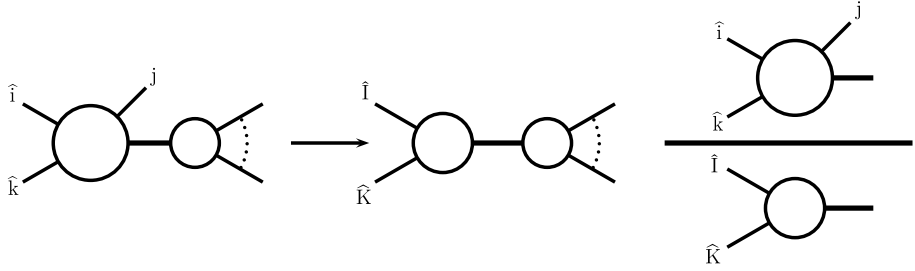}
\caption[Initial-initial antenna factorisation]
{Illustration of NLO antenna factorisation representing the
factorisation of both the squared matrix elements and the $(m+1)$-particle 
phase
space when the unresolved particle $j$ is colour connected between two initial state radiators $\hat{i}$ and $\hat{k}$.}
\label{fig:initial-initial}
\end{center}
\end{figure}

The last situation to be considered is when the two hard radiators are in the
initial state. The contribution from 
\begin{equation}
|{\cal M}_{m+3}(\hdots,\hat{i},j,\hat{k},\hdots)|^2
\end{equation}
contains the infrared singularities due to the unresolved parton $j$ when it is colour connected to both of the initial state partons $\hat{i}$ and $\hat{k}$. As sketched in \ref{fig:initial-initial}, these singularities can be approximated by,
\begin{equation}
X_{ik,j}|{\cal M}_{m+2}(\hdots,\hat{I},\hat{K},\hdots)|^2
\end{equation}
where $X_{ik,j}$ is an initial-initial antenna function that describes configurations (for this colour-ordered amplitude) where parton $j$ is unresolved.
In this case, the momentum mapping is more complicated, and not only are 
incoming momentum rescaled by an amount such that
\begin{equation}
 {p_{\hat I}} = x_i p_i, \qquad\qquad
 {p_{\hat K}} = x_k p_k, 
\end{equation}
but all other spectator momenta must be Lorentz boosted to preserve momentum conservation,
\begin{equation}
p_\ell \to \tilde{p}_\ell, \qquad \ell \neq i,j,k.
\end{equation}
The definitions of $x_i$ and $x_k$ as well as the precise definition of the spectator momenta $\tilde{p}_\ell$ will be discussed in section \ref{sec:ii}. 

The full subtraction term for the initial-initial configuration can be written as:
\begin{eqnarray}\label{eq:subii}
{\rm d}\hat\sigma^{S,(ii)}&=&{\cal N}\sum_{\textrm{perms}}{\rm d}\Phi_{m+1}(p_3,\hdots,p_{m+3};p_1,p_2)
  \,\frac{1}{S_{m+1}}\nonumber \\
&& \sum_{i,k=1,2}\sum_{j}X^{0}_{ik,j}
  \left|{\cal M}_{m+2}(\hdots,{\hat I},{\hat K},\hdots)\right|^2  J^{(m)}_{m}(\tilde{p}_3,\ldots,\tilde{p}_{m+3})\,.  
\end{eqnarray}
In order to fulfill overall momentum conservation all the momenta in the arguments of the reduced matrix elements and the jet functions have been redefined. 

The phase space factorises into the convolution of an $m$ particle phase space, involving only the redefined momenta, 
with the phase space of parton $j$ \cite{Daleo:2006xa} so that when 
$i=1$ and $k=2$,
\begin{eqnarray}
{\rm d}\Phi_{m+1}(p_3,\dots,p_{m+3};p_1,p_2)&=&
{\rm d}\Phi_{m}(\tilde{p}_3,\ldots,\tilde{p}_{m+3};x_1 p_1, x_2 p_2)
\nonumber\\
&&\times\delta(x_1-\hat{x}_1)\,\delta(x_2-\hat{x}_2)\,[{\rm d} p_j]\,{\rm d} x_1\,{\rm d} x_2\,,
\end{eqnarray}
where the single particle phase space measure is $[{\rm d} p_j]={{\rm d}^{d-1}p_j}/{2E_j(2\pi)^{d-1}}$.

Inserting the factorised expression for the phase space measure in eq.
(\ref{eq:subii}), the subtraction terms can be integrated over
the antenna phase space~\cite{Daleo:2006xa}. The integrated form of the
subtraction terms must be, then, combined with the virtual
and mass factorisation terms to cancel the explicit poles in $\epsilon$.
In the case of initial-initial subtraction terms, the antenna 
phase space is trivial: the two remaining Dirac delta functions
can be combined with the one particle phase space, such that there
are no integrals left. We define the initial-initial
integrated antenna functions by,
\begin{equation}
{\cal X}_{ik,j}(x_i,x_k)=\frac{1}{C(\epsilon)}\int
[{\rm d} p_j]\,  x_i\,  x_k\,\delta(x_i-\hat{x}_i)\,\delta(x_k-\hat{x}_k)\,X_{ik,j}\;
\end{equation}
where $C(\epsilon)$ is given in (\ref{eq:ceps}).

\subsection{Antenna functions}
In the previous subsections we have seen that the subtraction term is constructed from products of 
antenna functions with reduced matrix elements (with fewer final state 
partons than the original matrix element), and integrated over 
a phase space which is factorised into an antenna phase space (involving all
unresolved partons and the two radiators) multiplied with a reduced phase 
space (where the momenta of radiators and unresolved radiation are replaced 
by two redefined momenta).
The full subtraction term is obtained by summing over all antennae 
required for the problem under consideration. In the most general case
(two partons in the initial state, and two or more hard partons in the final 
state), this sum includes final-final, initial-final and initial-initial 
antennae. We will see an example of this in section \ref{sec:NLOreal} when we compute the NLO corrections to the dijet cross section.

The relevant antenna is determined by both the external state and the pair of hard
partons it collapses to.   In general we denote the antenna function as $X$.
For antennae that collapse onto a hard quark-antiquark pair, 
$X = A$ for $qg \bar q$.  Similarly, for quark-gluon antenna, we have 
$X = D$ for $qg g$ and $X=E$ for $qq^\prime
\bar q^\prime$ final states.  Finally, we characterise the gluon-gluon antennae as
$X=F$ for $ggg$,  $X=G$ for $gq\bar q$ final states.

At NLO, we only need to consider tree level three-particle antennae involving only one unresolved parton. At NNLO we will need four-particle antennae involving two unresolved partons and one-loop three-particle antennae.

In all cases the antenna functions are derived from physical matrix elements: 
the quark-antiquark antenna functions from 
$\gamma^* \to q\bar q~+$~(partons)~\cite{GehrmannDeRidder:2004tv}, the quark-gluon antenna 
functions from $\tilde\chi \to \tilde g~+$~(partons)~\cite{GehrmannDeRidder:2005hi} and 
the gluon-gluon antenna functions from $H\to$~(partons)~\cite{GehrmannDeRidder:2005aw}. The
tree-level antenna functions are obtained by normalising the 
colour-ordered three- and four-parton tree-level 
squared matrix elements to the squared matrix element 
for the basic two-parton process,
\begin{eqnarray}
X_{ijk}^0 = S_{ijk,IK}\, \frac{|{\cal M}^0_{ijk}|^2}{|{\cal M}^0_{IK}|^2}\;,\nonumber\\
X_{ijkl}^0 = S_{ijkl,IL}\, \frac{|{\cal M}^0_{ijkl}|^2}{|{\cal M}^0_{IL}|^2}\;,
\end{eqnarray}
where $S$ denotes the symmetry factor associated with the antenna, which accounts
both for potential identical particle symmetries and for the presence 
of more than one antenna in the basic two-parton process. 

Note that in the colour-ordered quark-gluon and gluon-gluon antenna functions derived 
from physical matrix elements for neutralino decay~\cite{GehrmannDeRidder:2005hi} and 
Higgs boson decay~\cite{GehrmannDeRidder:2005aw}, it is in general not possible to identify the 
hard radiators and the unresolved partons in a unique manner. The reason for 
this ambiguity is in the cyclic nature of the colour orderings, which 
is already evident in the three-parton antenna functions: each pair of 
two partons can in principle act as hard radiators, resulting in more than 
one antenna configuration present in a single antenna function. This means that we have to disentangle the multiple singularities of these antennae into sub-antennae where an appropriate mapping can be used.

\subsection{NLO antennae decomposition for numerical implementation}
\label{sec:NLOmap}
We will concentrate on the pure gluon channel and describe the numerical implementation of the gluon-gluon antenna function with a pure gluonic final state $F_3^0$. For that we will develop in the next subsections a further decomposition of the three parton gluon-gluon-gluon antenna function into different sub-antenna configurations for the different configurations of two emitters in the final-state (final-final emitters configuration), one emitter in the final state and one emitter in the initial state (initial-final emitters configuration) and both emitters in the initial state (initial-initial emitters configuration). The unresolved particle is always in the final state. 

These three configurations are needed so that we can subtract the IR singularities from final-state radiation and initial state radiation. The different sub-antennae derived then will each contain singular limits appropriate
to the phase-space mapping used in the three different configurations mentioned and all are extensively used in the subtraction terms at NLO and NNLO.

\subsubsection{Final-Final emitters}
\label{sec:ff}
The tree level three parton antenna corresponding to the gluon-gluon-gluon final state is \cite{GehrmannDeRidder:2005cm}:
\begin{eqnarray}
F_3^0(1_g,2_g,3_g)&=&\frac{2}{s_{123}^2}\Bigg(\frac{s_{123}^2s_{12}}{s_{13}s_{23}}
+\frac{s_{123}^2s_{13}}{s_{12}s_{23}}+\frac{s_{123}^2s_{23}}{s_{12}s_{13}}
+\frac{s_{12}s_{13}}{s_{23}}+\frac{s_{12}s_{23}}{s_{13}}+\frac{s_{13}s_{23}}{s_{12}}\nonumber\\
&&+4s_{123}+ {\cal O}(\epsilon)\ \Bigg)
\label{eq:bigFff}
\end{eqnarray}
where $
s_{ij}=(p_i+p_j)^2$.
As can be seen from the pole structure, this tree level antenna function contains three antenna configurations, corresponding 
to the three possible configurations of emitting a gluon in between a gluon pair. We make the decomposition \cite{GehrmannDeRidder:2005cm} 
\begin{equation}
F_3^0(1,2,3)=f_3^0(1,3,2)+f_3^0(3,2,1)+f_3^0(2,1,3)
\label{eq:decomp3FF}
\end{equation}
where
\begin{eqnarray}
f_3^0(1,3,2)&=&\frac{1}{s_{123}^2}\Bigg(2\frac{s_{123}^2s_{12}}{s_{13}s_{23}}+\frac{s_{12}s_{13}}{s_{23}}
+\frac{s_{12}s_{23}}{s_{13}}+\frac{8}{3}s_{123}+ {\cal O}(\epsilon)\ \Big).
\label{eq:smallFff}
\end{eqnarray}
The sub-antenna $f_3^0(i,j,k)$ has the full $j$ soft limit and part of the $i\parallel j$ and $j\parallel k$ limits of the full antenna (\ref{eq:bigFff}), such that $i$ and $k$ can be identified as hard radiators.  Therefore this is the antenna we use in the numerical implementation with a unique \{3$\to$2\} momenta mapping,  $(i,j,k)\to(I,K)$~\cite{Kosower:1997zr,Kosower:2003bh}
\begin{eqnarray}
p_I^{\mu} \equiv p_{\widetilde{(ij)}}^{\mu}&=&x\,p_i^{\mu}+r\,p_j^{\mu}+z\,p_k^{\mu}\nonumber\\
p_K^{\mu} \equiv p_{\widetilde{(jk)}}^{\mu}&=&(1-x)\,p_i^{\mu}+(1-r)\,p_j^{\mu}+(1-z)\,p_k^{\mu}\;
\label{3to2FFmap}
\end{eqnarray}
where, 
\begin{eqnarray}
x&=&\frac{1}{2(s_{ij}+s_{ik})}\Big[(1+\rho)\,s_{ijk} -2\,r\,s_{jk}     \Big],\nonumber\\
z&=&\frac{1}{2(s_{jk}+s_{ik})}\Big[(1-\rho)\,s_{ijk} -2\,r\,s_{ij}     \Big],\nonumber\\
&&\nonumber\\
\rho^2&=&1+\frac{4\,r(1-r)\,s_{ij}s_{jk}}{s_{ijk}s_{ik}}.\;
\end{eqnarray}
The parameter $r$ can be chosen conveniently~\cite{Kosower:1997zr,Kosower:2003bh} and we use $ r=s_{jk}/(s_{ij}+s_{jk}).$
The mapping \eqref{3to2FFmap} implements momentum conservation $p_{\widetilde{(ij)}}+p_{\widetilde{(jk)}}=p_i+p_j+p_k$ and satisfies the following properties:
\begin{eqnarray}
p_{\widetilde{(ij)}}^2=0,\qquad&&\qquad p_{\widetilde{(jk)}}^2=0,\nonumber\\
p_{\widetilde{(ij)}}\to p_i,\qquad&&\qquad p_{\widetilde{(jk)}}\to p_k \qquad\qquad\textrm{when \textit{j} is soft},\nonumber\\
p_{\widetilde{(ij)}}\to p_i+p_j,\qquad&&\qquad p_{\widetilde{(jk)}}\to p_k \qquad\qquad\textrm{when \textit{i} becomes collinear with \textit{j}},\nonumber\\
p_{\widetilde{(ij)}}\to p_i,\qquad&&\qquad p_{\widetilde{(jk)}}\to p_j+p_k \qquad\textrm{when \textit{j} becomes collinear with \textit{k}}.\nonumber
\end{eqnarray}
This guarantees the proper subtraction of infrared singularities.

As mentioned in the description of the antenna formulation, we also need the analytic integral of the subtraction term to combine it with the virtual corrections and obtain the cancellation of the singularities analytically. That necessarily implies that we need the integrated form of the antenna over the antenna phase space (final state kinematics) which has been calculated and documented in~\cite{GehrmannDeRidder:2005cm}.

\subsubsection{Initial-Final emitters}
\label{sec:if}
The initial-final gluon-gluon-gluon antenna function can be obtained by the appropriate crossing of one of the particles from the final to the initial state \cite{Daleo:2006xa}. Its unintegrated form can then be obtained from (\ref{eq:bigFff}) by making the replacements,
$s_{23}\to(p_2+p_3)^2$,
$s_{12}\to(p_1-p_2)^2$,
$s_{13}\to(p_1-p_3)^2$ and $Q^2=s_{12}+s_{13}+s_{23}$
and it reads \cite{Daleo:2006xa},
\begin{eqnarray}
&&F_3^0(\hat{1}_g,2_g,3_g)=\frac{1}{2(Q^2)^2}\Bigg(\frac{8s_{12}^2}{s_{13}}+\frac{8s_{12}^2}{s_{23}}
+\frac{8s_{13}^2}{s_{12}}+\frac{8s_{13}^2}{s_{23}}+\frac{8s_{23}^2}{s_{12}}+\frac{8s_{23}^2}{s_{13}}\nonumber\\
&&
+\frac{12s_{12}s_{13}}{s_{23}}+\frac{12s_{23}s_{13}}{s_{12}}
+\frac{12s_{12}s_{23}}{s_{13}}
+\frac{4s_{12}^3}{s_{23}s_{13}}+\frac{4s_{13}^3}{s_{23}s_{12}}+\frac{4s_{23}^3}{s_{12}s_{13}}\nonumber\\
&&
+24s_{23}+24s_{12}+24s_{13} + {\cal O}(\epsilon)\ \Bigg)
\label{eq:bigFif}
\end{eqnarray}
where the hat identifies the gluon crossed to the initial state.
Because the initial gluon can never be soft, it is convenient to decompose this antenna into two contributions, that each contain the soft singularities of one of the final state gluons,
\begin{equation}
F_3^0(\hat{1},2,3)=f_3^0(\hat{1},2,3)+f_3^0(\hat{1},3,2).
\end{equation}
Here,
\begin{eqnarray}
f_3^0(\hat{1},2,3)&=&\frac{1}{2(Q^2)^2}\Big(\frac{8s_{13}^2}{s_{12}}+\frac{8s_{23}^2}{s_{12}}
+\frac{12s_{23}s_{13}}{s_{12}}+\frac{4s_{13}^3}{s_{23}s_{12}}+\frac{4s_{23}^3}{s_{12}(s_{12}+s_{13})}\nonumber\\
&&+\frac{8s_{13}^2}{s_{23}}+\frac{6s_{12}s_{13}}{s_{23}}
+12s_{23}+12s_{12}+12s_{13}+ {\cal O}(\epsilon)\ \Big). 
\end{eqnarray}
The sub-antenna $f_3^0(\hat{1},j,k)$ has the full $j$ soft limit, the full $1\parallel j$ limit and part of the $j\parallel k$ limit of the full antenna (\ref{eq:bigFif}), such that we can identify $\hat{1}$ as the initial state radiator and $k$ the final state radiator. To numerically implement this antenna we use the following \{3$\to$2\} mapping: $(\hat{i},j,k)\to(\hat{I},K) \equiv ( \hat {\bar{i}},K)$ \cite{Daleo:2006xa}
where the bar denotes a rescaling of the initial state parton
\begin{eqnarray} 
p_{\hat I}^\mu &\equiv& \bar{p}_{i}^{\mu}=x_i\,p_i^{\mu},\nonumber\\
p_{K}^\mu &\equiv& p_{\widetilde{(jk)}}^{\mu}=p_j^{\mu}+\,p_k^{\mu}-(1-x_i)\,p_i^{\mu}\;,
\label{3to2IFmap}
\end{eqnarray}
with $p_{\hat I}^2=p_{ K}^2=0$ and where $x_i$ is given by,
\begin{eqnarray}
x_i&=&\frac{s_{ij}+s_{ik}+s_{jk}}{s_{ij}+s_{ik}}.
\end{eqnarray}
Proper subtraction of infrared singularities requires that the momenta mapping satisfies,
\begin{eqnarray}
&&x_i p_i\rightarrow p_i\,,\qquad\qquad p_{\widetilde{(jk)}}\rightarrow p_k\qquad\qquad\mbox{when $j$ becomes soft}\,,\nonumber\\
&&x_i p_i\rightarrow p_i\,,\qquad\qquad p_{\widetilde{(jk)}}\rightarrow p_j+p_k
\qquad\mbox{when $j$ becomes collinear with $k$}\,,\nonumber\\
&&x_i p_i\rightarrow p_i-p_j\,,\qquad p_{\widetilde{(jk)}}\rightarrow p_k
\qquad\qquad\mbox{when $j$ becomes collinear with $\hat{i}$}\,.\nonumber
\end{eqnarray}
In this way, infrared singularities are subtracted locally, except for 
angular correlations, {\em before convoluting with the
parton distributions}. That is, 
matrix elements and subtraction terms are convoluted together with PDFs.
In addition, the redefined momentum $p_K$ is on shell and 
momentum is conserved, $p_i-p_j-p_k=x_i p_i-p_K$.

The integrated form of the antenna (\ref{eq:bigFif}) for initial state kinematics has been calculated and documented in \cite{Daleo:2006xa}.

\subsubsection{Initial-Initial emitters}
\label{sec:ii}

The initial-initial gluon-gluon-gluon antenna function is obtained by crossing from the corresponding initial-final antenna function (\ref{eq:bigFif}), with the replacements 
$s_{12}\to(p_1+p_2)^2$, 
$s_{13}\to(p_1-p_3)^2$, $s_{23}\to(p_2-p_3)^2$ and $Q^2=s_{12}+s_{13}+s_{23}$. It reads \cite{Daleo:2006xa},
\begin{eqnarray}
&&F_3^0(\hat{1},3,\hat{2})=\frac{1}{2(Q^2)^2}\Bigg(\frac{8s_{13}^2}{s_{23}}+\frac{8s_{13}^2}{s_{12}}
+\frac{8s_{23}^2}{s_{13}}+\frac{8s_{23}^2}{s_{12}}+\frac{8s_{12}^2}{s_{13}}
+\frac{8s_{12}^2}{s_{23}}\nonumber\\
&&+\frac{12s_{13}s_{23}}{s_{12}}+\frac{12s_{12}s_{23}}{s_{13}}
+\frac{12s_{13}s_{12}}{s_{23}}
+\frac{4s_{13}^3}{s_{12}s_{23}}+\frac{4s_{23}^3}{s_{12}s_{13}}
+\frac{4s_{12}^3}{s_{13}s_{23}}\nonumber\\
&&+24s_{12}+24s_{13}+24s_{23}+ {\cal O}(\epsilon)\ \Bigg)
\label{eq:bigFii}
\end{eqnarray}
where the hat identifies the gluons crossed to the initial state.
In $F_3^0(\hat{i},j,\hat{k})$, the only gluon that can be soft is $j$, because the initial state gluons are not allowed to be soft by kinematics, and it can also be collinear with the initial state gluons $\hat{i}$ or $\hat{k}$. This antenna can then be used with a single initial-initial mapping where $j$ is unresolved and $i$ and $k$ act as the initial state radiators and therefore does not need to be further decomposed in sub-antennae. The mapping used in this configuration is the following $\{\hat{i},j,\hat{k},\ldots,l,m,\ldots\}\to\{\hat{I},\hat{K},\ldots,\tilde{l},\tilde{m},\ldots\}$ \cite{Daleo:2006xa},
\begin{eqnarray}
p_{\hat I}^\mu \equiv \bar{p}_i^{\mu}&=&x_i\,p_i^{\mu},\nonumber\\
p_{\hat K}^\mu \equiv \bar{p}_k^{\mu}&=&x_k\,p_k^{\mu},\nonumber\\
\tilde{p}_\ell^{\mu}&=&p_\ell^{\mu}-\frac{2p_\ell\cdot(q+\tilde{q})}{(q+\tilde{q})^2}\;(q^{\mu}+\tilde{q}^{\mu})+\frac{2p_\ell\cdot q}{q^2}\;\tilde{q}^{\mu},  \label{3to2IImap}
\end{eqnarray}
where $p_{\hat I}^2=p_{\hat K}^2=\tilde{p}_\ell^2=0$ and  
$$
q^{\mu}=p_i^{\mu}+p_k^{\mu}-p_j^{\mu},\qquad\qquad
\tilde{q}^{\mu}=\bar{p}_i^{\mu}+\bar{p}_k^{\mu}
$$
and where tilde momenta $\tilde{p}_\ell$ are all the momenta in the final that are not actually part of the antenna but require boosting in order to restore momentum conservation. This is because $\tilde{q}\equiv\bar{p}_1+\bar{p}_{2}$ lies along the beam axis but the vector component of $q\equiv p_1+p_2-p_j$ is in general not along the beam axis. The $x_i$ and $x_k$ are given by \cite{Daleo:2006xa},
\begin{eqnarray}
x_i&=&\sqrt{\frac{s_{ik}+s_{jk}}{s_{ik}+s_{ij}}}\sqrt{\frac{s_{ik}+s_{ij}+s_{jk}}{s_{ik}}},\nonumber\\
x_k&=&\sqrt{\frac{s_{ik}+s_{ij}}{s_{ik}+s_{jk}}}\sqrt{\frac{s_{ik}+s_{jk}+s_{ij}}{s_{ik}}}.
\end{eqnarray}
Proper subtraction of infrared singularities requires that the momenta mapping satisfies,
\begin{eqnarray}
\bar{p}_i\rightarrow p_i\,,&&\qquad\bar{p}_k\rightarrow p_k\qquad\mbox{when $j$ becomes soft}\,,\nonumber\\
\qquad  \bar{p}_i\rightarrow (1-z_i)p_i\,,&&
\qquad\bar{p}_k\rightarrow p_k\qquad\mbox{when $j$ becomes collinear with $\hat{i}$}\,,\nonumber\\
\bar{p}_k\rightarrow (1-z_k)p_k\,,&&
\qquad\bar{p}_i\rightarrow p_i\qquad~\mbox{when $j$ becomes collinear with $\hat{k}$}\,.\nonumber
\end{eqnarray}
This transformation is not unique, however alternative
transformations are strongly constrained~\cite{Daleo:2006xa}.

The integrated form of the antenna (\ref{eq:bigFii}) for initial-initial state kinematics has been calculated and documented in \cite{Daleo:2006xa}.

\subsection{NLO corrections to dijet production}
\label{sec:NLOdijet}
The single-jet inclusive and di-jet cross sections have been studied at next-to-leading order (NLO) ~\cite{Ellis:1988hv,Ellis:1990ek,Ellis:1992en,Giele:1993dj,Giele:1994gf} and successfully compared with data from the TEVATRON. 

The two-jet contribution to the cross section from the gluon-gluon channel can be written as,
\begin{equation}
{\rm d}\sigma=\int\frac{d\xi_1}{\xi_1}\frac{d\xi_2}{\xi_2}f_g(\xi_1)f_g(\xi_2){\rm d}\hat{\sigma}_{gg}
\end{equation}
where ${\rm d}\hat{\sigma}_{gg}$ is the partonic cross section that can be calculated perturbatively in a power series of the strong coupling constant $\alpha_s$ and $f_g$ is the gluon distribution function. 
In this section, we consider the real radiative corrections to dijet production from the pure gluon channel where the antennae decompositions of the previous sections can be immediately applied. 

Following the notation of eq.~\eqref{eq:LOcross}, the Born cross section is given by,
\begin{eqnarray}
{\rm d}\hat\sigma^{B}_{LO}={\cal N}
{\rm d}\Phi_2(p_3,p_4;p_1,p_2)\frac{1}{2!}
\sum_{\sigma\in S_4/Z_4}
{A}^0_4(\sigma(1),\ldots,\sigma(4)) J_2^{(2)}(p_3,p_4)
\label{eq:born}
\end{eqnarray}
where ${A}^0_4$ is the square of the four-gluon tree-level colour ordered amplitude.
The normalisation factor ${\cal N}$ is given by,
\begin{equation}
{\cal N} =  \frac{1}{2s} \times \left(g^2N\right)^2 (N^2-1)
\times \frac{1}{4(N^2-1)^2}
\end{equation}
where $\sqrt{s}$ is the hadron-hadron centre of mass energy.

It is convenient to rewrite \eqref{eq:born} as,
\begin{eqnarray}
{\rm d}\hat\sigma^{B}_{LO}&=&{\cal N}
{\rm d}\Phi_2(p_3,p_4;p_1,p_2)\frac{1}{2!}J_2^{(2)}(p_i,p_j) \nonumber \\
&\times& 
\sum_{P(i,j)\in(3,4)}
\Bigg(2{A}^0_4(\hat{1}_g,\hat{2}_g,i_g,j_g) 
+{A}^0_4(\hat{1}_g,i_g,\hat{2}_g,j_g)\Bigg).
\label{eq:LOborn}
\end{eqnarray}

\subsubsection{NLO real corrections to dijet production}
\label{sec:NLOreal}

Similarly, we can write the five gluon real radiation cross section in the following form,
\begin{eqnarray}
&{\rm d}\hat\sigma^R_{NLO}&={\cal N}
\left(\frac{\alpha_sN}{2\pi}\right)\frac{\bar{C}(\epsilon)}{C(\epsilon)}
 {\rm d}\Phi_{3}(p_3,p_4,p_5;p_1,p_2) \,\frac{2}{3!}\nonumber\\
&&\sum_{P(i,j,k)\in(3,4,5)} \,
\Bigg(
{ A}^{0}_{5}(\hat{1}_g,\hat{2}_g,i_g,j_g,k_g)\,J_{2}^{(3)}(p_i,p_j,p_k)\;\nonumber\\
&&
\phantom{\sum_{P(i,j,k)\in(3,4,5)} \,}+{ A}^{0}_{5}(\hat{1}_g,i_g,\hat{2}_g,j_g,k_g)\, J_{2}^{(3)}(p_i,p_j,p_k)\Bigg)
\label{eq:RNLO}
\end{eqnarray}
where we absorbed the factor of $g^2$, 
\begin{eqnarray}
g^2C(\epsilon)&=&\left(\frac{\alpha_s}{2\pi}\right)\bar{C}(\epsilon)
\end{eqnarray}
and introduced useful factors of 
$C(\epsilon)$ and $\bar{C}(\epsilon) = (4\pi)^{\epsilon}e^{-\epsilon\gamma}$.

With the help of the antennae functions defined in the previous sections the subtraction term for the single unresolved configurations of the $gg\to ggg$ matrix element of (\ref{eq:RNLO}) reads,
\begin{eqnarray}
&{\rm d}\hat\sigma_{NLO}^S=&\,{\cal N}
\left(\frac{\alpha_sN}{2\pi}\right)\frac{\bar{C}(\epsilon)}{C(\epsilon)}
{\rm d}\Phi_{3}(p_{3},p_{4},p_{5};p_{1},p_{2})\,\frac{2}{3!}\, \nonumber\\
&&
\sum_{P(i,j,k)\in(3,4,5)}\Bigg\{
\phantom{+}f_3^0(\hat{2}_g,i_g,j_g){ A}_4^0(\hat{1}_g,\hat{\bar{2}}_g,(\widetilde{ij})_g,k_g)\,
{J}_{2}^{(2)}(\widetilde{p_{ij}},p_k)\nonumber\\
&&\phantom{\sum_{P(i,j,k)\in(3,4,5)} \{}+f_3^0(i_g,j_g,k_g){ A}_4^0(\hat{1}_g,\hat{2}_g,(\widetilde{ij})_g,(\widetilde{jk})_g)\,
{J}_{2}^{(2)}(\widetilde{p_{ij}},\widetilde{p_{jk}})\nonumber\\
&&\phantom{\sum_{P(i,j,k)\in(3,4,5)} \{}+f_3^0(j_g,k_g,\hat{1}_g){ A}_4^0(\hat{\bar{1}}_g,\hat{2}_g,i_g,(\widetilde{kj})_g)\,
{J}_{2}^{(2)}(p_i,\widetilde{p_{kj}})\nonumber\\
&&\nonumber\\
&&\phantom{\sum_{P(i,j,k)\in(3,4,5)} \{}+F_3^0(\hat{1}_g,i_g,\hat{2}_g){ A}_4^0(\hat{\bar{1}}_g,\hat{\bar{2}}_g,\tilde{j}_g,\tilde{k}_g)\,
{J}_{2}^{(2)}(\tilde{p_j},\tilde{p_k})\nonumber\\
&&\phantom{\sum_{P(i,j,k)\in(3,4,5)} \{}+f_3^0(\hat{2}_g,j_g,k_g){ A}_4^0(\hat{1}_g,i_g,\hat{\bar{2}}_g,(\widetilde{jk})_g)\,
{J}_{2}^{(2)}(p_i,\widetilde{p_{jk}})\nonumber\\
&&\phantom{\sum_{P(i,j,k)\in(3,4,5)} \{}+f_3^0(j_g,k_g,\hat{1}_g){ A}_4^0(\hat{\bar{1}}_g,i_g,\hat{2}_g,(\widetilde{kj})_g)\,
{J}_{2}^{(2)}(p_i,\widetilde{p_{kj}})\Bigg\}
\label{eq:SNLO}
\end{eqnarray}
where we have used a combination of gluon-gluon-gluon antenna functions with two emitters in the final state, initial state and one emitter in the final state and one in the initial state. Each of these antennae has a reduced matrix element evaluated with hard momenta given by the momentum mappings of the previous sections. The total number of antenna functions used is equal to the total number of unresolved particles in the final state per colour ordered amplitude.

This subtraction term has been checked with phase space trajectories generated with {\tt RAMBO} \cite{Stirling:1986ab}. In each singular region the ratio of the matrix element with the subtraction approaches unity.

\subsubsection{NLO virtual corrections to dijet production}
\label{sec:NLOvirtual}
The virtual contribution has the following form,
\begin{eqnarray}
{\rm d}\hat\sigma^{V}_{NLO}={\cal N}\left(\frac{\alpha_sN}{2\pi}\right)
{\rm d}\Phi_2(p_3,p_4;p_1,p_2)\frac{1}{2!}
\sum_{\sigma\in S_4/Z_4}
{A}^1_4(\sigma(1),\ldots,\sigma(4)) J_2^{(2)}(p_3,p_4).
\label{eq:NLOvirtual}
\end{eqnarray}
Introducing the colour ordered infrared singularity operator~\cite{Catani:1998bh}
\begin{eqnarray}
{\bom I}_{gg}^{(1)}(s_{gg})=-\frac{e^{\epsilon\gamma}}{2\Gamma(1-\epsilon)}\left[\frac{1}{\epsilon^2}+
\frac{11}{6\epsilon}\right]\Re\left(-\frac{s_{gg}}{\mu^2}\right)^{-\epsilon}
\end{eqnarray}
we can rewrite the singular part
of the renormalised loop amplitude as~\cite{Bern:1991aq},
\begin{eqnarray}
{\rm d}\hat\sigma_{NLO}^V&=&{\cal N}\left(\frac{\alpha_sN}{2\pi}\right){\rm d}\Phi_2(p_3,p_4;p_1,p_2)\bar{C}(\epsilon)
2\Bigg\{\nonumber\\
&&\phantom{+}\left[{\bom I}_{gg}^{(1)}(s_{12})+{\bom I}_{gg}^{(1)}(s_{23})+{\bom I}_{gg}^{(1)}(s_{34})+{\bom I}_{gg}^{(1)}(s_{14})\right] 
2{  A}_4^0(\hat{1}_g,\hat{2}_g,3_g,4_g)J_2^{(2)}(p_3,p_4)\nonumber\\
&&+\left[{\bom I}_{gg}^{(1)}(s_{13})+{\bom I}_{gg}^{(1)}(s_{23})+{\bom I}_{gg}^{(1)}(s_{24})+{\bom I}_{gg}^{(1)}(s_{14})\right] 
{  A}_4^0(\hat{1}_g,3_g,\hat{2}_g,4_g)J_2^{(2)}(p_3,p_4)\Bigg\}\nonumber\\
&&+{\cal O}(\epsilon^0).
\label{eq:NLOvirtualpoles}
\end{eqnarray}

\subsubsection{Cancellation of infrared divergences}
We will now collect the leading poles of the integrated antenna functions used to write down 
the subtraction term ${\rm d}\hat\sigma^S$~\cite{GehrmannDeRidder:2005cm,Daleo:2006xa}:
\begin{eqnarray}
&&\frac{1}{C(\epsilon)}\int {\rm d}\Phi_{123}\;f_3^0(1,2,3)=-2{\bom I}_{gg}^{(1)}(s_{123})+{\cal O}(\epsilon^0)\nonumber\\
&&\frac{1}{C(\epsilon)}\int {\rm d}\Phi_{1,23}\;f_3^0(\hat{1},2,3)=-2{\bom I}_{gg}^{(1)}(Q^2)\delta(1-x_1)-\left(\frac{Q^2}{\mu^2}\right)^{-\epsilon}\frac{1}{2\epsilon}p_{gg}^{(0)}(x_1)
+{\cal O}(\epsilon^0)\nonumber\\
&&\frac{1}{C(\epsilon)}\int {\rm d}\Phi_{12,3}\;F_3^0(\hat{1},3,\hat{2})=-{\bom I}_{gg}^{(1)}(Q^2)\delta(1-x_1)\delta(1-x_2)
-{\bom I}_{gg}^{(1)}(Q^2)\delta(1-x_2)\delta(1-x_1)\nonumber\\
&&-\left(\frac{Q^2}{\mu^2}\right)^{-\epsilon}\frac{1}{2\epsilon}p_{gg}^{(0)}(x_1)\delta(1-x_2)
-\left(\frac{Q^2}{\mu^2}\right)^{-\epsilon}\frac{1}{2\epsilon}
p_{gg}^{(0)}(x_2)\delta(1-x_1)+{\cal O}(\epsilon^0)
\label{eq:Xint}
\end{eqnarray}
where the colour ordered splitting kernel is,
\begin{eqnarray}
p_{gg}^{(0)}(x)=\frac{11}{6}\delta(1-x)+\left(\frac{2}{1-x}\right)_++\frac{2}{x}-4+2x-2x^2.
\end{eqnarray}
The analytic integration of the subtraction term \eqref{eq:SNLO} over the factorised phase space can be carried out using the results of \eqref{eq:Xint}. Using the symmetry of the gluon phase space the 3! permutations in (\ref{eq:SNLO}) precisely cancels the identical particle factor of 1/3!. After relabeling the final state particles we obtain the following pole structure for the integrated counterterm:
\begin{eqnarray}
\lefteqn{\int_1 {\rm d}\hat\sigma^{S}_{NLO}={\cal N}
\left(\frac{\alpha_sN}{2\pi}\right)\bar{C}(\epsilon)
2\,J_2^{(2)}(p_3,p_4)\,\Bigg\{}\nonumber\\
&&\phantom{+}{\rm d}\Phi_2(p_3,p_4;p_1,p_2)2{  A}_4^{(0)}(\hat{1}_g,\hat{2}_g,3_g,4_g) \left[-{\bom I}_{gg}^{(1)}(s_{23})-{\bom I}_{gg}^{(1)}(s_{34})-{\bom I}_{gg}^{(1)}(s_{14})-{\bom I}_{gg}^{(1)}(s_{12})\right]\nonumber\\
&&-{\rm d}\Phi_2(p_3,p_4;\bar{p}_1,p_2)2{  A}_4^{0}(\hat{\bar{1}}_g,\hat{2}_g,3_g,4_g) 
\left(\frac{Q^2}{\mu^2}\right)^{-\epsilon}\int \frac{dx_1}{x_1}\frac{1}{2\epsilon}p_{gg}^{(0)}(x_1)\nonumber\\
&&-{\rm d}\Phi_2(p_3,p_4;p_1,\bar{p}_2)2{  A}_4^{0}(\hat{1}_g,\hat{\bar{2}}_g,3_g,4_g) 
\left(\frac{Q^2}{\mu^2}\right)^{-\epsilon}\int \frac{dx_2}{x_2}\frac{1}{2\epsilon}p_{gg}^{(0)}(x_2)\nonumber\\
&&+{\rm d}\Phi_2(p_3,p_4;p_1,p_2){ A}_4^{0}(\hat{1}_g,3_g,\hat{2}_g,4_g) \left[-{\bom I}_{gg}^{(1)}(s_{24})-{\bom I}_{gg}^{(1)}(s_{14})-{\bom I}_{gg}^{(1)}(s_{13})-{\bom I}_{gg}^{(1)}(s_{23})\right]\nonumber\\
&&-{\rm d}\Phi_2(p_3,p_4;\bar{p}_1,p_2){  A}_4^{0}(\hat{\bar{1}}_g,3_g,\hat{2}_g,4_g) 
\left(\frac{Q^2}{\mu^2}\right)^{-\epsilon}\int \frac{dx_1}{x_1}\frac{1}{2\epsilon}p_{gg}^{(0)}(x_1)\nonumber\\
&&-{\rm d}\Phi_2(p_3,p_4;p_1,\bar{p}_2){  A}_4^{0}(\hat{1}_g,3_g,\hat{\bar{2}}_g,4_g) 
\left(\frac{Q^2}{\mu^2}\right)^{-\epsilon}\int \frac{dx_2}{x_2}\frac{1}{2\epsilon}p_{gg}^{(0)}(x_2). 
\label{eq:dSint}
\end{eqnarray}
The terms containing the singularity operator ${\bom I}_{gg}^{(1)}$ in eq.~\eqref{eq:dSint} match exactly those appearing with opposite sign in the virtual contribution in eq.~\eqref{eq:NLOvirtualpoles}. The remaining poles correspond to the mass factorisation contribution and are cancelled by the NLO mass factorisation counter term given by \cite{Catani:1996vz},
\begin{eqnarray}
{\rm d}\hat\sigma_{NLO}^{MF}&=&
-\left(\frac{\alpha_sN}{2\pi}\right)\bar{C}(\epsilon) 2
\int \frac{{\rm d}x_1}{x_1}\frac{{\rm d}x_2}{x_2}
\,{\rm d}\hat\sigma^B_{LO}(\bar{p}_1,\bar{p}_2)\Bigg\{\nonumber\\
&&\phantom{+}\delta(1-x_2)\left[-\frac{1}{2\epsilon}\left(\frac{\mu^2}{\mu_F^2}\right)^{\epsilon}
p_{gg}^{(0)}(x_1)+K_{F.S.}^{gg}(x_1)\right]\nonumber\\
&&+\delta(1-x_1)\left[-\frac{1}{2\epsilon}\left(\frac{\mu^2}{\mu_F^2}\right)^{\epsilon}
p_{gg}^{(0)}(x_2)+K_{F.S.}^{gg}(x_2)\right] + {\cal O}(\epsilon)\Bigg\}
\label{eq:MFNLO}
\end{eqnarray}
where ${\rm d}\hat\sigma^B_{LO}(\bar{p}_1,\bar{p}_2)$ is obtained by evaluating eq.~\eqref{eq:LOborn} with the scaled initial momenta $\bar{p}_1$ and $\bar{p}_2$. The actual form of the kernel $K_{F.S.}^{gg}(x)$ specifies the factorisation scheme. Setting $K_{F.S.}^{gg}(x)=0$ defines the $\overline{\rm{ MS}}$ factorisation scheme. Combining eqs.~\eqref{eq:NLOvirtualpoles}, \eqref{eq:dSint} and \eqref{eq:MFNLO} we achieve a complete cancellation of the infrared singularities to NLO.

\section{Antenna subtraction for double real radiation at NNLO}
\label{sec:NNLOantenna}

Let us now consider the construction of the subtraction term 
for the double radiation
contribution ${\rm d}\hat\sigma^{S}_{NNLO}$, which shall correctly subtract all 
single and double unresolved singularities contained in the $(m+4)$-parton 
real 
radiation contribution to $m$-jet final states in $pp$ collisions,
\begin{eqnarray}
\lefteqn{{\rm d}\hat\sigma^{R}_{NNLO}
= {\cal N}\,\sum_{\textrm{perms}}{\rm d}\Phi_{m+2}(p_3,\ldots,p_{m+4};p_1,p_2)
\frac{1}{S_{{m+2}}} }\nonumber \\ &\times&
|{\cal M}_{m+4}(p_1,\ldots,p_{m+4})|^{2}\;
J_{m}^{(m+2)}(p_3,\ldots,p_{m+4})\;.
\label{eq:nnloreal}
\end{eqnarray}
Single real radiation 
singularities correspond to one parton becoming soft or collinear, while 
double real radiation singularities occur if two partons become soft or
collinear simultaneously. Singular terms in these limits can be identified 
by requiring a minimum number of invariants tending to zero in a given 
kinematical configuration. This number depends on the limit under 
consideration and follows from the phase space volume available to a given 
configuration. A detailed discussion of the 
kinematical definition of double unresolved 
limits is available in \cite{GehrmannDeRidder:1997gf,Campbell:1997hg,Catani:1998nv,Catani:1999ss}. 

We must distinguish the following 
configurations according to the colour connection of the unresolved partons:
\begin{itemize}
\item[(a)] One unresolved parton but the experimental observable selects only
$m$ jets.
\item[(b)] Two colour-connected unresolved partons (colour-connected).
\item[(c)] Two unresolved partons that are not colour connected but share a common
radiator (almost colour-unconnected).
\item[(d)] Two unresolved partons that are well separated from each other 
in the colour 
chain (colour-unconnected).
\end{itemize}

For each configuration mentioned the subtraction formula has a characteristic antenna structure. Therefore in the following subsections we will discuss the individual formulae for each of the configurations (a) to (d).

The first configuration  
was treated already in the context of antenna subtraction at NLO in 
sections~\ref{sec:NLOff}, \ref{sec:NLOif}, \ref{sec:NLOii}. In the context of the construction of 
 ${\rm d}\hat\sigma^{S}_{NNLO}$, the same single-particle subtraction terms 
can be used. These do however not yet guarantee a 
finite $(m+4)$-parton contribution in all 
single unresolved regions for two reasons:
\begin{enumerate}
\item[(1)] while the jet function in 
${\rm d}\hat\sigma^{S}_{NLO}$ ensures that the subtraction term is non-zero only
in the single unresolved limit it was constructed for, this is no
longer the case for single unresolved 
radiation at NNLO; 
\item[(2)] the subtraction terms for the remaining three double 
unresolved configurations will in general be singular in the single unresolved
regions, where they do not match the matrix element. 
\end{enumerate}
Both of these problems will be addressed below.

\begin{figure}[t!] 
\epsfig{file=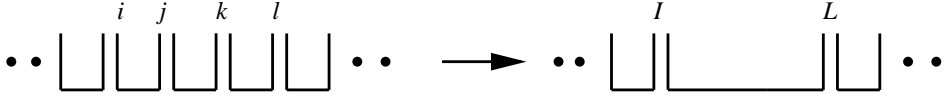,height=1.2cm}
\caption[Colour connected configuration]{Colour connection of the partons showing the parent and daughter
partons for the double unresolved antenna.}
\label{fig:double}
\end{figure}

\begin{figure}[t!] 
\epsfig{file=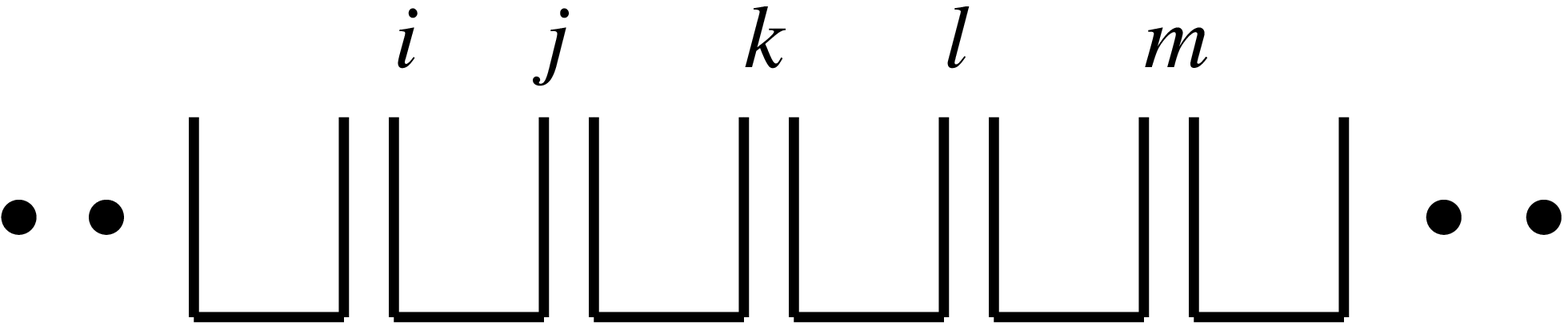,height=1.2cm}
\caption[Almost colour connected configuration]{Colour connection of the partons showing the parent and daughter
partons for two adjacent single unresolved antennae.}
\label{fig:overlap}
\end{figure}

\begin{figure}[t!]
\epsfig{file=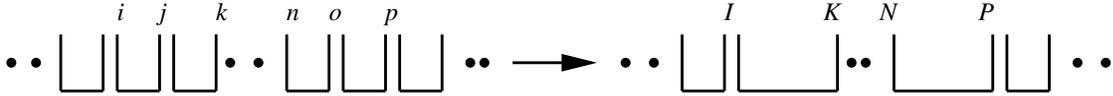,height=1.2cm}
\caption[Colour unconnected configuration]{Colour connection of the partons showing the parent and daughter
partons for two disconnected single unresolved antennae.}
\label{fig:nonoverlap}
\end{figure}

The remaining
 three configurations (b)--(d) are illustrated in Figures~\ref{fig:double}, 
\ref{fig:overlap} and \ref{fig:nonoverlap}. The singular behaviour 
of the full $(m+4)$-parton matrix 
element in these configurations is the product
of double unresolved factors and 
reduced $(m+2)$-parton matrix elements. Subtraction terms for all these 
configurations can be constructed using either 
single four-parton antenna functions or products of two three-parton antenna 
functions. In all cases, attention has to be paid to the matching of 
different double and single unresolved regions. 
This problem has been addressed already in publications 
on subtraction at 
NNLO~\cite{Kosower:2002su,Weinzierl:2003fx,GehrmannDeRidder:2004tv,Kilgore:2004ty,Frixione:2004is,Somogyi:2005xz}, and a concise discussion can be found in~\cite{Somogyi:2005xz}.

In the following, we construct the subtraction
terms for all four configurations.

\subsection{Subtraction terms for single unresolved partons ${\rm d}\hat\sigma_{NNLO}^{S,a}$}
The starting point for the subtraction terms for single unresolved partons 
are the NLO single unresolved antenna subtraction terms (\ref{eq:sub1}), (\ref{eq:subif}), (\ref{eq:subii}) with one additional particle, 
\begin{eqnarray}
\lefteqn{{\rm d}\hat\sigma_{NNLO}^{S,a,(ff)}
=  {\cal N}\,\sum_{\textrm{perms}}{\rm d}\Phi_{m+2}(p_{3},\ldots,p_{m+4};p_1,p_2)
\frac{1}{S_{{m+2}}} }\nonumber \\
&\times& \,  \sum_{j}\;X^0_{ijk}\,
|{\cal M}_{m+3}(\ldots,I,K,\ldots)|^2 J_{m}^{(m+1)}(p_{3},\ldots,p_I,p_K,\ldots,p_{m+4})\;
,  \\
\label{eq:sub2aff}
\lefteqn{{\rm d}\hat\sigma_{NNLO}^{S,a,(if)}
={\cal N}\,\sum_{\textrm{perms}}{\rm d}\Phi_{m+2}(p_3,\ldots,p_{m+4};p_1,p_2)
  \frac{1}{S_{m+2}} }\nonumber\\
&\times&\, \sum_{{i}=1,2}\sum_{j} X^{0}_{i,jk}\,
 |{\cal M}_{m+3}(\ldots,\hat{I},K,\ldots)|^2 J^{(m+1)}_{m}(p_{3},\ldots,p_K,\ldots,p_{m+4})\;,\\
\label{eq:sub2aif}
\lefteqn{{\rm d}\hat\sigma_{NNLO}^{S,a,(ii)}
={\cal N}\, \sum_{\textrm{perms}}{\rm d}\Phi_{m+2}(p_{3},\ldots,p_{m+4};p_1,p_2)
  \frac{1}{S_{m+2}}}\nonumber \\
&\times&\, \sum_{i,k=1,2} \sum_{j} X^{0}_{ik,j}
  |{\cal M}_{m+3}(\ldots,\hat{I},\hat{K},\ldots)|^2
  \,J^{(m+1)}_{m}(\tilde{p}_3,\ldots,\tilde{p}_{m+4})\;, 
\label{eq:sub2aii}
\end{eqnarray}
where the NLO jet function $J_{m}^{(m)}$ is now replaced by 
$J_{m}^{(m+1)}$. The sum over $j$ is the sum over all unresolved partons in a colour ordered amplitude between radiators $i$ and $k$ which can be both located in the final state (\ref{eq:sub2aff}), $\hat{i}$ in the initial state and $k$ in the final state (\ref{eq:sub2aif}) or both in the initial state (\ref{eq:sub2aii}). Their position defines the type of the three-parton antenna which is used $X_{ijk}$, $X_{i,jk}$ or $X_{ij,k}$ respectively and the mapping used to generate the momenta in the reduced colour ordered amplitude.

When $j$ is unresolved, ${\rm d}\hat\sigma_{NNLO}^{S,a}$ coincides with the matrix element (\ref{eq:nnloreal}). However at NNLO the jet function $J_m^{(m+1)}$ allows one of the $(m+1)$ momenta to become unresolved. In this limit ${\rm d}\hat\sigma_{NNLO}^{S,a}$ does not coincide with the matrix element (\ref{eq:nnloreal}). We distinguish two cases: (1) when one of the new momenta, $p_I$ or $p_K$, becomes unresolved and (2) where any other momentum $p_o$ or $\tilde{p}_o$ becomes unresolved.

Case (1) is necessarily a double unresolved limit since the new momenta, $p_I$ or $p_K$, are linear combinations of two momenta and we discuss it below.

In case (2), ${\rm d}\hat\sigma_{NNLO}^{S,a}$ becomes singular as $p_o$ or $\tilde{p}_o$ become unresolved and it does not coincide with the limit of the full $(m+4)$-parton matrix element. However if we take this limit, we find that ${\rm d}\hat\sigma_{NNLO}^{S,a}$ collapses into the product of two almost colour-connected or colour-unconnected antenna functions multipled by reduced $(m+2)$-parton matrix elements which coincide with the structures (c) and (d) that we will define below. This means that this spurious limit cancels exactly against ${\rm d}\hat\sigma_{NNLO}^{S,c}$ and ${\rm d}\hat\sigma_{NNLO}^{S,d}$.

For the double unresolved limits, we have on the one hand the limit where one of the new momenta, $p_I$ or $p_K$, is unresolved, or the colour-neighbouring limit, where two pairs of momenta become independently collinear but one pair lies inside the antenna while the other pair consists of the remaining antenna momentum and its colour-connected neighbour. Each of these appears twice in the sum over $j$ giving the two possibilities of attributing the inside/outside pair. These spurious limits will cancel exactly against similar terms that also appear twice in the structure (b) of ${\rm d}\hat\sigma_{NNLO}^{S,b}$ that we will define below. Any other colour connected unresolved configuration vanishes.

On the other hand, double unresolved limits involving $p_j$ and any other momenta $p_o$ in the reduced matrix element that are almost colour connected or colour unconnected are not vanishing in ${\rm d}\hat\sigma_{NNLO}^{S,a}$. In fact they yield exactly twice these double unresolved limits of the $(m+4)$-parton matrix element because the role of $p_j$ and $p_o$ can be interchanged and having $j=o$ in the sum results in two identical terms contributing to the same limit. The formulae for ${\rm d}\hat\sigma_{NNLO}^{S,c}$ and ${\rm d}\hat\sigma_{NNLO}^{S,d}$ will be defined to compensate this oversubtraction.
 
The analytic integration of this contribution proceeds by using the formulae for the phase space factorisations given in sections \ref{sec:NLOff}, \ref{sec:NLOif}, \ref{sec:NLOii} and leads to the same results for the integrated antennae as at NLO. 

\subsection{Subtraction terms for two colour-connected unresolved partons ${\rm d}\hat\sigma_{NNLO}^{S,b}$}
When two unresolved partons $j$ and $k$ are adjacent, we construct the subtraction term starting from the four-particle tree-level antennae $X_{ijkl}$, $X_{i,jkl}$, $X_{il,jk}$. By construction they contain all colour connected double unresolved limits of the $(m+4)$-parton matrix element associated with partons $j$ and $k$ unresolved between radiators $i$ and $l$. However this antenna can also become singular in \textit{single} unresolved limits associated with $j$ or $k$ where it does not coincide with limits of the matrix element. To ensure a finite 
subtraction term in all these single unresolved limits, we therefore
subtract the 
appropriate limits of the four-particle tree antennae, which are products of 
two tree-level three-particle antennae.  As in the single unresolved case, we replace the original hard radiators with new particles, $I$ and $L$. When one of the hard radiators is in the initial state, $p_{\hat{I}} \equiv \hat{\bar{p}}_i = x_i p_i$ and when both are in the initial state,
$p_{\hat{I}} \equiv \hat{\bar{p}}_i = x_i p_i$, $p_{\hat{L}} \equiv \hat{\bar{p}}_l = x_l p_l$ and all other momenta have to be Lorentz boosted.

The colour-connected double 
subtraction term reads:
\begin{eqnarray}
\lefteqn{{\rm d}\hat\sigma_{NNLO}^{S,b,(ff)}
=  {\cal N}\,\sum_{\textrm{perms}}{\rm d}\Phi_{m+2}(p_{3},\ldots,p_{m+4};p_1,p_2)
\frac{1}{S_{{m+2}}}\,}\nonumber \\
&\times&  \sum_{jk}\left( X^0_{ijkl}
- X^0_{ijk} X^0_{IKl} - X^0_{jkl} X^0_{iJL} \right)\nonumber\\
&\times&|{\cal M}_{m+2}(\ldots,I,L,\ldots)|^2\,
J_{m}^{(m)}(\ldots,p_{I},p_{L},\ldots),\;
\label{eq:sub2bff}\\
\lefteqn{{\rm d}\hat\sigma_{NNLO}^{S,b,(if)}
=  {\cal N}\,\sum_{\textrm{perms}}{\rm d}\Phi_{m+2}(p_{3},\ldots,p_{m+4};p_1,p_2)
\frac{1}{S_{{m+2}}} }\nonumber \\
&\times&\sum_{i=1,2} \sum_{jk}\;\left( X^0_{i,jkl}
- X^0_{i,jk} X^0_{I,Kl} - X^0_{jkl} X^0_{i,JL} \right)\nonumber\\
&\times&|{\cal M}_{m+2}(\ldots,\hat{I},L,\ldots)|^2\,
J_{m}^{(m)}(\ldots,p_{L},\ldots) 
 \;,
\label{eq:sub2bif}\\
\lefteqn{{\rm d}\hat\sigma_{NNLO}^{S,b,(ii)}
=  {\cal N}\,\sum_{\textrm{perms}}{\rm d}\Phi_{m+2}(p_{3},\ldots,p_{m+4};p_1,p_2)
\frac{1}{S_{{m+2}}} }\nonumber \\
&\times& \,\sum_{il=1,2} \sum_{jk}\;\left( X^0_{il,jk}
- X^0_{l,jk} X^0_{iL,K} - X^0_{i,kj} X^0_{Il,J} \right)\nonumber \\
&\times&
|{\cal M}_{m+2}(\ldots,\hat{I},\hat{L},\ldots)|^2\,
J_{m}^{(m)}(\tilde{p}_3,\ldots,\tilde{p}_{m+4})\;,
\label{eq:sub2bii}
\end{eqnarray}
where the sum runs over all colour-adjacent pairs $j,k$ and implies the 
appropriate selection of hard momenta $i,l$ which as usual have three possible assignments of radiators. In all cases the $(m+2)$-parton matrix element is evaluated with new on-shell momenta given by a momentum mapping that we will discuss in section \ref{sec:NNLOnimplemtation} when we describe the numerical implementation of this formula.

The products of three-parton antenna functions in ${\rm d}\hat\sigma_{NNLO}^{S,b}$ subtract the singular limits of the associated four parton antenna and each contribute equally in the colour-neighbouring configuration and spurious limits of ${\rm d}\hat\sigma_{NNLO}^{S,a}$ discussed in the previous section. In all genuinely colour-connected limits, the four-parton antenna functions correctly match the singularity structure of the $(m+4)$-parton matrix element (\ref{eq:nnloreal}). Singularities in the $(m+2)$-parton matrix element itself are forbidden by the jet function. 

\begin{figure}[t!]
\begin{center}
\includegraphics[width=0.8\textwidth]{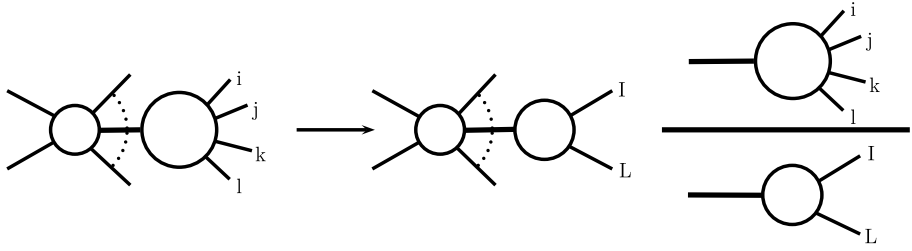}
\caption[Colour connected final-final antenna/phase space factorisation]{Illustration of NNLO antenna factorisation representing the
factorisation of both the squared matrix elements and the $(m+2)$-particle 
phase
space when the unresolved particles $j$ and $k$ are colour connected between two final state radiators $i$ and $l$.}
\label{fig:bfffact}
\end{center}
\end{figure}

\begin{figure}[t!]
\begin{center}
\includegraphics[width=0.8\textwidth]{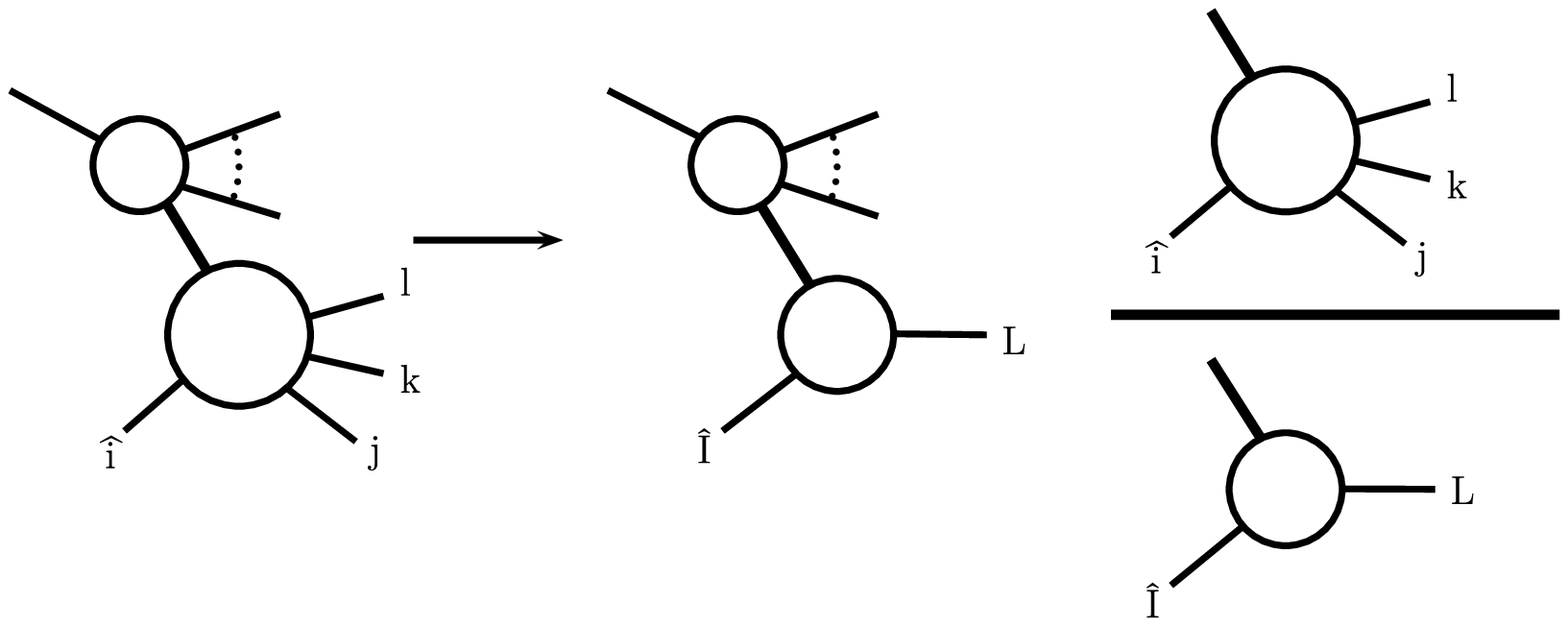}
\caption[Colour connected initial-final antenna/phase space factorisation]{Illustration of NNLO antenna factorisation representing the
factorisation of both the squared matrix elements and the $(m+2)$-particle 
phase
space when the unresolved particles $j$ and $k$ are colour connected between an initial state radiator $\hat{i}$ and a final state radiator $l$.}
\label{fig:biffact}
\end{center}
\end{figure}

\begin{figure}[t!]
\begin{center}
\includegraphics[width=0.8\textwidth]{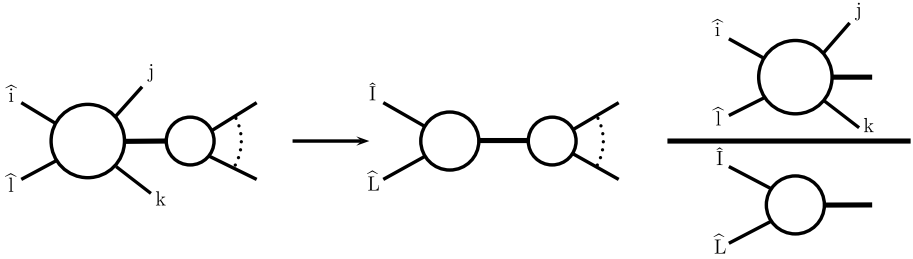}
\caption[Colour connected initial-initial antenna/phase space factorisation]{Illustration of NNLO antenna factorisation representing the
factorisation of both the squared matrix elements and the $(m+2)$-particle 
phase
space when the unresolved particles $j$ and $k$ are colour connected between two initial state radiators $\hat{i}$ and $\hat{l}$.}
\label{fig:biifact}
\end{center}
\end{figure}

The analytic integration of this counterterm follows from the antenna factorisation of both the squared matrix elements and the $(m+2)$-particle phase space shown in figures \ref{fig:bfffact}, \ref{fig:biffact} and \ref{fig:biifact}.

The factorisation of the phase space for final-final, initial-final and initial-initial antennae reads respectively,
\begin{eqnarray}
\label{eq:psx4}
{\rm d} \Phi_{m+2}(p_3,\ldots,p_{m+4};p_1,p_2)&=& 
{\rm d} \Phi_{m}(p_{3},\ldots,p_I,p_L,\ldots,p_{m+4};p_1,p_2)\nonumber\\
&\times&{\rm d} \Phi_{X_{ijkl}} (p_i,p_j,p_k,p_l;p_I+p_L)\;,\\
\label{eq:psif4}
{\rm d}\Phi_{m+2}(p_3,\ldots,p_{m+4};p_1,p_2)&=&{\rm d}\Phi_{m}(p_3,\dots,p_L,\dots,p_{m+4};p_1,x_2p_2)\nonumber\\
&\times&\frac{Q^2}{2\pi}{\rm d}\Phi_{3}(p_j,p_k,p_l;p_2,q)\frac{{\rm d} x_2}{x_2}\,,\\
\label{eq:psii4}
{\rm d}\Phi_{m+2}(p_3,\ldots,p_{m+4};p_1,p_2)&=&
{\rm d}\Phi_{m}(\tilde{p}_3,\ldots,\tilde{p}_{m+4};x_1p_1,x_2p_2)
\nonumber\\
&&\times\delta(x_1-\hat{x}_1)\,\delta(x_2-\hat{x}_2)\,[{\rm d}p_j][{\rm d}p_k]\, {\rm d}x_1\, {\rm d}x_2\,,
\end{eqnarray}
where in (\ref{eq:psif4}) $Q^2=-q^2$, $q=p_j+p_k+p_l-p_2$. A similar factorisation holds  with $(1\leftrightarrow2)$  for initial state singularities with parton 1. 
Using (\ref{eq:psx4}), (\ref{eq:psif4}), (\ref{eq:psii4}) we can rewrite each of the genuine four-particle subtraction terms in the form,
\begin{eqnarray}
&&|{\cal M}_{m+2}|^2\,
J_{m}^{(m)}\; 
{\rm d}\Phi_{m}
\int {\rm d} \Phi_{X_{ijkl}}\;X^0_{ijkl},\\
&&|{\cal M}_{m+2}|^2\,
J_{m}^{(m)}\; 
{\rm d}\Phi_{m}
\int \frac{Q^2}{2\pi} {\rm d}\Phi_{3}(p_j,p_k,p_l;p,q)\;X^0_{i,jkl}\frac{{\rm d}x}{x},\\
&&|{\cal M}_{m+2}|^2\,
J_{m}^{(m)}\; 
{\rm d}\Phi_{m}
\int [{\rm d} p_j][{\rm d} p_k] \delta(x_1-\hat{x}_1)\,\delta(x_2-\hat{x}_2)\;X^0_{il,jk}  {\rm d}x_1\,  {\rm d}x_2.
\end{eqnarray}
The antennae integrals can be worked out separately once and for all to become universal building blocks for subtraction at NNLO. The integrated antenna is the antenna function integrated over the fully inclusive antenna phase space including a normalisation factor to account for powers of the QCD coupling constant,
\begin{eqnarray}
\label{eq:x4intff}
&&{\cal X}^0_{ijkl} = \frac{1}{[C(\epsilon)]^2}
\int {\rm d} \Phi_{X_{ijkl}}\;X^0_{ijkl},\\
\label{eq:x4intif}
&&{\cal X}^0_{i,jkl}(x_i)=\frac{1}{[C(\epsilon)]^2}\int {\rm d}\Phi_3 \frac{Q^2}{2\pi} X^0_{i,jkl},\\
\label{eq:x4intii}
&&{\cal X}^0_{il,jk}(x_i,x_l)=\frac{1}{[C(\epsilon)]^2}\int [{\rm d} p_j][{\rm d} p_k]\;x_i\;x_l\; \delta(x_i-\hat{x}_i)\,\delta(x_l-\hat{x}_l)\,X^0_{il,jk},\,
\end{eqnarray}
where $C(\epsilon)$ is given in (\ref{eq:ceps}). These integrations are performed analytically in $d$ dimensions to make the infrared singularities explicit. Using the techniques in \cite{GehrmannDeRidder:2003bm} all integrated antennae in (\ref{eq:x4intff}) were obtained and are documented in \cite{GehrmannDeRidder:2005cm}. The integrated initial-final antennae of (\ref{eq:x4intif}) were computed recently in \cite{Daleo:2009yj,Daleo:2010tz}. The remaining initial-initial integrated antennae functions of (\ref{eq:x4intii}) are presently unknown. Work on their analytic evaluation is underway~\cite{Boughezal:2010ty}.

\subsection{Subtraction terms for two almost colour-unconnected unresolved partons ${\rm d}\hat\sigma_{NNLO}^{S,c}$}
There are double unresolved configurations where the unresolved partons are separated  by a hard radiator parton, for example, $i,j,k,l,m$ where $j$ and $l$ are unresolved. In this case we take the strongly ordered approach where $i,j,k$ form an antenna with hard partons $I$ and $K$ yielding an ordered amplitude involving $I,K,l,m$. As usual, the momenta of the hard radiator partons $I$ and $K$ are constructed from $p_i$, $p_j$, $p_k$. The cases where $l$ is unresolved are then treated using an antenna $K,l,m$  with hard partons $K^\prime$ and $M^\prime$. The momenta of the hard radiator partons $K^\prime$ and $M^\prime$ are made from $p_K$, $p_l$ and $p_m$. The other case where first $k,l,m$ form an antenna followed by $i,j,K$ is also included where the momenta are obtained by iterative use of the NLO momentum mappings of sec.~\ref{sec:NLOmap}.

In this configuration there is a common radiator that can be in the final or the initial state. The subtraction terms for the almost colour-connected configuration read,
\begin{eqnarray}
\lefteqn{{\rm d}\hat\sigma_{NNLO}^{S,c,(ff)}
= - {\cal N}\,\sum_{\textrm{perms}}{\rm d}\Phi_{m+2}(p_{3},\ldots,p_{m+4};p_1,p_2)
\frac{1}{S_{{m+2}}}} \nonumber \\
&\times& \,\Bigg [  \sum_{j,l}\;X^0_{ijk}\;x^0_{mlK}\,
|{\cal M}_{m+2}(\ldots,I,K^\prime,M^\prime,\ldots)|^2\,
J_{m}^{(m)}(p_{3},\ldots,p_{I},p_{K^\prime},p_{M^\prime},\ldots,p_{m+4})\;
\phantom{\Bigg]}
\nonumber \\
&& \,+ \sum_{j,l}\;X^0_{klm}\;x^0_{ijK}\,
|{\cal M}_{m+2}(\ldots,I^\prime,K^\prime,M,\ldots)|^2\,
J_{m}^{(m)}(p_{3},\ldots,p_{I^\prime},p_{K^\prime},p_{M},\ldots,p_{m+4})\;
\Bigg
]\;,\nonumber\\
\label{eq:sub2cff}\\
\lefteqn{{\rm d}\hat\sigma_{NNLO}^{S,c,(if)}
= - {\cal N}\,\sum_{\textrm{perms}}{\rm d}\Phi_{m+2}(p_{3},\ldots,p_{m+4};p_1,p_2)
\frac{1}{S_{{m+2}}}} \nonumber \\
&\times& \,\Bigg[  \sum_{i=1,2}
\sum_{j,l}\;X^0_{i,jk}\;x^0_{mlK}\,
|{\cal M}_{m+2}(\ldots,\hat{I},K^\prime,M^\prime,\ldots)|^2\,
J_{m}^{(m)}(p_{3},\ldots,p_{K^\prime},p_{M^\prime},\ldots,p_{m+4})\;
\phantom{\Bigg]}
\nonumber \\
&& \,+ \sum_{i=1,2}\sum_{j,l}\;X^0_{klm}\;x^0_{i,jK}\,
|{\cal M}_{m+2}(\ldots,\hat{I}^\prime,K^\prime,M,\ldots)|^2\,
J_{m}^{(m)}(p_{3},\ldots,p_{K^\prime},p_{M},\ldots,p_{m+4})\;
\phantom{\Bigg]},\nonumber\\ 
&&\,+  \sum_{k=1,2}\sum_{j,l}\;X^0_{k,ji}\;x^0_{K,lm}\,
|{\cal M}_{m+2}(\ldots,I,\hat{K}^\prime,M^\prime,\ldots)|^2\,
J_{m}^{(m)}(p_{3},\ldots,p_{I},p_{M^\prime},\ldots,p_{m+4})\;
\phantom{\Bigg]}
\nonumber \\
&& \,+ \sum_{k=1,2} \sum_{j,l}\;X^0_{k,lm}\;x^0_{K,ji}\,
|{\cal M}_{m+2}(\ldots,I^\prime,\hat{K}^\prime,M,\ldots)|^2\,
J_{m}^{(m)}(p_{3},\ldots,p_{I^\prime},p_{M},\ldots,p_{m+4})\;
\Bigg],\nonumber\\
\label{eq:sub2cif}\\
\lefteqn{{\rm d}\hat\sigma_{NNLO}^{S,c,(ii)}
= - {\cal N}\,\sum_{\textrm{perms}}{\rm d}\Phi_{m+2}(p_{3},\ldots,p_{m+4};p_1,p_2)
\frac{1}{S_{{m+2}}} }\nonumber \\
&\times& \,\Bigg[  \sum_{k,m=1,2} \sum_{j,l}\;X^0_{k,ji}\;x^0_{Km,l}\,
|{\cal M}_{m+2}(\ldots,I,\hat{K}^\prime,\hat{M}^\prime,\ldots)|^2\,
J_{m}^{(m)}(\tilde{p}_3,\ldots,p_I,\ldots,\tilde{p}_{m+4})\;
\phantom{\Bigg]}
\nonumber \\
&& \,+ \sum_{k,m=1,2}\sum_{j,l}\;X^0_{km,l}\;x^0_{K,ji}\,
|{\cal M}_{m+2}(\ldots,I^\prime,\hat{K}^\prime,\hat{M},\ldots)|^2\,
J_{m}^{(m)}(\tilde{p}_3,\ldots,p_{I^\prime},\ldots,\tilde{p}_{m+4})\;
\phantom{\Bigg]}
\nonumber\\
&& \,+ \sum_{i,m=1,2}\sum_{j,l}\;X^0_{i,jk}\;x^0_{m,lK}\,
|{\cal M}_{m+2}(\ldots,\hat{I},{K}^\prime,\hat{M}^\prime,\ldots)|^2\,
J_{m}^{(m)}({p}_3,\ldots,p_{K^\prime},\ldots,{p}_{m+4})\;
\phantom{\Bigg]} \nonumber\\
&& \,+ \sum_{i,m=1,2}\sum_{j,l}\;X^0_{m,lk}\;x^0_{i,jK}\,
|{\cal M}_{m+2}(\ldots,\hat{I}^\prime,{K}^\prime,\hat{M},\ldots)|^2\,
J_{m}^{(m)}({p}_3,\ldots,p_{K^\prime},\ldots,{p}_{m+4})\;
\Bigg],\nonumber\\
\label{eq:sub2cii}
\end{eqnarray}
where $x^0_{mlK}$ denotes a sub-antenna that contains only the collinear limit of $m$ with $l$ and not the collinear limit of $l$ with $K$. In the soft limit of $l$, this sub-antenna yields half the soft eikonal factor. In eq.~(\ref{eq:sub2cif}), the first two terms apply if the common radiator is in the final state, while the latter two terms correspond to the common radiator being in the initial state. Conversely, in eq.~(\ref{eq:sub2cii}), the first two terms apply if the common radiator is in the initial state, while the last two terms correspond to the common radiator being in the final state.

In the almost colour connected configuration ${\rm d}\hat\sigma_{NNLO}^{S,c}$ yields minus the double unresolved limit of the matrix element and therefore cancels the oversubtraction of ${\rm d}\hat\sigma_{NNLO}^{S,a}$ in the same configuration. In the single unresolved limits when either $j$ or $l$ is unresolved ${\rm d}\hat\sigma_{NNLO}^{S,c}$ exactly cancels the spurious single unresolved singularities that are produced in ${\rm d}\hat\sigma_{NNLO}^{S,a}$ when one of the particles appearing in $(m+3)$-parton matrix element becomes unresolved.

To obtain the integrated form of this counterterm, we exploit the factorisation of the $(m+2)$-parton phase space into a $m$-parton phase space and the phase space for the product of the two antennae. The integrated form of the inner antennae can be found in \cite{GehrmannDeRidder:2005cm,Daleo:2006xa}. The integrals over the outer antenna are of a similar type and are expected to be straightforward. Work on their analytic evaluation is in progress.

\subsection{Subtraction terms for two colour-unconnected unresolved partons ${\rm d}\hat\sigma_{NNLO}^{S,d}$}
When two unresolved partons $j$ and $o$ are completely disconnected i.e. for colour ordered amplitudes of the type  ${\cal M}(\hdots,i,j,k,\hdots,n,o,p,\hdots)$, the $(m+4)$-parton matrix element factorises into the product of two uncorrelated single unresolved factors with hard partons $I,K$ and $N,P$ respectively multiplied by a reduced $(m+2)$-parton matrix element. The subtraction terms for the colour-unconnected configuration read,
\begin{eqnarray}
\lefteqn{{\rm d}\hat\sigma_{NNLO}^{S,d,(ff)}
= - {\cal N}\,\sum_{\textrm{perms}}{\rm d}\Phi_{m+2}(p_{3},\ldots,p_{m+4};p_1,p_2)
\frac{1}{S_{{m+2}}}} \nonumber \\
&\times& \,\Bigg [ \sum_{j,o}\;X^0_{ijk}\;X^0_{nop}\,
|{\cal M}_{m+2}(\ldots,I,K,\ldots,N,P,\ldots)|^2\,
\nonumber \\ &&\hspace{3cm}\times
J_{m}^{(m)}(p_3,\ldots,p_I,p_K,\ldots,p_N,p_P,\ldots,p_{m+4})\;\Bigg
]\;,\nonumber\\  
\label{eq:sub2dff}\\
\lefteqn{ {\rm d}\hat\sigma_{NNLO}^{S,d,(if)}
= - {\cal N}\,\sum_{\textrm{perms}}{\rm d}\Phi_{m+2}(p_{3},\ldots,p_{m+4};p_1,p_2)
\frac{1}{S_{{m+2}}} }\nonumber \\
&\times &\Bigg [ \sum_{i=1,2}
\sum_{j,o}\;X^0_{i,jk}\;X^0_{nop}\,
|{\cal M}_{m+2}(\ldots,\hat{I},K,\ldots,N,P,\ldots)|^2\,
\nonumber \\ &&\hspace{3cm}\times
J_{m}^{(m)}(p_3,\ldots,p_K,\ldots,p_N,p_P,\ldots,p_{m+4})\;\Bigg
],\nonumber\\
\label{eq:sub2dif}\\
\lefteqn{ {\rm d}\hat\sigma_{NNLO}^{S,d,(ii)}
= - {\cal N}\,\sum_{\textrm{perms}}{\rm d}\Phi_{m+2}(p_{3},\ldots,p_{m+4};p_1,p_2)
\frac{1}{S_{{m+2}}}}\nonumber \\
&\times&\Bigg [ \sum_{i,n=1,2}\sum_{j,o}\;X^0_{i,jk}\;X^0_{n,op}\, \,|{\cal M}_{m+2}(\ldots,\hat{I},K,\ldots,\hat{N},P,\ldots)|^2\,
\nonumber \\ &&\hspace{3cm}\times
J_{m}^{(m)}( {p}_3,\ldots,p_K,\ldots,p_P,\ldots, {p}_{m+4})\;
\phantom{\Bigg]}
\nonumber\\
&&+\sum_{k,n=1,2}\sum_{j,o}\;X^0_{k,ji}\;X^0_{n,op}\, \,|{\cal M}_{m+2}(\ldots,I,\hat{K},\ldots,\hat{N},P,\ldots)|^2\,
\nonumber \\ &&\hspace{3cm}\times
J_{m}^{(m)}({p}_3,\ldots,p_I,\ldots,p_P,\ldots,{p}_{m+4})\;
\phantom{\Bigg]}
\nonumber\\
&&+\, \sum_{i,k=1,2}\sum_{j,o}\;X^0_{ik,j}\;X^0_{nop}\,
|{\cal M}_{m+2}
(\ldots,\hat{I},\hat{K},\ldots,N,P,\ldots)|^2\,
\nonumber \\ &&\hspace{3cm}\times
J_{m}^{(m)}(\tilde{p}_3,\ldots,p_N,p_P,\ldots,\tilde{p}_{m+4})\;\Bigg
]\;,
\label{eq:sub2dii}
\end{eqnarray}
where the summation over $o$ is such that it only includes 
two antenna configurations with no common momenta. The nature of the radiator pairs $i$,$k$ and $n$,$p$ defines the formula to be used.

In the colour unconnected configuration ${\rm d}\hat\sigma_{NNLO}^{S,d}$ yields minus the double unresolved limit of the matrix element and therefore cancels the oversubtraction of ${\rm d}\hat\sigma_{NNLO}^{S,a}$ in the same configuration. In the single unresolved limits when either $j$ or $o$ is unresolved ${\rm d}\hat\sigma_{NNLO}^{S,d}$ exactly cancels the spurious single unresolved singularities that are produced in ${\rm d}\hat\sigma_{NNLO}^{S,a}$ when one of the particles appearing in $(m+3)$-parton matrix element becomes unresolved.

To obtain the integrated form of this counterterm we exploit the factorisation of the $(m+2)$-parton phase space into a $m$-parton phase space multiplied by two independent phase space factors for each of the two antennae. The integrated form is thus simply the product of two integrated three-parton antennae.   Explicit expressions for individual integrated three-parton antennae can be found in \cite{GehrmannDeRidder:2005cm,Daleo:2006xa}.  

\subsection{Subtraction terms for large angle soft emission}
\label{sec:LAST}
It was shown in \cite{Weinzierl:2008iv,Weinzierl:2009nz} that the  antenna subtraction terms, ${\rm d}\hat\sigma_{NNLO}^{S,a},\ldots,{\rm d}\hat\sigma_{NNLO}^{S,c}$, result in an oversubtraction of large-angle soft gluon radiation. If we take a single soft gluon limit $j\to0$ of the formulae of the previous sections we obtain the following contributions of the type:\newline
final-final:
\begin{eqnarray}
X_{ilk}\hspace{-0.2cm}&&|{\cal M}_{m+2}(\hdots,a,I,K,b,\ldots)|^2\nonumber\\
&\times&(-S_{IjK}+S_{ijk}+S_{ajI}-S_{aji}+S_{Kjb}-S_{kjb}).
\label{eq:Ssoft1}
\end{eqnarray}
final-initial:
\begin{eqnarray}
  X_{i,lk}\hspace{-0.2cm}&&|{\cal M}_{m+2}(\hdots,a,\hat{I},K,b,\ldots)|^2\nonumber\\
&\times&
(-S_{\hat{I}jK}+S_{\hat{i}jk}+S_{aj\hat{I}}-S_{aj\hat{i}}+S_{Kjb}-S_{kjb}).
\label{eq:Ssoft2} 
\end{eqnarray}
initial-initial:
\begin{eqnarray}
  X_{ik,l}\hspace{-0.2cm}&&|{\cal M}_{m+2}
(\hdots,a,\hat{I},\hat{K},b,\ldots)|^2\nonumber\\
&\times&(-S_{\hat{I}j\hat{K}}+S_{\hat{i}j\hat{k}}+S_{\tilde{a}j\hat{I}}-S_{aj\hat{i}}+S_{\hat{K}j\tilde{b}}-S_{\hat{k}jb}).
\label{eq:Ssoft3} 
\end{eqnarray}
Here 
\begin{equation}
S_{abc}=2\frac{s_{ac}}{s_{ab}s_{bc}}
\end{equation}
is an eikonal factor related to the remnant soft behaviour of the phase space mappings.

To account for this large angle soft radiation, a new (process dependent) subtraction term ${\rm d}\hat\sigma_{NNLO}^A$ is added to the $(m+4)$-parton piece ${\rm d}{\hat\sigma_{NNLO}^S}$.  For example, the appropriate subtraction term for eq.~\eqref{eq:Ssoft1} is given by,
\begin{eqnarray}
&&{\cal N}\,\sum_{\textrm{perms}}{\rm d}\Phi_{m+2}(p_{3},\ldots,p_{m+4};p_1,p_2)
\frac{1}{S_{{m+2}}} \nonumber \\
&\times&  \Bigg [ X^0_{IlK} \, (S_{I^\prime jK^\prime}-S_{IjK}-S_{ajI^\prime}+S_{ajI}-S_{K^\prime jb}+S_{Kjb})
\nonumber \\ &&\times
|{\cal M}_{m+2}(\ldots,a,I^\prime,K^\prime,b,\ldots)|^2\, J_{m}^{(m)}(p_3,\ldots,p_a,p_{I^\prime},p_{K^\prime},p_b,\ldots,p_{m+4})\;\Bigg
]\;.  
\label{eq:Seiks1} 
\end{eqnarray}
The large-angle soft subtraction term contains 
soft antenna functions of the form $S_{ajb}$
which is simply the eikonal factor for a soft gluon $j$ emitted between hard
partons $a$ and $b$. 
The soft factors are associated with an 
NLO antenna phase space mapping $(i,j,k)\to(I,K)$, followed by a second
NLO antenna phase space mapping $(I,l,K)\to(I^\prime,K^\prime)$.
In the $j$ soft limit, $I \to i$ and $K \to k$ so that $I^\prime\to I$ and 
$K^\prime \to K$  so that eq.~(\ref{eq:Seiks1}) precisely cancels the behaviour in eq.~(\ref{eq:Ssoft1}).
In the $l$ soft limit, the eikonal factors cancel between each other such that no new spurious limits are introduced and eq.~(\ref{eq:Seiks1}) vanishes.

Similarly, the wide angle soft emission contributions corresponding to the initial-final  contribution of eq.~(\ref{eq:Ssoft2}) can be cancelled by,
\begin{eqnarray}
&& {\cal N}\,\sum_{\textrm{perms}}{\rm d}\Phi_{m+2}(p_{3},\ldots,p_{m+4};p_1,p_2)
\frac{1}{S_{{m+2}}} \nonumber \\
&\times& \,\Bigg [ X^0_{I,lK} \, (S_{\hat{I}^\prime jK^\prime}-S_{\hat{I}jK}-S_{aj\hat{I}^\prime}+S_{aj\hat{I}}-S_{K^\prime jb}+S_{Kjb})
\nonumber \\ &&\times
|{\cal M}_{m+2}(\ldots,a,\hat{I}^\prime,K^\prime,b,\ldots)|^2\, J_{m}^{(m)}(p_3,\ldots,p_a,p_{K^\prime},p_b,\ldots,p_{m+4})\;\Bigg
]\;, 
\label{eq:Seiks2} 
\end{eqnarray}
while the initial-initial contribution of eq.~(\ref{eq:Ssoft3}) can be cancelled by,
\begin{eqnarray}
&&{\cal N}\,\sum_{\textrm{perms}}{\rm d}\Phi_{m+2}(p_{3},\ldots,p_{m+4};p_1,p_2)
\frac{1}{S_{{m+2}}} \nonumber \\
&\times& \,\Bigg [ X^0_{IK,\tilde{l}} \, (S_{\hat{I}^\prime \tilde{j}\hat{K}^\prime}-S_{\hat{I}j\hat{K}}-S_{\tilde{\tilde{a}}\tilde{j}\hat{I}^\prime}+S_{\tilde{a}j\hat{I}}-S_{\hat{K}^\prime \tilde{j}\tilde{\tilde{b}}}+S_{\hat{K}j\tilde{b}})
\nonumber \\ &&\times
|{\cal M}_{m+2}(\ldots,a,\hat{I}^\prime,\hat{K}^\prime,b,\ldots)|^2\, J_{m}^{(m)}(\tilde{\tilde{p}}_3,\ldots,\tilde{\tilde{p}}_a,\tilde{\tilde{p}}_b,\ldots,\tilde{\tilde{p}}_{m+4})\;\Bigg
]\;.  
\label{eq:Seiks3} 
\end{eqnarray}

\subsection{Correction terms in the $m$-jet region}
\label{sec:NNLOfullRR}
The full double radiation subtraction term is the sum of the
subtraction terms constructed above:
\begin{equation}
{\rm d}\hat\sigma_{NNLO}^{S} = {\rm d}\hat\sigma_{NNLO}^{S,a} 
+{\rm d}\hat\sigma_{NNLO}^{S,b}+{\rm d}\hat\sigma_{NNLO}^{S,c}
+{\rm d}\hat\sigma_{NNLO}^{S,d}+{\rm d}\hat\sigma_{NNLO}^{A} \;.
\label{eq:sub2}
\end{equation}
As outlined in the previous subsections, this subtraction term 
correctly approximates the $(m+4)$-parton matrix element contribution to 
$m$-jet final states as defined in (\ref{eq:nnloreal}) in all double and 
single unresolved regions. Although individual terms in (\ref{eq:sub2}) 
contain spurious singularities in these limits, they cancel among each
other in the sum. 

The integrated form of (a) corresponds to an ($m+3$)-parton configuration, 
while the integrated forms of (b), (c) and (d) are either ($m+3$)-parton 
or $(m+2)$-parton configurations (for all but the four-parton antenna terms 
in (b), we can actually choose which type of configuration we want to
integrate). 
 They must be combined with the two-loop 
$(m+2)$-parton and the one-loop $(m+3)$-parton contributions to $m$-jet final 
states to yield an integrand free of explicit infrared poles.

\subsection{NNLO antennae decomposition for numerical implementation}
\label{sec:NNLOnimplemtation}
Having looked at the general formula for the double real radiation piece it is important to discuss its numerical implementation and to do so, we focus on a specific example. If we concentrate on the pure gluon channel contributing to the two-jet cross section we find that the four gluon antenna $F_4^0$ (given in \cite{GehrmannDeRidder:2005cm}) is the genuinely new ingredient at NNLO. In $F_4^0$, derived from  $H\to gggg$, the gluonic emissions are colour ordered. The colour structure is a trace over the gluon indices and $F_4^0$ is symmetric under cyclic interchanges of momenta. We will take this into account when we discuss the numerical implementation of $F_4^0$ for the final-final, initial-final and initial-initial arrangements of radiators.
 
\subsubsection{Final-Final emitters}
\label{sec:F40FFdecomp}
A decomposition of $F_4^0(1_g,2_g,3_g,4_g)$ is needed since any pair of gluons can become soft. Its unintegrated and integrated form has been written down in \cite{GehrmannDeRidder:2005cm}. In the case of the final-final configuration, this antenna has all the partons in the final state and can be used to subtract double unresolved final state singularities of colour ordered matrix elements when the unresolved gluons are colour connected between two final state gluons. We consider eight different mappings to achieve the decomposition:
\begin{eqnarray}
 \mbox{(a): } (1,2,3,4) \to (\widetilde{123},\widetilde{432})\,,&\qquad&
 \mbox{(b): } (1,2,4,3) \to (\widetilde{124},\widetilde{342})\,, \nonumber\\
 \mbox{(c): } (1,4,3,2) \to (\widetilde{143},\widetilde{234})\,,&\qquad&
 \mbox{(d): } (1,4,2,3) \to (\widetilde{142},\widetilde{324})\,, \nonumber\\ 
 \mbox{(e): } (2,3,1,4) \to (\widetilde{231},\widetilde{413})\,,&\qquad&
 \mbox{(f): } (2,1,4,3) \to (\widetilde{214},\widetilde{341})\,, \nonumber\\ 
 \mbox{(g): } (4,3,1,2) \to (\widetilde{431},\widetilde{213})\,,&\qquad&
 \mbox{(h): } (4,1,2,3) \to (\widetilde{412},\widetilde{321})\,.
\label{eq:8maps}
\end{eqnarray}

In each mapping of the type $({i_1},{i_2},{i_3},{i_4}) \to 
(\widetilde{{i_1i_2i_3}},\widetilde{{i_4i_3i_2}})$, $i_1$ and $i_4$ are the hard radiators and partons $i_2$ and $i_3$ are unresolved. Here we interchangeably use the notation $I_1 \equiv \widetilde{i_1i_2i_3}$ and $I_4 \equiv
\widetilde{i_4i_3i_2}$ for the remaining hard partons.
The new momenta are given by:
\begin{eqnarray}
p_{I_1}^\mu \equiv \widetilde{p_{(i_1i_2i_3)}}&=&x\,p_{i_1}+r_1\,p_{i_2}+r_2\,p_{i_3}+z\,p_{i_4},\nonumber\\
p_{I_4}^\mu \equiv\widetilde{p_{(i_4i_3i_2)}}&=&(1-x)\,p_{i_1}+(1-r_1)\,p_{i_2}+(1-r_2)\,p_{i_3}+(1-z)\,p_{i_4}\;,
\end{eqnarray}
with $p_{I_1}^2=p_{I_4}^2=0$.
Defining $s_{kl}=(p_{i_k}+p_{i_l})^2$, 
the coefficients are given by \cite{Kosower:2002su}:
\begin{eqnarray}
r_1&=&\frac{s_{23}+s_{24}}{s_{12}+s_{23}+s_{24}}\nonumber\\
r_2&=&\frac{s_{34}}{s_{13}+s_{23}+s_{34}}\nonumber\\
x&=&\frac{1}{2(s_{12}+s_{13}+s_{14})}\Big[(1+\rho)\,s_{1234}-r_1\,(s_{23}+2\,s_{24})   -r_2\,(s_{23}+2\,s_{34})  \nonumber\\
&&+(r_1-r_2)\frac{s_{12}s_{34}-s_{13}s_{24}}{s_{14}}     \Big]\nonumber\\
z&=&\frac{1}{2(s_{14}+s_{24}+s_{34})}\Big[(1-\rho)\,s_{1234} -r_1\,(s_{23}+2\,s_{12})   -r_2\,(s_{23}+2\,s_{13})  \nonumber\\
&&-(r_1-r_2)\frac{s_{12}s_{34}-s_{13}s_{24}}{s_{14}}     \Big]\nonumber\\
\rho&=&\Big[1+\frac{(r_1-r_2)^2}{s_{14}^2\,s_{1234}^2}\,\lambda(s_{12}\,
s_{34},s_{14}\,s_{23},s_{13}\,s_{24})\nonumber\\
&&  +\frac{1}{s_{14}\,s_{1234}}\Big\{
2\,\big(r_1\,(1-r_2)+r_2(1-r_1)\big)\big( s_{12}s_{34}+s_{13}s_{24}-s_{23}s_{14} \big)\nonumber\\
&&\qquad\qquad +\,4\,r_1\,(1-r_1)\,s_{12} s_{24}+4\,r_2\,(1-r_2)\,s_{13}s_{34}\Big\}\Big]^{\frac{1}{2}}\;,
\nonumber\\
\lambda(u,v,w)&=&u^2+v^2+w^2-2(uv+uw+vw)\;.\nonumber
\end{eqnarray}
This mapping smoothly interpolates all colour connected double unresolved singularities. It satisfies the following properties:
\begin{align} 
&\widetilde{p_{(i_1i_2i_3)}}\to p_{i_1}, 
&\widetilde{p_{(i_4i_3i_2)}}\to p_{i_4}\hspace{2.5cm} 
&\textrm{when $i_2,i_3\to0$,}\nonumber\\
&\widetilde{p_{(i_1i_2i_3)}}\to p_{i_1}+p_{i_2}+p_{i_3},
&\widetilde{p_{(i_4i_3i_2)}}\to p_{i_4}\hspace{2.5cm}  
&\textrm{when $i_1//i_2//i_3$,}\nonumber\\
&\widetilde{p_{(i_1i_2i_3)}}\to p_{i_1}, 
&\widetilde{p_{(i_4i_3i_2)}}\to p_{i_4}+p_{i_3}+p_{i_2}\hspace{0.6cm}   
&\textrm{when $i_2//i_3//i_4$,}\nonumber\\
&\widetilde{p_{(i_1i_2i_3)}}\to p_{i_1}, 
&\widetilde{p_{(i_4i_3i_2)}}\to p_{i_4}+p_{i_3}\hspace{1.5cm}  
&\textrm{when $i_2\to 0 +i_3//i_4$,}\nonumber\\
&\widetilde{p_{(i_1i_2i_3)}}\to p_{i_1}+p_{i_2}, 
&\widetilde{p_{(i_4i_3i_2)}}\to p_{i_2}\hspace{2.4cm} 
&\textrm{when $i_3\to 0 +i_1//i_2$,}\nonumber\\
&\widetilde{p_{(i_1i_2i_3)}}\to p_{i_1}+p_{i_2}, 
&\widetilde{p_{(i_4i_3i_2)}}\to p_{i_3}+p_{i_4} \hspace{1.5cm}
&\textrm{when $i_1//i_2+i_3//i_4$.}\nonumber\\
\end{align}
Moreover in single unresolved limits, the momentum mapping collapses into an NLO mapping, thereby allowing the subtraction of single unresolved limits of $F_4^0(1_g,2_g,3_g,4_g)$ with products of three parton antenna functions as in eq.~(\ref{eq:sub2bff}). 

The task left now is to disentangle the various double and single unresolved limits of the full antenna $F_4^0(1_g,2_g,3_g,4_g)$ into eight sub-antennae such that each sub-antenna (a),(b),...,(h) contains only those singularities appropriate to the mapping (a),(b),...,(h).

These numerous double and single unresolved limits can be disentangled very
elegantly by repeatedly exploiting the ${\cal N} =1$ supersymmetry relation \cite{Campbell:1997hg} among the 
different triple collinear splitting functions \cite{Campbell:1997hg,GehrmannDeRidder:1997gf,DelDuca:1999ha,Catani:1998nv,Catani:1999ss}.
Using this relation to introduce the antenna functions introduced in \cite{GehrmannDeRidder:2005cm}; $A_4^0(1_q,2_g,3_g,4_{\bar q})$,
 $\tilde{A}_4^0(1_q,2_g,3_g,4_{\bar q})$  and $A_3^0(1_q,2_g,3_{\bar q})$ for   quark-antiquark antennae,
$D_4^0(1_q,2_g,3_g,4_g)$, $D_3^0(1_q,2_g,3_g)$, $d_3^0(1_q,2_g,3_g)$
$E_4^0(1_q,2_q,3_{\bar q},4_g)$, $E_3^0(1_q,2_q,3_{\bar q})$ for quark-gluon antennae 
and 
$H_4^0(1_q,2_{\bar q},3_{q^\prime},4_{\bar q^\prime})$ and $G_3^0(1_g,2_q,3_{\bar q})$ for gluon-gluon antennae,
 one can show that the following 
left-over combination is 
finite in all single unresolved and double unresolved limits:
\begin{eqnarray}
F_{4,l}^0(1,2,3,4) &=& F_{4}^0(1,2,3,4) - \nonumber \\
&& \Bigg[ D_4^0(1,2,3,4)  + D_4^0(2,3,4,1) + D_4^0(3,4,1,2) + D_4^0(4,1,2,3)\nonumber \\
&&  -A_4^0(1,2,3,4)  - A_4^0(2,3,4,1) - A_4^0(3,4,1,2) - A_4^0(4,1,2,3)\nonumber \\
&&
- \tilde{A}_4^0(1,2,4,3) - \tilde{A}_4^0(2,3,1,4) + H_4^0(2,1,4,3) + H_4^0(4,1,2,3)\nonumber \\
&&
+ A_3^0(4,1,2)\, J_{3}^0 (\widetilde{(12)},3,\widetilde{(14)})
+ A_3^0(1,2,3)\, J_{3}^0 (\widetilde{(12)},\widetilde{(23)},4)\nonumber \\
&&
+ A_3^0(2,3,4)\, J_{3}^0 (1,\widetilde{(23)},\widetilde{(34)})
+ A_3^0(3,4,1)\, J_{3}^0 (\widetilde{(14)},2,\widetilde{(34)})\nonumber \\
&&
- \frac{1}{2}G_3^0(4,1,2)\, K_{3}^0 (\widetilde{(12)},\widetilde{(14)},3)
- \frac{1}{2}G_3^0(1,2,3)\, K_{3}^0 (\widetilde{(23)},\widetilde{(12)},4)\nonumber \\
&&
- \frac{1}{2}G_3^0(2,3,4)\, K_{3}^0 (\widetilde{(34)},\widetilde{(23)},1)
- \frac{1}{2}G_3^0(3,4,1)\, K_{3}^0 (\widetilde{(14)},\widetilde{(34)},2)\nonumber \\
&&
- \frac{1}{2}G_3^0(2,1,4)\, K_{3}^0 (\widetilde{(14)},\widetilde{(12)},3)
- \frac{1}{2}G_3^0(3,1,2)\, K_{3}^0 (\widetilde{(12)},\widetilde{(23)},4)\nonumber \\
&&
- \frac{1}{2}G_3^0(4,3,2)\, K_{3}^0 (\widetilde{(23)},\widetilde{(34)},1)
- \frac{1}{2}G_3^0(1,4,3)\, K_{3}^0 (\widetilde{(34)},\widetilde{(14)},2) \Bigg] \;\nonumber\\
\label{eq:F40l}
\end{eqnarray}
where $J_3^0$ and $K_3^0$ are useful combinations of the following three parton antenna functions:
\begin{eqnarray}
J^0_{3} (1,2,3) &=& F_3^0 (1,2,3) + A_3^0(3,1,2) + A_3^0(1,2,3) + A_3^0(1,3,2)\nonumber \\
&&-D_3^0(1,2,3)-D_3^0(2,3,1)-D_3^0(3,1,2),\;\\
K^0_{3} (1,2,3) &=& F^0_{3}(1,2,3) - D^0_{3}(2,3,1) - D^0_{3}(3,1,2) + A^0_{3}(3,1,2)\nonumber \\
&&+ G^0_{3}(1,2,3).
\end{eqnarray}
Neither $J^0_{3}$ or $K^0_{3}$ contains any soft or collinear limit, but to distribute the single unresolved limits 
among the momentum mappings it is convenient to introduce the following antennae:
\begin{eqnarray}
T^0_{3} (1,2,3) &=& f_3^0 (1,2,3) + A_3^0(1,2,3) - d_3^0(1,2,3) - d_3^0(3,2,1),\;\\
U^0_{3} (1,2,3) &=& f^0_{3}(1,2,3) - d^0_{3}(1,2,3),
\end{eqnarray}
where $T^0_3$ is finite in all single unresolved limits but $U^0_3$ contains the $1\parallel 2$ limit.
We can now rewrite $J_3^0$ and $K_3^0$ as,
\begin{eqnarray}
J^0_{3} (1,2,3) &=& T_3^0 (1,2,3) + T_3^0(1,3,2) + T_3^0(2,1,3),\;\\
K^0_{3} (1,2,3) &=& U^0_{3}(2,3,1) + U^0_{3}(3,2,1) + T^0_{3}(2,1,3) + G^0_{3}(1,2,3).
\end{eqnarray}
Starting from the terms in the expression (\ref{eq:F40l}) the following sub-antennae can be constructed,
\begin{eqnarray}
F_{4,a}^0(1,2,3,4) &=& \frac{1}{4} F_{4,l}^0(1,2,3,4)+ D_{4,a}^0(1,2,3,4) 
+ D_{4,a}^0(4,3,2,1)
\nonumber \\
&&-A_4^0(1,2,3,4) + \frac{1}{2} H_4^0(1,2,3,4)\nonumber\\
&&+A_3^0(1,2,3)\,T_3^0(\widetilde{(12)},\widetilde{(23)},4)
+ A_3^0(2,3,4)\, T_3^0 (1,\widetilde{(23)},\widetilde{(34)})\nonumber\\
&&- \frac{1}{2}G_3^0(1,2,3)\,T_3^0(\widetilde{(12)},\widetilde{(23)},4)\,
- \frac{1}{2}G_3^0(4,3,2)\,T_3^0(\widetilde{(34)},\widetilde{(23)},1)\;, \nonumber \\
F_{4,b}^0(1,2,3,4) &=& D_{4,c}(1,2,3,4)+D_{4,c}(3,4,1,2)-\tilde{A}_{4,a}^0(1,2,4,3)\nonumber\\
&&
+ a_3^0(1,2,3)\,T_3^0(\widetilde{(12)},4,\widetilde{(23)}) 
+ a_3^0(3,4,1)\,T_3^0(\widetilde{(14)},2,\widetilde{(34)})\nonumber\\
&&
-\frac{1}{2}G_3^0(1,4,3)\left(U_3^0(\widetilde{(14)},2,\widetilde{(34)})
+U_3^0(2,\widetilde{(14)},\widetilde{(34)})
+G_3^0(\widetilde{(34)},\widetilde{(14)},2)\right)\nonumber\\
&&
-\frac{1}{2}G_3^0(3,2,1)\left(U_3^0(\widetilde{(23)},4,\widetilde{(12)})
+U_3^0(4,\widetilde{(23)},\widetilde{(12)})
+G_3^0(\widetilde{(12)},\widetilde{(23)},4)\right)\nonumber\\
&&
+\frac{1}{2}G_3^0(4,1,2)\left(U_3^0(\widetilde{(41)},3,\widetilde{(12)})
-U_3^0(3,\widetilde{(41)},\widetilde{(12)})\right)\nonumber\\
&&
-\frac{1}{2}G_3^0(1,4,3)\left(U_3^0(\widetilde{(14)},2,\widetilde{(43)})
-U_3^0(2,\widetilde{(14)},\widetilde{(43)})\right),\nonumber\\
\label{eq:smallff40}
\end{eqnarray}
where the definition of $D_{4,a}^0$ and $D_{4,c}^0$ is given by decomposition,
\begin{eqnarray}
D_4^0=D_{4,a}^0+D_{4,b}^0+D_{4,c}^0+D_{4,d}^0, \nonumber
\end{eqnarray}
obtained in \cite{GehrmannDeRidder:2007jk}.
$F_{4,a}^0(1,2,3,4)$ contains no singularities when gluons 1 or 4 are soft.  Likewise,
$F_{4,b}^0(1,2,3,4)$ contains no singularities when gluons 1 or 3 are soft.
The remaining six sub-antennae are obtained by permutations,
\begin{eqnarray}
F_{4,c}^0(1,2,3,4) &=& F_{4,a}^0(1,4,3,2),\nonumber\\
F_{4,d}^0(1,2,3,4) &=& F_{4,b}^0(1,4,3,2),\nonumber\\
F_{4,e}^0(1,2,3,4) &=& F_{4,b}^0(2,3,4,1),\nonumber\\
F_{4,f}^0(1,2,3,4) &=& F_{4,a}^0(2,1,4,3),\nonumber\\
F_{4,g}^0(1,2,3,4) &=& F_{4,b}^0(4,3,2,1),\nonumber\\
F_{4,h}^0(1,2,3,4) &=& F_{4,a}^0(4,1,2,3).
\end{eqnarray}

The sum of the sub-antennae $F_{4,i}^0$ produces the full gluon-gluon antenna $F_4^0$,
\begin{eqnarray}
F_4^0(1,2,3,4)=\sum_{i=a,b,c,d,e,f,g,h} F_{4,i}^0(1,2,3,4), \nonumber
\end{eqnarray}
and we can organise the calculation such that only $F_4^0$ needs to be integrated analytically over the antenna phase space. That integral has been calculated and documented in \cite{GehrmannDeRidder:2005cm}. For the numerical implementation we only need to implement the sub-antennae $F_{4,a}$ and $F_{4,b}$ because we can reconstruct the full $F_4^0(1_g,2_g,3_g,4_g)$ by adding four $F_{4,a}$ and four $F_{4,b}$ with different orderings of the gluon indices,
\begin{eqnarray}
F_4^0(1,2,3,4)&&=F_{4,a}^0(1^h,2,3,4^h)+F_{4,b}^0(1^h,2,3^h,4)\nonumber\\&&+F_{4,a}^0(1^h,4,3,2^h)+F_{4,b}^0(1^h,4,3^h,2)\nonumber\\
&&\hspace{0.1cm}+F_{4,b}^0(2^h,3,4^h,1)+F_{4,a}^0(2^h,1,4,3^h)\nonumber\\&&+F_{4,b}^0(4^h,3,2^h,1)+F_{4,a}^0(4^h,1,2,3^h).
\label{eq:F40decomp}
\end{eqnarray}
The label $h$ identifies the hard momenta within each sub-antennae. This means that the sub-antenna vanishes if we take the $h$ soft limit and therefore each sub-antenna has only the singularities appropriate to the mappings (\ref{eq:8maps}). This disentanglement of the different double and single unresolved limits is the following:
\begin{eqnarray}
F_{4,h}^0(1,2,3,4)&\stackrel{ 1_g \to 0, 2_g \to 0}
{\longrightarrow}&\; S_{4123}\;,\nonumber\\
F_{4,e}^0(1,2,3,4)+F_{4,g}(1,2,3,4)&\stackrel{ 1_g \to 0, 3_g \to 0}
{\longrightarrow}&\; S_{412}S_{234}\;,\nonumber\\
F_{4,f}^0(1,2,3,4)&\stackrel{ 1_g \to 0, 4_g \to 0}
{\longrightarrow}&\; S_{3412}\;,\nonumber\\
F_{4,a}^0(1,2,3,4)&\stackrel{ 2_g \to 0, 3_g \to 0}
{\longrightarrow}&\; S_{1234}\;,\nonumber\\
F_{4,b}^0(1,2,3,4)+F_{4,d}(1,2,3,4)&\stackrel{ 2_g \to 0, 4_g \to 0}
{\longrightarrow}&\; S_{123}S_{341}\;,\nonumber\\
F_{4,c}^0(1,2,3,4)&\stackrel{ 3_g \to 0, 4_g \to 0}
{\longrightarrow}&\; S_{2341}\;,\nonumber\\
F_{4,a}^0(1,2,3,4)+F_{4,e}^0(1,2,3,4)+F_{4,g}^0(1,2,3,4)&\stackrel{ 1_g\parallel 2_g, 3_g \to 0}
{\longrightarrow}& S_{4;312}(z) \;
\frac{1}{s_{12}}\; P_{gg\to G}(z)
\;,\nonumber \\
F_{4,b}^0(1,2,3,4)+F_{4,d}^0(1,2,3,4)+F_{4,f}^0(1,2,3,4)&\stackrel{ 1_g\parallel 2_g, 4_g \to 0}
{\longrightarrow}& S_{3;412}(z) \;
\frac{1}{s_{12}}\; P_{gg\to G}(z)
\;,\nonumber \\
F_{4,e}^0(1,2,3,4)+F_{4,g}^0(1,2,3,4)+F_{4,h}^0(1,2,3,4)&\stackrel{ 2_g\parallel 3_g, 1_g \to 0}
{\longrightarrow}& S_{4;123}(z) \;
\frac{1}{s_{23}}\; P_{gg\to G}(z)
\;,\nonumber \\
F_{4,b}^0(1,2,3,4)+F_{4,c}^0(1,2,3,4)+F_{4,d}^0(1,2,3,4)&\stackrel{ 2_g\parallel 3_g, 4_g \to 0}
{\longrightarrow}& S_{1;432}(z) \;
\frac{1}{s_{23}}\; P_{gg\to G}(z)
\;,\nonumber \\
F_{4,e}^0(1,2,3,4)+F_{4,f}^0(1,2,3,4)+F_{4,g}^0(1,2,3,4)&\stackrel{ 3_g\parallel 4_g, 1_g \to 0}
{\longrightarrow}& S_{2;143}(z) \;
\frac{1}{s_{34}}\; P_{gg\to G}(z)
\;,\nonumber \\
F_{4,a}^0(1,2,3,4)+F_{4,b}^0(1,2,3,4)+F_{4,d}^0(1,2,3,4)&\stackrel{ 3_g\parallel 4_g, 2_g \to 0}
{\longrightarrow}& S_{1;234}(z) \;
\frac{1}{s_{34}}\; P_{gg\to G}(z)
\;,\nonumber \\
F_{4,b}^0(1,2,3,4)+F_{4,d}^0(1,2,3,4)+F_{4,h}^0(1,2,3,4)&\stackrel{ 4_g\parallel 1_g, 2_g \to 0}
{\longrightarrow}& S_{3;214}(z) \;
\frac{1}{s_{14}}\; P_{gg\to G}(z)
\;,\nonumber \\
F_{4,c}^0(1,2,3,4)+F_{4,e}^0(1,2,3,4)+F_{4,g}^0(1,2,3,4)&\stackrel{ 4_g\parallel 1_g, 3_g \to 0}
{\longrightarrow}& S_{2;341}(z) \;
\frac{1}{s_{14}}\; P_{gg\to G}(z)
\;,\nonumber
\end{eqnarray}
where $S_{abcd}$ is the double soft factor and $S_{a;bcd}$ the soft-collinear factor given in eqs. (8.3) and (8.5) of ref.~\cite{GehrmannDeRidder:2005cm}.
In the triple-collinear and independent double-collinear limits we find,
\begin{eqnarray}
F_{4,a}^0(1,2,3,4)+F_{4,e}^0(1,2,3,4)+F_{4,g}^0(1,2,3,4)+F_{4,h}^0(1,2,3,4)&\stackrel{ 1_g\parallel 2_g  \parallel 3_g }
{\longrightarrow}& P_{123 \to G}(w,x,y)\;,\nonumber\\
F_{4,a}^0(1,2,3,4)+F_{4,b}^0(1,2,3,4)+F_{4,c}^0(1,2,3,4)+F_{4,d}^0(1,2,3,4)&\stackrel{ 2_g\parallel 3_g  \parallel 4_g }
{\longrightarrow}& P_{234 \to G}(w,x,y)\;,\nonumber\\
F_{4,c}^0(1,2,3,4)+F_{4,e}^0(1,2,3,4)+F_{4,f}^0(1,2,3,4)+F_{4,g}^0(1,2,3,4)&\stackrel{ 3_g\parallel 4_g  \parallel 1_g }
{\longrightarrow}& P_{341 \to G}(w,x,y)\;,\nonumber\\
F_{4,b}^0(1,2,3,4)+F_{4,d}^0(1,2,3,4)+F_{4,f}^0(1,2,3,4)+F_{4,h}^0(1,2,3,4)&\stackrel{ 4_g\parallel 1_g  \parallel 2_g }
{\longrightarrow}& P_{412 \to G}(w,x,y)\;,\nonumber\\
F_{4,a}^0(1,2,3,4)+F_{4,b}^0(1,2,3,4)+F_{4,f}^0(1,2,3,4)+F_{4,g}^0(1,2,3,4)
&\nonumber\\\stackrel{ 1_g \parallel 2_g,  3_g \parallel 4_g}
{\longrightarrow} \frac{1}{s_{12} s_{34}}\;
P_{gg\to G}(z)\; P_{g g\to G}(y)\;,\nonumber \\
F_{4,c}^0(1,2,3,4)+F_{4,d}^0(1,2,3,4)+F_{4,e}^0(1,2,3,4)+F_{4,h}^0(1,2,3,4)\nonumber\\
\stackrel{ 2_g \parallel 3_g,  4_g \parallel 1_g}
{\longrightarrow} 
\frac{1}{s_{23} s_{14}}\;
P_{gg\to G}(z)\; P_{g g\to G}(y)\;,\nonumber
\end{eqnarray}
where $P_{abc \to G}$ is the triple collinear factor given in eqs. (8.6) of ref.~\cite{GehrmannDeRidder:2005cm}.

Finally in the single unresolved limits,
\begin{eqnarray}
F_{4,f}^0(1,2,3,4)&\stackrel{{1}_g\to 0}
{\longrightarrow}&S_{412}\; f_3^0(3,4,2)\;,\nonumber\\
F_{4,h}^0(1,2,3,4)&\stackrel{{1}_g\to 0}
{\longrightarrow}&S_{412}\; f_3^0(3,2,4)\;,\nonumber\\
F_{4,e}^0(1,2,3,4)+F_{4,g}^0(1,2,3,4)&\stackrel{{1}_g\to 0}
{\longrightarrow}&S_{412}\; f_3^0(2,3,4)\;,\nonumber\\
F_{4,a}^0(1,2,3,4)&\stackrel{{2}_g\to 0}
{\longrightarrow}&S_{123}\; f_3^0(1,3,4)\;,\nonumber\\
F_{4,h}^0(1,2,3,4)&\stackrel{{2}_g\to 0}
{\longrightarrow}&S_{123}\; f_3^0(3,1,4)\;,\nonumber\\
F_{4,b}^0(1,2,3,4)+F_{4,d}^0(1,2,3,4)&\stackrel{{2}_g\to 0}
{\longrightarrow}&S_{123}\; f_3^0(3,4,1)\;,\nonumber\\
F_{4,a}^0(1,2,3,4)&\stackrel{{3}_g\to 0}
{\longrightarrow}&S_{234}\; f_3^0(1,2,4)\;,\nonumber\\
F_{4,c}^0(1,2,3,4)&\stackrel{{3}_g\to 0}
{\longrightarrow}&S_{234}\; f_3^0(1,4,2)\;,\nonumber\\
F_{4,e}^0(1,2,3,4)+F_{4,g}^0(1,2,3,4)&\stackrel{{3}_g\to 0}
{\longrightarrow}&S_{234}\; f_3^0(2,1,4)\;,\nonumber\\
F_{4,c}^0(1,2,3,4)&\stackrel{{4}_g\to 0}
{\longrightarrow}&S_{341}\; f_3^0(1,3,2)\;,\nonumber\\
F_{4,f}^0(1,2,3,4)&\stackrel{{4}_g\to 0}
{\longrightarrow}&S_{341}\; f_3^0(2,1,3)\;,\nonumber\\
F_{4,b}^0(1,2,3,4)+F_{4,d}^0(1,2,3,4)&\stackrel{{4}_g\to 0}
{\longrightarrow}&S_{341}\; f_3^0(3,2,1)\;,\nonumber\\
F_{4,a}^0(1,2,3,4)+F_{4,g}^0(1,2,3,4)&\stackrel{1_g \parallel 2_g}
{\longrightarrow}&\frac{1}{s_{12}}\;P_{gg\to G}(z)\; 
f_3^0(4,3,(12))\;+ {\rm ang.}\;,\nonumber\\
F_{4,b}^0(1,2,3,4)+F_{4,f}^0(1,2,3,4)&\stackrel{1_g \parallel 2_g}
{\longrightarrow}&\frac{1}{s_{12}}\;P_{gg\to G}(z)\; 
f_3^0((12),4,3)\;+ {\rm ang.}\;,\nonumber\\
F_{4,h}^0(1,2,3,4)&\stackrel{1_g \parallel 2_g}
{\longrightarrow}&\frac{1}{s_{12}}\;P_{gg\to G}(z)\; 
f_3^0(3,(12),4)\;+ {\rm ang.}\;,\nonumber\\
F_{4,c}^0(1,2,3,4)+F_{4,d}^0(1,2,3,4)&\stackrel{2_g \parallel 3_g}
{\longrightarrow}&\frac{1}{s_{23}}\;P_{gg\to G}(z)\; 
f_3^0(1,4,(23))\;+ {\rm ang.}\;,\nonumber\\
F_{4,e}^0(1,2,3,4)+F_{4,h}^0(1,2,3,4)&\stackrel{2_g \parallel 3_g}
{\longrightarrow}&\frac{1}{s_{23}}\;P_{gg\to G}(z)\; 
f_3^0((23),1,4)\;+ {\rm ang.}\;,\nonumber\\
F_{4,a}^0(1,2,3,4)&\stackrel{2_g \parallel 3_g}
{\longrightarrow}&\frac{1}{s_{12}}\;P_{gg\to G}(z)\; 
f_3^0(4,(23),1)\;+ {\rm ang.}\;,\nonumber\\
F_{4,a}^0(1,2,3,4)+F_{4,b}^0(1,2,3,4)&\stackrel{3_g \parallel 4_g}
{\longrightarrow}&\frac{1}{s_{34}}\;P_{gg\to G}(z)\; 
f_3^0(1,2,(34))\;+ {\rm ang.}\;,\nonumber\\
F_{4,f}^0(1,2,3,4)+F_{4,g}^0(1,2,3,4)&\stackrel{3_g \parallel 4_g}
{\longrightarrow}&\frac{1}{s_{34}}\;P_{gg\to G}(z)\; 
f_3^0((34),1,2)\;+ {\rm ang.}\;,\nonumber\\
F_{4,c}^0(1,2,3,4)&\stackrel{3_g \parallel 4_g}
{\longrightarrow}&\frac{1}{s_{34}}\;P_{gg\to G}(z)\; 
f_3^0(2,(34),1)\;+ {\rm ang.}\;,\nonumber\\
F_{4,c}^0(1,2,3,4)+F_{4,e}^0(1,2,3,4)&\stackrel{4_g \parallel 1_g}
{\longrightarrow}&\frac{1}{s_{14}}\;P_{gg\to G}(z)\; 
f_3^0(2,3,(14))\;+ {\rm ang.}\;,\nonumber\\
F_{4,d}^0(1,2,3,4)+F_{4,h}^0(1,2,3,4)&\stackrel{4_g \parallel 1_g}
{\longrightarrow}&\frac{1}{s_{14}}\;P_{gg\to G}(z)\; 
f_3^0((14),2,3)\;+ {\rm ang.}\;,\nonumber\\
F_{4,f}^0(1,2,3,4)&\stackrel{4_g \parallel 1_g}
{\longrightarrow}&\frac{1}{s_{14}}\;P_{gg\to G}(z)\; 
f_3^0(3,(14),2)\;+ {\rm ang.}\;.\nonumber\\
\label{eq:F40dsntlgd}
\end{eqnarray}
All other limits are vanishing. It can be seen that certain limits are shared 
among several antenna functions, which can be largely understood for two 
reasons: 
\begin{enumerate}
\item[(1)] In a gluon-gluon collinear splitting, either gluon can become 
soft, and the gluon-gluon splitting function is always shared between two 
sub-antennae, as in (\ref{eq:decomp3FF}) to disentangle the two 
soft limits.
\item[(2)] The unresolved emission of gluon pairs $1_g$ and $3_g$ and also $2_g$ and $4_g$
is shared between the mappings (e) and (g) and (b) and (d) respectively according to the decomposition 
of the non-ordered antenna function $\tilde{A}_4^0$, which distributes 
the soft limit of both gluons between both mappings. 
\end{enumerate}

\subsubsection{Angular terms}
\label{sec:angterms}
The angular terms in the single unresolved limits 
are associated with a gluon splitting into two gluons or 
into a quark-antiquark pair. In this collinear configuration the four-parton antenna functions factorise into the corresponding tensorial splitting functions and tensorial three parton antenna functions \cite{Catani:1996vz}, e.g.,
\begin{eqnarray}
F_4^0(1,2,3,4)&\stackrel{i_g \parallel j_g}{\longrightarrow}&\frac{1}{s_{ij}}
P_{gg\to G}^{\mu\nu}(z)(F_3^0)_{\mu\nu}((ij),k,l)\nonumber\\
&=&\frac{1}{s_{ij}}P_{gg\to G}(z)F_3^0((ij),k,l)+{\rm ang.}
\label{eq:angtermlimit}
\end{eqnarray}
$P_{ij\to(ij)}^{\mu\nu}$ stands for the spin dependent gluon splitting function given 
by \cite{Catani:1996vz},
\begin{eqnarray}
P_{gg}^{\mu\nu}=2\left[-g^{\mu\nu}\left(\frac{z}{1-z}+\frac{1-z}{z}\right)-2(1-\epsilon)z(1-z)
\frac{k_{\perp}^\mu k_{\perp}^\nu}{k_{\perp}^2}\right],
\end{eqnarray}
while $P_{ij\to(ij)}$ stands for the spin averaged gluon splitting function:
\begin{equation}
P_{gg\to g}(z)=2\left(\frac{z}{1-z}+\frac{1-z}{z}+z(1-z)\right).
\end{equation}
The tensorial structure of the three-parton antenna function $(F_3^0)_{\mu\nu}$ is obtained by leaving the polarisation index of the gluon associated with momentum $P^\mu$ uncontracted and can be derived by analogy with the scalar three-parton antenna functions from physical matrix elements. 

Since we use scalar three parton antenna functions to remove the single unresolved limits of the four parton antenna function (this was discussed in eqs.~(\ref{eq:sub2bff}), (\ref{eq:sub2bif}), (\ref{eq:sub2bii})) we are left with angular terms in (\ref{eq:angtermlimit}) that do not cancel. 
However, these angular terms  average to zero after integration over the antenna 
phase space. The angular 
average in single collinear limits can be made using the standard 
momentum parametrisation \cite{Catani:1996vz,Altarelli:1977cd} for the $i_g\parallel j_g$  limit:
\begin{eqnarray}
p_i^\mu = z p^\mu + k_\perp^\mu - \frac{k_\perp^2}{z}\frac{n^\mu}{2p\cdot n}
\;, &\qquad&
 p_j^\mu = (1-z) p^\mu - k_\perp^\mu 
- \frac{k_\perp^2}{1-z}\frac{n^\mu}{2p\cdot n}\;, \nonumber\\
\mbox{with } 2p_i\cdot p_j = -\frac{k_\perp^2}{z(1-z)}\;, &\qquad& p^2 
= n^2=k_\perp.p=k_\perp.n=0\;.\label{eq:colmap}
\end{eqnarray}
Here $p^\mu$ denotes the collinear momentum direction, and $n^\mu$ is an 
auxiliary vector. The collinear limit is approached as $k_\perp^2\to 0$. 

In the simple collinear $i\parallel j$ limit of the four-parton antenna function 
$F_4^0(l_g,i_g,j_g,k_g)$, 
one chooses $n=p_k$ to be one of the non-collinear momenta, such that the 
antenna function can be expressed in terms of $p$, $n$, $k_\perp$ and $p_l$. 
Expanding in $k_\perp^\mu$ yields only non-vanishing  scalar products of the 
form $p_l\cdot k_\perp$. Expressing the integral over the antenna phase 
space in the $(p,n)$ centre-of-mass frame, the angular average can be 
carried out as
\begin{equation}
\frac{1}{2\pi} \int_0^{2\pi} {\rm d} \phi\, (p_l\cdot k_\perp) =0 \;, \qquad 
 \frac{1}{2\pi} \int_0^{2\pi} {\rm d} \phi\, (p_l\cdot k_\perp)^2 = - 
k_\perp^2\, \frac{p\cdot p_l\, n\cdot p_l}{p\cdot n}\;.\label{eq:azymav}
\end{equation}

Higher powers of $k_\perp^\mu$ are not sufficiently singular to contribute to 
the collinear limit. Using \eqref{eq:azymav},  the angular terms 
analytically vanish for each single phase space mapping, and this is
independent of the choice of the reference vector $n_\mu$. 
This means that for a collinear limit which is distributed between two mappings as in (\ref{eq:F40dsntlgd}), the angular terms vanish separately within each single phase space mapping that contributes to the limit.

This cancellation is obtained only after performing the azimuthal integration analytically. 
On a point-by-point check, the subtraction terms do not correctly reproduce the 
azimuthal terms. One solution is to add a local
counterterm to every four parton antenna containing gluon-gluon collinear splittings.  This local counterterm should yield 
the correct behaviour in this particular collinear limit and integrate to zero over the corresponding unresolved phase-space.
The following replacement would provide a local counter-term for
the azimuthal terms associated with a $i_g \parallel j_g$ splitting,
\begin{eqnarray}
F_4^0(1,i,j,2)\to F_4^0(1,i,j,2) - \Theta_{F_3^0}(i,j,z,k_\perp),
\end{eqnarray}
where  $\Theta_{F_3^0}(i,j,z,k_\perp)$ is given by,
\begin{eqnarray}
\Theta_{F_3^0}(i,j,z,k_\perp)&=&\left[\frac{1}{s_{ij}}P^{\mu\nu}_{ij\to(ij)}(z,k_\perp)(F_3^0)_{\mu\nu}
-\frac{1}{s_{ij}}P_{ij\to(ij)}(z)F_3^0(1,(ij),2)\right]\nonumber\\
&=&\frac{4}{s_{ij}^2s_{1p2}^2}\left(\frac{s_{12}^2s_{1p2}^2+s_{1p}^2s_{p2}^2}{s_{12}^2s_{1p}^2s_{p2}^2}\right)
\Bigg[s_{12}s_{1p}s_{p2}\;k_\perp\cdot k_\perp\nonumber\\
&&-4p_1\cdot k_\perp p_2\cdot k_\perp s_{1p}s_{p2}
+2(p_1\cdot k_\perp)^2 s_{p2}^2+2(p_2\cdot k_\perp)^2 s_{1p}^2\Bigg],\nonumber\\
\label{eq:angfunction}
\end{eqnarray}
with $p$ and $k_{\perp}$ defined in (\ref{eq:colmap}). Using (\ref{eq:azymav}), we can easily see that (\ref{eq:angfunction}) integrates to zero. 

However, due to the decomposition of $F_4$ in eight sub-antennae (\ref{eq:F40decomp}), the angular terms of the full $F_4^0$ antenna function given by (\ref{eq:angfunction}) are distributed across the sub-antennae that contribute in the singular limit. To make a local subtraction we have to compute this singular limit for each sub-antennae and subtract it explicitly. This is allowed since the angular terms integrate to zero for each phase space mapping. For the $F_4^0$ initial-final and initial-initial antennae functions that we discuss in the following subsections, the azimuthal counterterm (\ref{eq:angfunction}) can be implemented with a single phase space mapping
providing we make an appropriate crossing of particles from the final to the initial state.

After these replacements, the resulting four-parton antenna is locally free from angular terms in the single collinear $(s_{ij}\to0)$ unresolved region. However, in the regions of the phase space where other invariants in the denominator of (\ref{eq:angfunction}) vanish, new singularities are introduced. For example since,
\begin{equation}
p^{\mu}=p_{i}^{\mu}+p_{j}^{\mu}+\frac{s_{ij}}{s_{in}+s_{jn}}n^{\mu}
\end{equation}
the invariants $s_{1p}$ and $s_{2p}$ vanish in the $1//i//j$ and $2//i//j$ triple collinear limits respectively producing additiona singularities in eq.~(\ref{eq:angfunction}). This means that introducing the azimuthal counterterm $\Theta_{F_3^0}(i,j,z,k_\perp)$ may not be the best strategy to achieve the cancellation of the angular terms. 
Nevertheless, in subsection \ref{sec:colllim} we will discuss the numerical impact of introducing an azimuthal counterterm for the single collinear limit.

A second approach is to try to cancel the angular terms by combining phase space points related to each other by a rotation of the system of unresolved partons \cite{Weinzierl:2006wi,GehrmannDeRidder:2007jk}.  In the $(p,n)$ centre-of-mass frame, it can be shown that 
\begin{equation}
\Theta_{F_3^0}(i,j,z,k_\perp) \sim A\cos(2\phi+\alpha)
\end{equation}
where $\phi$ is the same azimuthal angle as in \eqref{eq:azymav}.  Therefore by combining two phase space points with azimuthal angles $\phi$ and $\phi+\pi/2$ and all other coordinates equal, the azimuthal correlations drop out.  When particles $i$ and $j$ are in the final state, this can be achieved by rotating particles $i$ and $j$ by $\pi/2$ about the $p^\mu$ direction.
In the initial-final configurations produced when $p_i^\mu \to p^\mu+p_j^\mu$ for $i=1,2$ and with $i||j$ and $p^2=0$, the phase space points should again be related by rotations of $\pi/2$ about the direction of $p^\mu$.  This has the consequence of rotating $p_i^\mu$ off the beam axis and therefore has to be compensated by a Lorentz boost.

\subsubsection{Initial-Final emitters}
The NNLO antenna with one parton in the initial state $F_4^0(\hat{1}_g,2_g,3_g,4_g)$ is obtained by crossing one gluon from the final state antenna $F_4^0(1_g,2_g,3_g,4_g)$ to the initial state. As usual, the hat denotes the initial state particle. This antenna is used to subtract double unresolved initial state singularities when both the unresolved gluons are colour connected between an initial and a final state gluon. The unintegrated form is obtained by making the replacements,
\begin{eqnarray}
&&s_{1i}\to(p_1-p_i)^2,\nonumber\\
&&s_{ij}\to(p_i+p_j)^2\qquad i,j=2,3,4,\nonumber
\end{eqnarray}
in the expression for $F_4^0(1_g,2_g,3_g,4_g)$ ~\cite{GehrmannDeRidder:2005cm}.

In all single (double) unresolved limits, this antenna collapses into a three (two) parton antenna with a gluon in the initial state. There is no need to further split this antenna since the reduced matrix elements that accompany it have a gluon in the initial state and can be convoluted with a gluon parton distribution function. However special care would have to be taken with three parton or four parton gluon initiated antennae with quarks in the final state since the splitting $g\to qg$ looks like either $\bar{q}\to g$ or $g\to g$ depending on the collinear limit. 

The mapping relevant for the initial-final configuration $F_4^0(\hat{1},i,j,k)$ is the following $\{4\to2\}$ mapping: $(\hat{i},j,k,l)\to (\hat{I},L)$ \cite{Daleo:2006xa},
\begin{eqnarray}
p_{\hat I}^\mu \equiv \bar{p}_i^{\mu}&=&x\,p_i^{\mu},\nonumber\\
p_L^\mu \equiv \widetilde{p_{(jkl)}}^{\mu}&=&p_j^{\mu}+\,p_k^{\mu}+\,p_l^{\mu}-(1-x)\,p_i^{\mu},\;
\label{4to2IFmap}
\end{eqnarray}
with $p_{\hat I}^2=p_L^2=0$ and where the bar denotes a rescaling of the initial state parton and $x$ is given by \cite{Daleo:2006xa}:
\begin{eqnarray}
x&=&\frac{s_{ij}+s_{ik}+s_{il}+s_{jk}+s_{jl}+s_{kl}}{s_{ij}+s_{ik}+s_{il}}.
\end{eqnarray}
It satisfies the appropriate limits in all double singular configurations:
\begin{enumerate}
\item[(1)] $j$ and $k$ soft: $x\rightarrow 1$, $\tilde{p}_{(jkl)}\rightarrow p_l$,
\item[(2)] $j$ soft and $p_k\parallel p_l$: $x\rightarrow 1$, $\tilde{p}_{(jkl)}\rightarrow p_k+p_l$,
\item[(3)] $p_j=zp_i\parallel p_i$ and $k$ soft: $x\rightarrow 1-z$, $\tilde{p}_{(jkl)}\rightarrow
  p_l$,
\item[(4)] $p_j=zp_i\parallel p_i$ and $p_k\parallel p_l$: $x\rightarrow 1-z$, $\tilde{p}_{(jkl)}\rightarrow
  p_k+p_l$,
\item[(5)] $p_j\parallel p_k\parallel p_l$: $x\rightarrow 1$, $\tilde{p}_{(jkl)}\rightarrow
  p_j+p_k+p_l$,
\item[(6)] $p_j+p_k=zp_i\parallel p_i$: $x\rightarrow 1-z$, $\tilde{p}_{(jkl)}\rightarrow
  p_l$,
\end{enumerate}  
where partons $j$ and $k$ can be interchanged.
In the single unresolved limits this mapping collapses into the NLO mapping (\ref{3to2IFmap}), thereby allowing the subtraction of these limits from the four-parton antenna $F_4^0(\hat{1}_g,2_g,3_g,4_g)$ with products of three parton antennae.

We can identify the initial state parton as the hard radiator and because of the symmetry under $j\leftrightarrow k$,$j\leftrightarrow l$,$k\leftrightarrow l$ of the mapping (\ref{4to2IFmap}) any of final state partons $j$, $k$, $l$ can act as a hard radiator. We could in principle decompose 
$F_4^0(\hat{1}_g,2_g,3_g,4_g)$ into separate terms where the final state radiator is uniquely identified.  However,  we would end up with three sub-antennae where we would use the same mapping for each. Therefore,  we simply use the full $F_4^0(\hat{1}_g,2_g,3_g,4_g)$ in the numerical implementation. The integral of $F_4^0(\hat{1}_g,2_g,3_g,4_g)$ over the antenna phase space with initial-state kinematics  is given in \cite{Daleo:2009yj,Daleo:2010tz}.

\subsubsection{Initial-Initial emitters}
The gluonic NNLO antenna function with two partons in the initial state is obtained from the corresponding initial-final antenna of the previous section by crossing one final-state gluon to the initial state by making the replacements,
\begin{eqnarray}
&&s_{12}\to(p_1+p_2)^2,\nonumber\\
&&s_{1i}\to(p_1-p_i)^2,\nonumber\\
&&s_{2i}\to(p_2-p_i)^2,\nonumber\\
&&s_{ij}\to(p_i+p_j)^2,\qquad\qquad i,j=3,4.\nonumber
\end{eqnarray}

We have to distinguish the cases where the two initial state partons are adjacent $F_4^0(\hat{1}_g,\hat{2}_g,3_g,4_g)$ or non-adjacent $F_4^0(\hat{1}_g,3_g,\hat{2}_g,4_g)$.  
Since in the single (double) unresolved limits these antennae collapse into a three (two) parton antenna with gluons in the initial state no further splitting is required. The reduced matrix elements accompanying this antenna have gluons in the initial state and can be doubly convoluted with a gluon parton distribution function. In both cases, the initial state gluons become the hard radiators, and there is no need to further decompose the antenna.

The mapping used in the numerical implementation when $j$ and $k$ are the unresolved partons in the final state is the following mapping: $(\hat{i},j,k,\hat{l},\ldots,m,\ldots)\to (\hat{I},\hat{L},...\tilde{m},\ldots)$ \cite{Daleo:2006xa}:
\begin{eqnarray}
p_{\hat I}^\mu \equiv \bar{p}_i^{\mu}&=&x_i\,p_i^{\mu},\nonumber\\
p_{\hat L}^\mu \equiv \bar{p}_l^{\mu}&=&x_l\,p_l^{\mu},\nonumber\\
\tilde{p}_m^{\mu}&=&p_m^{\mu}-\frac{2p_m\cdot (q+\tilde{q})}{(q+\tilde{q})^2}\;(q^{\mu}+\tilde{q}^{\mu})+\frac{2p_m\cdot q}{q^2}\;\tilde{q}^{\mu},\nonumber\\
q^{\mu}&=&p_i^{\mu}+p_l^{\mu}-p_j^{\mu}-p_k^{\mu}, \nonumber\\
\tilde{q}^{\mu}&=&\bar{p}_i^{\mu}+\bar{p}_l^{\mu}
\label{4to2IImap}
\end{eqnarray}
with $p_{\hat I}^2=p_{\hat K}^2=\tilde{p}_m^2=0$ and where the bar denotes rescaling of both of the initial state partons and the $m$ runs over all the particles in the final state that are not actually part of the antenna but require boosting in order to restore momentum conservation.

 The $x_i$ and $x_l$ are given by \cite{Daleo:2006xa}:
\begin{eqnarray}
x_i&=&
\sqrt{\frac{s_{il}+s_{jl}+s_{kl}}{s_{il}+s_{ij}+s_{ik}}}
\sqrt{\frac{s_{ij}+s_{ik}+s_{il}+s_{jk}+s_{jl}+s_{kl}}{s_{il}}},\nonumber\\
x_l&=&
\sqrt{\frac{s_{il}+s_{ij}+s_{ik}}{s_{il}+s_{jl}+s_{kl}}}
\sqrt{\frac{s_{ij}+s_{ik}+s_{il}+s_{jk}+s_{jl}+s_{kl}}{s_{il}}}.\nonumber\\
\end{eqnarray}
In the double unresolved limits:
\begin{enumerate}
\item[(1)] $j$ and $k$ soft: $x_i\rightarrow 1$, $x_l\rightarrow 1$,
\item[(2)] $j$ soft and $p_{k}=z_ip_i\parallel p_i$: $x_i\rightarrow (1-z_i)$, $x_l\rightarrow 1$,
\item[(3)] $p_{j}=z_ip_i\parallel p_i$ and $p_{k}=z_lp_l\parallel p_l$:
  $x_i\rightarrow (1-z_i)$, $x_l\rightarrow (1-z_l)$,
\item[(4)] $p_{j}+p_{k}=z_ip_i\parallel p_i$: $x_i\rightarrow (1-z_i)$,
  $x_l\rightarrow 1$,
\end{enumerate}
together with the limits obtained from these by the exchange of $i$ and
$l$ and of $j$ with $k$. Moreover in the single unresolved limits this mapping collapses into a NLO mapping (\ref{3to2IImap}), thereby allowing the subtraction of the single resolved limits from the four-parton antennae functions.

The first results for the integrated form of the initial-initial four parton antenna over the antenna phase space (initial state kinematics) are given in ref.~\cite{Boughezal:2010ty}.

\section{Construction of the NNLO subtraction term for gluon scattering}
\label{sec:DsigmaSNNLO}
In this section we will consider the six-gluon contribution to the NNLO di-jet cross section.  This double real radiation process requires subtraction of all double unresolved and single unresolved singularities. At leading colour, the double real radiation contribution is given by,
 \begin{eqnarray}
{\rm d}\hat\sigma_{NNLO}^R&=& {\cal N} \left(\frac{\alpha_s N}{2\pi}\right)^2
{\rm d}\Phi_{4}(p_{3},\ldots,p_{6};p_1,p_2)\, \JET_{2}^{(4)}(p_{3},\ldots,p_{6})
 \, \nonumber\\
&\times &\frac{1}{4!} \left[\sum_{\sigma\in S_6/Z_6}
A_{6}^0 (\sigma(1),\ldots,\sigma(6))
+{\cal O}\left(\frac{1}{N^2}\right)\right].\label{eq:6gNNLO} 
\end{eqnarray}
We systematically drop these sub-leading colour terms from the subsequent discussion.

Therefore, depending on the position of the initial state gluons in the colour ordered matrix elements, we have three different topologies. These are labelled by the the colour ordering of initial and final state gluons.   We denote the configurations where the two initial state gluons are colour connected (i.e. adjacent) as IIFFFF, those where the colour ordering allows one final state gluon to be sandwiched between the initial state gluons is denoted by IFIFFF, while the third configuration with two final state gluons inserted between the initial state gluons is labelled IFFIFF such that, 
\begin{equation}
{\rm d}\hat\sigma_{NNLO}^R
=
{\rm d}\hat\sigma_{NNLO}^{R,IIFFFF}+
{\rm d}\hat\sigma_{NNLO}^{R,IFIFFF}+
{\rm d}\hat\sigma_{NNLO}^{R,IFFIFF},
\end{equation}
where,
\begin{eqnarray}
{\rm d}\hat\sigma_{NNLO}^{R,IIFFFF}&=& {\cal N} \left(\frac{\alpha_s N}{2\pi}\right)^2
{\rm d}\Phi_{4}(p_{3},\ldots,p_{6};p_1,p_2)\, \JET_{2}^{(4)}(p_{3},\ldots,p_{6})
 \, \nonumber\\
&\times &
 \frac{2}{4!}\,
\sum_{P(i,j,k,l)\in(3,4,5,6)}
A_{6}^0 (\hat{1}_g,\hat{2}_g,i_g,j_g,k_g,l_g),\label{eq:RIIFFFF}\\
{\rm d}\hat\sigma_{NNLO}^{R,IFIFFF}&=& {\cal N} \left(\frac{\alpha_s N}{2\pi}\right)^2
{\rm d}\Phi_{4}(p_{3},\ldots,p_{6};p_1,p_2)\, \JET_{2}^{(4)}(p_{3},\ldots,p_{6})
 \, \nonumber\\
&\times &
 \frac{2}{4!}\,
\sum_{P(i,j,k,l)\in(3,4,5,6)}
A_{6}^0 (\hat{1}_g,i_g,\hat{2}_g,j_g,k_g,l_g),\label{eq:RIFIFFF}\\
{\rm d}\hat\sigma_{NNLO}^{R,IFFIFF}&=& {\cal N} \left(\frac{\alpha_s N}{2\pi}\right)^2
{\rm d}\Phi_{4}(p_{3},\ldots,p_{6};p_1,p_2)\, \JET_{2}^{(4)}(p_{3},\ldots,p_{6})
 \, \nonumber\\
&\times &
 \frac{2}{4!}\,
\sum_{P_C(i,j,k,l)\in(3,4,5,6)}
A_{6}^0 (\hat{1}_g,i_g,j_g,\hat{2}_g,k_g,l_g).\label{eq:RIFFIFF}
\end{eqnarray}
where the first two sums are over the 4! permutations of the final state gluons 
and in the last sum only $4!/2$ cyclic permutations are summed. In the following subsections we will write down the counterterm that regularises the infrared divergences of the double real correction for each topology separately.

\subsection{IIFFFF topology}

It is convenient to rearrange the twenty-four six-gluon amplitudes present in eq.~\eqref{eq:RIIFFFF} into three terms,
\begin{eqnarray}
\sum_{(i,j,k,l)\in P(3,4,5,6) }
A^0_{6}(\hat{1}_g,\hat{2}_g,i_g,j_g,k_g,l_g)
&=& X^0_{6}(\hat{1}_g,\hat{2}_g,3_g,4_g,5_g,6_g)\nonumber\\ 
&+& X^0_{6}(\hat{1}_g,\hat{2}_g,3_g,5_g,4_g,6_g)\nonumber\\ 
&+& X^0_{6}(\hat{1}_g,\hat{2}_g,3_g,4_g,6_g,5_g),
\end{eqnarray}
where each $X_6^0$ contains eight colour ordered squared amplitudes given by the 4 cyclic permutations of the final state gluons plus their line reversals:
\begin{eqnarray}
X_6^0(\hat{1}_g,\hat{2}_g,3_g,4_g,5_g,6_g)&=&
\phantom{+}A_6^0(\hat{1}_g,\hat{2}_g,3_g,4_g,5_g,6_g)+A_6^0(\hat{1}_g,\hat{2}_g,6_g,5_g,4_g,3_g)\nonumber\\
&&+A_6^0(\hat{1}_g,\hat{2}_g,4_g,5_g,6_g,3_g)+A_6^0(\hat{1}_g,\hat{2}_g,3_g,6_g,5_g,4_g)\nonumber\\
&&+A_6^0(\hat{1}_g,\hat{2}_g,5_g,6_g,3_g,4_g)+A_6^0(\hat{1}_g,\hat{2}_g,4_g,3_g,6_g,5_g)\nonumber\\
&&+A_6^0(\hat{1}_g,\hat{2}_g,6_g,3_g,4_g,5_g)+A_6^0(\hat{1}_g,\hat{2}_g,5_g,4_g,3_g,6_g).\nonumber\\
\label{eq:IIFFFF}
\end{eqnarray}

For the numerical implementation we use the $X_6^0$ function because this form of the real correction is more appropriate for the construction of the subtraction term which will contain the final-final antenna $F_4^0(i_g,j_g,k_g,l_g)$ as well as the initial-final antenna $F_4^0(\hat{i}_g,j_g,k_g,l_g)$.
The final-final antenna has a cyclic ambiguity in the momentum arrangements, while the initial-final antenna has a line-reversal ambiguity.   $X_6^0$ therefore has the right symmetry to match onto both configurations.  

The remaining 16 colour orderings in this topology corresponding to the last two $X_6^0$ functions are obtained by permutations of the gluon indices when calling the IIFFFF routine. These actually yield the same contribution to the cross section and need not to be evaluated. However if we sum over all orderings the topology becomes symmetric with respect to all gluon indices and this is better for the Monte Carlo integration.

The real radiation subtraction term for $X_6^0(\hat{1}_g,\hat{2}_g,3_g,4_g,5_g,6_g)$ is explicitly symmetric under cyclic permutations of the gluons
\begin{eqnarray}
(i,j,k,l)\equiv (3,4,5,6),(4,5,6,3),(5,6,3,4),(6,3,4,5)\nonumber
\end{eqnarray}
and is given by,
\begin{eqnarray}
\lefteqn{{\rm d}\hat\sigma_{NNLO}^{S,X_6}= {\cal N} \left(\frac{\alpha_s N}{2\pi}\right)^2 
{\rm d}\Phi_{4}(p_{3},\ldots,p_{6};p_{1},p_{2})
 \, \frac{2}{4!}\,\sum_{(ijkl)}\,\Bigg\{ }\nonumber \\
\ph{(1a1)}&&\phantom{+}f_3^0(\hat{2}_g,i_g,j_g)\,A^0_5(\hat{1}_g,\hat{\bar{2}}_g,(\widetilde{ij})_g,k_g,l_g)\,
{J}_{2}^{(3)}(\widetilde{p_{ij}},p_k,p_l)\nonumber \\
\ph{(1a2)}&&+f_3^0(i_g,j_g,k_g)A^0_5(\hat{1}_g,\hat{2}_g,(\widetilde{ij})_g,(\widetilde{jk})_g,l_g)
{J}_{2}^{(3)}(\widetilde{p_{ij}},\widetilde{p_{jk}},p_l)\nonumber \\
\ph{(1a3)}&&+f_3^0(j_g,k_g,l_g)A^0_5(\hat{1}_g,\hat{2}_g,i_g,(\widetilde{jk})_g,(\widetilde{kl})_g)
{J}_{2}^{(3)}(p_i,\widetilde{p_{jk}},\widetilde{p_{kl}})\nonumber \\
\ph{(1a4)}&&+f_3^0(k_g,l_g,\hat{1}_g)A^0_5(\hat{\bar{1}}_g,\hat{2}_g,i_g,j_g,(\widetilde{kl})_g)
{J}_{2}^{(3)}(p_i,p_j,\widetilde{p_{kl}})\nonumber \\
&&\nonumber\\
\ph{(2a1)}&&+f_3^0(\hat{2}_g,l_g,k_g)\,A^0_5(\hat{1}_g,\hat{\bar{2}}_g,(\widetilde{lk})_g,j_g,i_g)\,
{J}_{2}^{(3)}(\widetilde{p_{lk}},p_j,p_i)\nonumber \\
\ph{(2a2)}&&+f_3^0(l_g,k_g,j_g)A^0_5(\hat{1}_g,\hat{2}_g,(\widetilde{lk})_g,(\widetilde{kj})_g,i_g)
{J}_{2}^{(3)}(\widetilde{p_{lk}},\widetilde{p_{kj}},p_i)\nonumber \\
\ph{(2a3)}&&+f_3^0(k_g,j_g,i_g)A^0_5(\hat{1}_g,\hat{2}_g,l_g,(\widetilde{kj})_g,(\widetilde{ji})_g)
{J}_{2}^{(3)}(p_l,\widetilde{p_{kj}},\widetilde{p_{ji}})\nonumber \\
\ph{(2a4)}&&+f_3^0(j_g,i_g,\hat{1}_g)A^0_5(\hat{\bar{1}}_g,\hat{2}_g,l_g,k_g,(\widetilde{ji})_g)
{J}_{2}^{(3)}(p_l,p_k,\widetilde{p_{ji}})\nonumber \\
&&\nonumber\\
\ph{b1}&&+\bigg(F_{4,a}^0(i_g,j_g,k_g,l_g)-f_3^0(i_g,j_g,k_g)f_3^0(\widetilde{(ij)}_g,\widetilde{(jk)}_g,l_g)\nonumber\\
&&\hspace{0.3cm}-f_3^0(j_g,k_g,l_g)f_3^0(i_g,\widetilde{(jk)}_g,\widetilde{(kl)}_g)\bigg) A^0_4(\hat{1}_g,\hat{2}_g,\widetilde{(ijk)}_g,\widetilde{(lkj)}_g)
J_2^{(2)}(\widetilde{p_{ijk}},\widetilde{p_{lkj}})\nonumber\\
\ph{b2}&&+\bigg(F_{4,b}^0(i_g,j_g,k_g,l_g)\nonumber\\
&&\hspace{0.3cm}-f_3^0(i_g,j_g,k_g)f_3^0(\widetilde{(ij)}_g,l_g,\widetilde{(jk)}_g)\bigg)
A_4^0(\hat{1}_g,\hat{2}_g,\widetilde{(ijl)}_g,\widetilde{(klj)}_g)
J_2^{(2)}(\widetilde{p_{ijl}},\widetilde{p_{klj}})\nonumber\\
\ph{b3}&&+\frac{1}{2}f_3^0(i_g,j_g,k_g)f_3^0(\widetilde{(ij)}_g,l_g,\widetilde{(jk)}_g)
A_4^0(\hat{1}_g,\hat{2}_g,\widetilde{((ij)l)}_g,\widetilde{((jk)l)}_g)
J_2^{(2)}(\widetilde{p_{(ij)l}},\widetilde{p_{(jk)l}})\nonumber\\
&&\nonumber\\
\ph{b4}&&+\bigg(F_{4,a}^0(l_g,k_g,j_g,i_g)-f_3^0(l_g,k_g,j_g)f_3^0(\widetilde{(lk)}_g,\widetilde{(kj)}_g,i_g)\nonumber\\
&&\hspace{0.3cm}-f_3^0(k_g,j_g,i_g)f_3^0(l_g,\widetilde{(kj)}_g,\widetilde{(ji)}_g)\bigg)
A_4^0(\hat{1}_g,\hat{2}_g,\widetilde{(lkj)}_g,\widetilde{(ijk)}_g)
J_2^{(2)}(\widetilde{p_{lkj}},\widetilde{p_{ijk}})\nonumber\\
\ph{b5}&&+\bigg(F_{4,b}^0(l_g,k_g,j_g,i_g)\nonumber\\
&&-f_3^0(l_g,k_g,j_g)f_3^0(\widetilde{(lk)}_g,i_g,\widetilde{(kj)}_g)\bigg)
A_4^0(\hat{1}_g,\hat{2}_g,\widetilde{(lki)}_g,\widetilde{(jik)}_g)
J_2^{(2)}(\widetilde{p_{lki}},\widetilde{p_{jik}})\nonumber\\
&&+\frac{1}{2}f_3^0(l_g,k_g,j_g)f_3^0(\widetilde{(lk)}_g,i_g,\widetilde{(kj)}_g)
A_4^0(\hat{1}_g,\hat{2}_g,\widetilde{((lk)i)}_g,\widetilde{((kj)i)}_g)
J_2^{(2)}(\widetilde{p_{(lk)i}},\widetilde{p_{(kj)i}})\nonumber\\
&&\nonumber\\
\ph{b6}&&+\bigg(F_4^0(\hat{1}_g,l_g,k_g,j_g)\nonumber\\
&&\hspace{0.3cm}-f_3^0(\hat{1}_g,l_g,k_g)F_3^0(\hat{\bar{1}}_g,\widetilde{(lk)}_g,j_g)
-f_3^0(l_g,k_g,j_g)F_3^0(\hat{1}_g,\widetilde{(lk)}_g,\widetilde{(kj)}_g)\nonumber\\
&&\hspace{0.3cm}-f_3^0(k_g,j_g,\hat{1}_g)F_3^0(\hat{\bar{1}}_g,l_g,\widetilde{(jk)}_g)\bigg)
A_4^0(\hat{\bar{1}}_g,\hat{2}_g,i_g,\widetilde{(lkj)}_g)J_2^{(2)}(p_i,\widetilde{p_{lkj}})\nonumber\\
\ph{b7}&&+\bigg(F_4^0(\hat{2}_g,i_g,j_g,k_g)\nonumber\\
&&\hspace{0.3cm}-f_3^0(\hat{2}_g,i_g,j_g)F_3^0(\hat{\bar{2}}_g,\widetilde{(ij)}_g,k_g)
-f_3^0(i_g,j_g,k_g)F_3^0(\hat{2}_g,\widetilde{(ij)}_g,\widetilde{(jk)}_g)\nonumber\\
&&\hspace{0.3cm}-f_3^0(j_g,k_g,\hat{2}_g)F_3^0(\hat{\bar{2}}_g,i_g,\widetilde{(kj)}_g)\bigg)
A_4^0(\hat{1}_g,\hat{\bar{2}}_g,\widetilde{(ijk)}_g,l_g)
J_2^{(2)}(p_{ijk},p_l)\nonumber\\
\ph{b8}&&+\frac{1}{2}f_3^0(\hat{1}_g,l_g,k_g)f_3^0(\hat{\bar{1}}_g,j_g,\widetilde{(lk)}_g)
A_4^0(\hat{\bar{\bar{1}}}_g,\hat{2}_g,i_g,\widetilde{((lk)j)}_g)
J_2^{(2)}(p_i,\widetilde{p_{(lk)j}})\nonumber\\
\ph{b9}&&+\frac{1}{2}f_3^0(\hat{1}_g,j_g,k_g)f_3^0(\hat{\bar{1}}_g,l_g,\widetilde{(jk)}_g)
A_4^0(\hat{\bar{\bar{1}}}_g,\hat{2}_g,i_g,\widetilde{((jk)l)}_g)
J_2^{(2)}(p_i,\widetilde{p_{(jk)l}})\nonumber\\
\ph{b10}&&+\frac{1}{2}f_3^0(\hat{2}_g,i_g,j_g)f_3^0(\hat{\bar{2}}_g,k_g,\widetilde{(ij)}_g)
A_4^0(\hat{1}_g,\hat{\bar{\bar{2}}}_g,\widetilde{((ij)k)}_g,l_g)
J_2^{(2)}(\widetilde{p_{(ij)k}},p_l)\nonumber\\
\ph{b11}&&+\frac{1}{2}f_3^0(\hat{2}_g,k_g,j_g)f_3^0(\hat{\bar{2}}_g,i_g,\widetilde{(kj)}_g)
A_4^0(\hat{1}_g,\hat{\bar{\bar{2}}}_g,\widetilde{((jk)i)}_g,l_g)
J_2^{(2)}(\widetilde{p_{(jk)i}},p_l)\nonumber\\
&&\nonumber\\
\ph{b12}&&-\bigg(F_4^0(\hat{1}_g,i_g,\hat{2}_g,k_g)
-F_3^0(\hat{1}_g,i_g,\hat{2}_g)F_3^0(\hat{\bar{1}}_g,\hat{\bar{2}}_g,\tilde{k}_g)\nonumber\\
&&\hspace{0.3cm}-F_3^0(\hat{1}_g,k_g,\hat{2}_g)F_3^0(\hat{\bar{1}}_g,\hat{\bar{2}}_g,\tilde{i}_g)\bigg)
A_4^0(\hat{\bar{1}}_g,\hat{\bar{2}}_g,\tilde{j}_g,\tilde{l}_g)
J_2^{(2)}(\tilde{p_j},\tilde{p_l})\nonumber\\
\ph{b13}&&-\frac{1}{2}F_3^0(\hat{1}_g,i_g,\hat{2}_g)F_3^0(\hat{\bar{1}}_g,\hat{\bar{2}}_g,\tilde{k}_g)
A_4^0(\hat{\bar{\bar{1}}}_g,\hat{\bar{\bar{2}}}_g,\tilde{\tilde{j}}_g,\tilde{\tilde{l}}_g)
J_2^{(2)}(\tilde{\tilde{p_j}},\tilde{\tilde{p_l}})\nonumber\\
\ph{b14}&&-\frac{1}{2}F_3^0(\hat{1}_g,k_g,\hat{2}_g)F_3^0(\hat{\bar{1}}_g,\hat{\bar{2}}_g,\tilde{i}_g)
A_4^0(\hat{\bar{\bar{1}}}_g,\hat{\bar{\bar{2}}}_g,\tilde{\tilde{j}}_g,\tilde{\tilde{l}}_g)
J_2^{(2)}(\tilde{\tilde{p_j}},\tilde{\tilde{p_l}})\nonumber\\
&&\nonumber\\
\ph{c1}&&-\frac{1}{2}f_3^0(\hat{2}_g,i_g,j_g)f_3^0(l_g,k_g,\widetilde{(ij)}_g)
A_4^0(\hat{1}_g,\hat{\bar{2}}_g,\widetilde{((ij)k)}_g,\widetilde{(kl)}_g)
J_2^{(2)}(\widetilde{p_{(ij)k}},\widetilde{p_{kl}})
\nonumber\\
\ph{c2}&&-\frac{1}{2}f_3^0(l_g,k_g,j_g)f_3^0(\hat{2}_g,i_g,\widetilde{(kj)}_g)
A_4^0(\hat{1}_g,\hat{\bar{2}}_g,\widetilde{((kj)i)}_g,\widetilde{(kl)}_g)
J_2^{(2)}(\widetilde{p_{(kj)i}},\widetilde{p_{kl}})
\nonumber\\
\ph{c3}&&-\frac{1}{2}f_3^0(\hat{1}_g,l_g,k_g)f_3^0(i_g,j_g,\widetilde{(lk)}_g)
A_4^0(\hat{\bar{1}}_g,\hat{2}_g,\widetilde{(ij)}_g,\widetilde{((lk)j)}_g)
J_2^{(2)}(\widetilde{p_{ij}},\widetilde{p_{(lk)j}})\nonumber\\
\ph{c4}&&-\frac{1}{2}f_3^0(i_g,j_g,k_g)f_3^0(\hat{1}_g,l_g,\widetilde{(jk)}_g)
A_4^0(\hat{\bar{1}}_g,\hat{2}_g,\widetilde{(ij)}_g,\widetilde{((jk)l)}_g)
J_2^{(2)}(\widetilde{p_{ij}},\widetilde{p_{(jk)l}})\nonumber\\
\ph{c4}&&-\frac{1}{2}f_3^0(\hat{2}_g,l_g,k_g)f_3^0(i_g,j_g,\widetilde{(lk)}_g)
A_4^0(\hat{1}_g,\hat{\bar{2}}_g,\widetilde{((lk)j)}_g,\widetilde{(ij)}_g)
J_2^{(2)}(\widetilde{p_{(lk)j}},\widetilde{p_{ij}})\nonumber\\
\ph{c5}&&-\frac{1}{2}f_3^0(i_g,j_g,k_g)f_3^0(\hat{2}_g,l_g,\widetilde{(jk)}_g)
A_4^0(\hat{1}_g,\hat{\bar{2}}_g,\widetilde{((jk)l)}_g,\widetilde{(ij)}_g)
J_2^{(2)}(\widetilde{p_{(jk)l},\widetilde{p_{ij}}})\nonumber\\
\ph{c6}&&-\frac{1}{2}f_3^0(\hat{1}_g,i_g,j_g)f_3^0(l_g,k_g,\widetilde{(ij)}_g)
A_4^0(\hat{\bar{1}}_g,\hat{2}_g,\widetilde{(kl)}_g,\widetilde{((ij)k)}_g)
J_2^{(2)}(\widetilde{p_{kl}},\widetilde{p_{(ij)k}})\nonumber\\
\ph{c7}&&-\frac{1}{2}f_3^0(l_g,k_g,j_g)f_3^0(\hat{1}_g,i_g,\widetilde{(kj)}_g)
A_4^0(\hat{\bar{1}}_g,\hat{2}_g,\widetilde{(kl)}_g,\widetilde{((kj)i)}_g)
J_2^{(2)}(\widetilde{p_{kl}},\widetilde{p_{(kj)i}})\nonumber\\
&&\nonumber\\
\ph{d1}&&-f_3^0(\hat{2}_g,i_g,j_g)f_3^0(k_g,l_g,\hat{1}_g)
A_4^0(\hat{\bar{1}}_g,\hat{\bar{2}}_g,\widetilde{(ij)}_g,\widetilde{(lk)}_g)
J_2^{(2)}(\widetilde{p_{ij}},\widetilde{p_{lk}})\nonumber\\
\ph{d2}&&-f_3^0(\hat{2}_g,l_g,k_g)f_3^0(j_g,i_g,\hat{1}_g)
A_4^0(\hat{\bar{1}}_g,\hat{\bar{2}}_g,\widetilde{(lk)}_g,\widetilde{(ij)}_g)
J_2^{(2)}(\widetilde{p_{lk}},\widetilde{p_{ij}})\nonumber\\
&&\nonumber\\
\ph{fc1}&&+\frac{1}{2}f_3^0(\hat{2}_g,k_g,j_g)F_3^0(\hat{1}_g,\hat{\bar{2}}_g,i_g)
A_4^0(\hat{\bar{1}}_g,\hat{\bar{\bar{2}}}_g,\widetilde{(kj)}_g,\tilde{l}_g)
J_2^{(2)}(\widetilde{p_{kl}},\tilde{p_l})\nonumber\\
&&-\frac{1}{2}F_3^0(\hat{1}_g,\hat{2}_g,i_g)f_3^0(\hat{\bar{2}}_g,\tilde{k}_g,\tilde{j}_g)
A_4^0(\hat{\bar{1}}_g,\hat{\bar{\bar{2}}}_g,\widetilde{(kj)}_g,\tilde{l}_g)
J_2^{(2)}(\widetilde{p_{kj}},\tilde{p_l})\nonumber\\
\ph{fc2}&&+\frac{1}{2}f_3^0(\hat{2}_g,i_g,j_g)F_3^0(\hat{1}_g,\hat{\bar{2}}_g,k_g)
A_4^0(\hat{\bar{1}}_g,\hat{\bar{\bar{2}}}_g,\widetilde{(ij)}_g,\tilde{l}_g)
J_2^{(2)}(\widetilde{p_{ij}},\widetilde{p_l})\nonumber\\
&&-\frac{1}{2}F_3^0(\hat{1}_g,\hat{2}_g,k_g)f_3^0(\hat{\bar{2}}_g,\tilde{i}_g,\tilde{j}_g)
A_4^0(\hat{\bar{1}}_g,\hat{\bar{\bar{2}}}_g,\widetilde{(ij)}_g,\tilde{l}_g)
J_2^{(2)}(\widetilde{p_{ij}},\tilde{p_l})\nonumber\\
\ph{fc3}&&+\frac{1}{2}f_3^0(\hat{1}_g,k_g,j_g)F_3^0(\hat{\bar{1}}_g,\hat{2}_g,i_g)
A_4^0(\hat{\bar{\bar{1}}}_g,\hat{\bar{2}}_g,\tilde{l}_g,\widetilde{(kj)}_g)
J_2^{(2)}(\tilde{p_l},\widetilde{p_{kj}})\nonumber\\
&&-\frac{1}{2}F_3^0(\hat{1}_g,\hat{2}_g,i_g)f_3^0(\hat{\bar{1}}_g,\tilde{k}_g,\tilde{j}_g)
A_4^0(\hat{\bar{\bar{1}}}_g,\hat{\bar{2}}_g,\tilde{l}_g,\widetilde{(kj)}_g)
J_2^{(2)}(\tilde{p_l},\widetilde{p_{kj}})\nonumber\\
\ph{fc4}&&+\frac{1}{2}f_3^0(\hat{1}_g,i_g,j_g)F_3^0(\hat{\bar{1}}_g,\hat{2}_g,l_g)
A_4^0(\hat{\bar{\bar{1}}}_g,\hat{\bar{2}}_g,\tilde{k}_g,\widetilde{(ij)}_g)
J_2^{(2)}(\tilde{p_k},\widetilde{p_{ij}})\nonumber\\
&&-\frac{1}{2}F_3^0(\hat{1}_g,\hat{\bar{2}}_g,l_g)f_3^0(\hat{\bar{1}}_g,\tilde{i}_g,\tilde{j}_g)
A_4^0(\hat{\bar{\bar{1}}}_g,\hat{\bar{2}}_g,\tilde{k}_g,\widetilde{(ij)}_g)
J_2^{(2)}(\tilde{p_k},\widetilde{p_{ij}})\Bigg\}.
\label{eq:nnlotop1} 
\end{eqnarray}

In the numerical study,  we explicitly implemented the summation over $(i,j,k,l)$. This is needed to reconstruct the full $F_4^0(i_g,j_g,k_g,l_g)$ from the $F_{4,a}$ and $F_{4,b}$ as in (\ref{eq:F40decomp}) so that only the full $F_4^0(i_g,j_g,k_g,l_g)$ needs to be analytically integrated over the antenna phase-space.
It is important to note ${\rm d}\hat\sigma_{NNLO}^{S,X_6}$ introduces spurious contributions from large
angle soft radiation which are cancelled by an additional subtraction term ${\rm d}\hat\sigma_{NNLO}^{A,X_6}$, as discussed in section~{\ref{sec:LAST}}, 
\begin{eqnarray}
\lefteqn{{\rm d}\hat\sigma_{NNLO}^{A,X_6}= {\cal N} \left(\frac{\alpha_s N}{2\pi}\right)^2 
{\rm d}\Phi_{4}(p_{3},\ldots,p_{6};p_{1},p_{2})
 \, \frac{2}{4!}\,\sum_{(ijkl)}\,\Bigg\{ }\nonumber \\
\ph{(ss2)}&&\phantom{+}\frac{1}{2}\Big(-S_{2l((il)j)}+S_{2l(il)}-S_{1l((kl)j)}+S_{1l(kl)}
+S_{((il)j)l((kl)j)}-S_{(il)l(kl)}\Big)\nonumber\\
&&\hspace{0.5cm}\times f_3^0(\widetilde{(il)}_g,j_g,\widetilde{(kl)}_g)
A_4^0(\hat{1}_g,\hat{2}_g,\widetilde{((il)j)}_g,\widetilde{((kl)j)}_g)
J_2^{(2)}(\widetilde{p_{(il)j}},\widetilde{p_{(kl)j}})\nonumber\\
\ph{ss1}&&+\frac{1}{2}\Big(-S_{2i((il)k)}+S_{2i(il)}-S_{1i((ij)k)}+S_{1i(ij)}
+S_{((il)k)i((ij)k)}-S_{(il)i(ij)}\Big)\nonumber\\
&&\hspace{0.5cm}\times f_3^0(\widetilde{(il)}_g,k_g,\widetilde{(ij)}_g)
A_4^0(\hat{1}_g,\hat{2}_g,\widetilde{((il)k)}_g,\widetilde{((ij)k)}_g)
J_2^{(2)}(\widetilde{p_{(il)k}},\widetilde{p_{(ij)k}})\nonumber\\
\ph{ss3}&&+\frac{1}{2}\Big(-S_{il((kl)j)}+S_{il(kl)}+S_{\bar{\bar{1}}l((kl)j)}-S_{\bar{1}l(kl)}
-S_{\bar{\bar{1}}l2}+S_{\bar{1}l2}\Big)\nonumber\\
&&\hspace{0.5cm}\times f_3^0(\hat{\bar{1}}_g,j_g,\widetilde{(kl)}_g)
A_4^0(\hat{\bar{\bar{1}}}_g,\hat{2}_g,i_g,\widetilde{((kl)j)}_g)
J_2^{(2)}(\tilde{p_i},\widetilde{p_{(kl)j}})\nonumber\\
\ph{ss4}&&+\frac{1}{2}\Big(-S_{ij((jk)l)}+S_{ij(jk)}+S_{\bar{\bar{1}}j((jk)l)}-S_{\bar{1}j(jk)}
-S_{\bar{\bar{1}}j2}+S_{\bar{1}j2}\Big)\nonumber\\
&&\hspace{0.5cm}\times f_3^0(\hat{\bar{1}}_g,l_g,\widetilde{(kj)}_g)
A_4^0(\hat{\bar{\bar{1}}}_g,\hat{2}_g,i_g,\widetilde{((kj)l)}_g)
J_2^{(2)}(p_i,\widetilde{p_{(kj)l}})\nonumber\\
\ph{ss5}&&+\frac{1}{2}\Big(-S_{lk((jk)i)}+S_{lk(jk)}+S_{\bar{\bar{2}}k((jk)i)}-S_{\bar{2}k(jk)}
-S_{1k\bar{\bar{2}}}+S_{1k\bar{2}}\Big)\nonumber\\
&&\hspace{0.5cm}\times f_3^0(\hat{\bar{2}}_g,i_g,\widetilde{(jk)}_g)
A_4^0(\hat{1}_g,\hat{\bar{\bar{2}}}_g,\widetilde{((jk)i)}_g,l_g)
J_2^{(2)}(\widetilde{p_{(jk)i}},p_l)\nonumber\\
\ph{ss6}&&+\frac{1}{2}\Big(-S_{li((ij)k)}+S_{li(ij)}+S_{\bar{\bar{2}}i((ij)k)}-S_{\bar{2}i(ij)}
-S_{1i\bar{\bar{2}}}+S_{1i\bar{2}}\Big)\nonumber\\
&&\hspace{0.5cm}\times f_3^0(\hat{\bar{2}}_g,k_g,\widetilde{(ij)}_g)
A_4^0(\hat{1}_g,\hat{\bar{\bar{2}}}_g,\widetilde{((ij)k)}_g,l_g)
J_2^{(2)}(\widetilde{p_{(ij)k}},p_l)\nonumber\\
\ph{ss7}&&+\frac{1}{2}\Big(-S_{\bar{\bar{1}}\tilde{i}\bar{\bar{2}}}+S_{\bar{1}i\bar{2}}-S_{\bar{2}i\tilde{j}}+S_{\bar{\bar{2}}\tilde{i}\tilde{\tilde{j}}}-S_{\bar{1}i\tilde{l}}+S_{\bar{\bar{1}}\tilde{i}\tilde{\tilde{l}}}\Big)\nonumber\\
&&\hspace{0.5cm}\times F_3^0(\hat{\bar{1}}_g,\tilde{k}_g,\hat{\bar{2}}_g)
A_4^0(\hat{\bar{\bar{1}}}_g,\hat{\bar{\bar{2}}}_g,\tilde{\tilde{j}}_g,\tilde{\tilde{l}}_g)
J_2^{(2)}(\tilde{\tilde{p_j}},\tilde{\tilde{p_l}})\nonumber\\
\ph{ss8}&&+\frac{1}{2}\Big(-S_{\bar{\bar{1}}\tilde{k}\bar{\bar{2}}}+S_{\bar{1}k\bar{2}}-S_{\bar{2}k\tilde{j}}+S_{\bar{\bar{2}}\tilde{k}\tilde{\tilde{j}}}-S_{\bar{1}k\tilde{l}}+S_{\bar{\bar{1}}\tilde{k}\tilde{\tilde{l}}}\Big)\nonumber\\
&&\hspace{0.5cm}\times F_3^0(\hat{\bar{1}}_g,\tilde{i}_g,\hat{\bar{2}}_g)
A_4^0(\hat{\bar{\bar{1}}}_g,\hat{\bar{\bar{2}}}_g,\tilde{\tilde{j}}_g,\tilde{\tilde{l}}_g)
J_2^{(2)}(\tilde{\tilde{p_j}},\tilde{\tilde{p_l}})\bigg \}.
\label{eq:LAST1}
\end{eqnarray}

\subsection{IFIFFF topology}
Just as for the IIFFFF topology, it is convenient to organise the twenty-four six-gluon amplitudes of eq.~\ref{eq:RIFIFFF} in groups.  In this case,

\begin{eqnarray}
\sum_{(i,j,k,l)\in P(3,4,5,6) }
A^0_{6}(\hat{1}_g,i_g,\hat{2}_g,j_g,k_g,l_g) &=&Y_6^0(\hat{1}_g,3_g,\hat{2}_g,4_g,5_g,6_g)
+Y_6^0(\hat{1}_g,4_g,\hat{2}_g,5_g,6_g,3_g)\nonumber\\
&+&
Y_6^0(\hat{1}_g,5_g,\hat{2}_g,6_g,3_g,4_g)
+Y_6^0(\hat{1}_g,6_g,\hat{2}_g,3_g,4_g,5_g),\nonumber\\
\end{eqnarray}
where each $Y_6^0$ contains 6 colour ordered squared amplitudes where the first final state gluon index is kept fixed and we sum the 3 cyclic permutations of the remaining final state gluons:
\begin{eqnarray}
Y_6^0(\hat{1}_g,3_g,\hat{2}_g,4_g,5_g,6_g)
&=&\phantom{+}A_6^0(\hat{1}_g,3_g,\hat{2}_g,4_g,5_g,6_g)
+A_6^0(\hat{1}_g,3_g,\hat{2}_g,4_g,6_g,5_g)\nonumber\\
&&
+A_6^0(\hat{1}_g,3_g,\hat{2}_g,5_g,4_g,6_g)
+A_6^0(\hat{1}_g,3_g,\hat{2}_g,5_g,6_g,4_g)\nonumber\\
&&
+A_6^0(\hat{1}_g,3_g,\hat{2}_g,6_g,4_g,5_g)
+A_6^0(\hat{1}_g,3_g,\hat{2}_g,6_g,5_g,4_g).\nonumber\\
\end{eqnarray}
This structure is more appropriate for the construction of 
the real radiation subtraction term, since it matches 
onto the symmetry of the initial-final and initial-initial antenna functions. $F_4^0(\hat{2},j,k,l)$
contains the unresolved limits of $j$ and $k$ between hard radiators $\hat{2}$ and $l$ and the unresolved limits 
of $k$ and $l$ between the radiators $\hat{2}$ and $j$. It is suitable to use $F_4^0(\hat{2},j,k,l)$ to subtract the singular limits 
of the combination of ordered amplitudes $A^0_{6}(\hat{1},i,\hat{2},j,k,l) + A^0_{6}(\hat{1},i,\hat{2},l,k,j)$.
Similarly, the initial-initial antenna function $F_4^0(\hat{1},i,\hat{2},j)$ subtracts double unresolved limits of the combination of ordered amplitudes $A_6^0(\hat{1},i,\hat{2},j,k,l)+A_6^0(\hat{1},i,\hat{2},k,l,j)$. Combining these symmetries brings the total number of ordered amplitudes present in $Y_6^0$ to six. 
The remaining eighteen amplitudes can be obtained by permutations of the gluon indices in $Y_6^0$.

The real radiation subtraction term for $Y_6^0(\hat{1},i,\hat{2},j,k,l)$ is,
\begin{eqnarray}
\lefteqn{{\rm d}\hat\sigma_{NNLO}^{S,Y_6}={\cal N} \left(\frac{\alpha_s N}{2\pi}\right)^2 
{\rm d}\Phi_{4}(p_{3},\ldots,p_{6};p_{1},p_{2})
 \, \frac{2}{4!}\,\sum_{P_C(j,k,l)}\Bigg\{ }\nonumber \\
\ph{(a1)}&&\phantom{+} F^0_3(\hat{1}_g,i_g,\hat{2}_g)\,
A^0_{5}(\hat{\bar{1}}_g,\hat{\bar{2}}_g,\tilde{j}_g,\tilde{k}_g,\tilde{l}_g) \,
{J}_{2}^{(3)}(\tilde{p}_j,\tilde{p}_k,\tilde{p}_l) 
\nonumber \\
\ph{(a2)}&&+ f^0_3(\hat{2}_g,j_g,k_g)\,
{A}^0_{5}(\hat{1}_g,i_g,\hat{\bar{2}}_g,\widetilde{(jk)}_g,l_g)\,
{J}_{2}^{(3)}(p_i,\widetilde{p_{jk}},p_l) 
\nonumber \\
\ph{(a3)}&&+ f^0_3(j_g,k_g,l_g)\,
{A}^0_{5}(\hat{1}_g,i_g,\hat{2}_g,\widetilde{(jk)}_g,\widetilde{(kl)}_g)\,
{J}_{2}^{(3)}(p_i,\widetilde{p_{jk}},\widetilde{p_{kl}}) 
\nonumber \\
\ph{(a4)}&&+ f^0_3(k_g,l_g,\hat{1}_g)\,
A^0_{5}(\hat{\bar{1}}_g,i_g,\hat{2}_g,j_g,\widetilde{(kl)}_g) \,
{J}_{2}^{(3)}(p_i,p_j,\widetilde{p_{kl}}) 
\nonumber \\ && \nonumber\\
\ph{(b1)}&&+ F^0_3(\hat{1}_g,i_g,\hat{2}_g)\,
A^0_{5}(\hat{\bar{1}}_g,\hat{\bar{2}}_g,\tilde{l}_g,\tilde{k}_g,\tilde{j}_g) \,
{J}_{2}^{(3)}(\tilde{p}_l,\tilde{p}_k,\tilde{p}_j) 
\nonumber \\
\ph{(b2)}&&+ f^0_3(\hat{2}_g,l_g,k_g)\,
{A}^0_{5}(\hat{1}_g,i_g,\hat{\bar{2}}_g,\widetilde{(lk)}_g,j_g)\,
{J}_{2}^{(3)}(p_i,\widetilde{p_{lk}},p_j) 
\nonumber \\
\ph{(b3)}&&+ f^0_3(l_g,k_g,j_g)\,
{A}^0_{5}(\hat{1}_g,i_g,\hat{2}_g,\widetilde{(lk)}_g,\widetilde{(kj)}_g)\,
{J}_{2}^{(3)}(p_i,\widetilde{p_{lk}},\widetilde{p_{kj}}) 
\nonumber \\
\ph{(b4)}&&+ f^0_3(k_g,j_g,\hat{1}_g)\,
A^0_{5}(\hat{\bar{1}}_g,i_g,\hat{2}_g,l_g,\widetilde{(jk)}_g) \,
{J}_{2}^{(3)}(p_i,p_l,\widetilde{p_{jk}}) 
\nonumber \\ && \nonumber\\
\ph{(g1,h1)}&&+
\Bigg( F_{4}^0(\hat{2}_g,j_g,k_g,l_g)\nonumber\\
&&\hspace{0.3cm} - f_3^0(\hat{2}_g,j_g,k_g)\,
 F_3^0(\hat{\bar{2}}_g,\widetilde{(jk)}_g,l_g)
 - f_3^0(j_g,k_g,l_g)\, F_3^0(\hat{2}_g,\widetilde{(jk)}_g,\widetilde{(kl)}_g)
\nonumber \\
\ph{(i1)}&& \hspace{0.3cm}
- f_3^0(k_g,l_g,\hat{2}_g)\,
 F_3^0(\hat{\bar{2}}_g,j_g,\widetilde{(kl)}_g)
\Bigg) A^0_{4}(\hat{1}_g,i_g,\hat{\bar{2}}_g,\widetilde{(jkl)}_g)\,
{J}_{2}^{(2)}(p_i,\widetilde{p_{jkl}})
  \nonumber \\
\ph{(g2,i2)}&&+
\Bigg( F_{4}^0(\hat{1}_g,l_g,k_g,j_g)\nonumber\\
&&\hspace{0.3cm} - f_3^0(\hat{1}_g,l_g,k_g)\,
 F_3^0(\hat{\bar{1}}_g,\widetilde{(kl)}_g,j_g)
 - f_3^0(l_g,k_g,j_g)\,F_3^0(\hat{1}_g,\widetilde{(lk)}_g,\widetilde{(kj)}_g)\nonumber\\
&&\hspace{0.3cm}- f_3^0(k_g,j_g,\hat{1}_g)\,F_3^0(\hat{\bar{1}}_g,l_g,\widetilde{(jk)}_g)
\Bigg) A^0_{4}(\hat{\bar{1}}_g,i_g,\hat{2}_g,\widetilde{(lkj)}_g)\,
{J}_{2}^{(2)}(p_i,\widetilde{p_{lkj}})
  \nonumber \\  && \nonumber\\
\ph{(r)}&& 
+ \frac{1}{2}f_3^0(\hat{2}_g,j_g,k_g) 
             f_3^0(\hat{\bar{2}}_g,l_g,\widetilde{(jk)}_g)\,
{A}^0_{4}(\hat{1}_g,i_g,\hat{\bar{\bar{2}}}_g,\widetilde{(jk)l}_g)\,
{J}_{2}^{(2)}(p_i,\widetilde{p_{(jk)l}}) 
  \nonumber \\
\ph{(s)}&& 
+ \frac{1}{2}f_3^0(\hat{2}_g,l_g,k_g) 
             f_3^0(\hat{\bar{2}}_g,j_g,\widetilde{(lk)}_g)\,
{A}^0_{4}(\hat{1}_g,i_g,\hat{\bar{\bar{2}}}_g,\widetilde{(lk)j}_g)\,
{J}_{2}^{(2)}(p_i,\widetilde{p_{(lk)j}}) 
  \nonumber \\
\ph{(t)}&& 
+ \frac{1}{2}f_3^0(\hat{1}_g,l_g,k_g) 
             f_3^0(\hat{\bar{1}}_g,j_g,\widetilde{(lk)}_g)\,
{A}^0_{4}(\hat{\bar{\bar{1}}}_g,i_g,\hat{2}_g,\widetilde{(lk)j}_g)\,
{J}_{2}^{(2)}(p_i,\widetilde{p_{(lk)j}}) 
  \nonumber \\
\ph{(u)}&& 
+ \frac{1}{2}f_3^0(\hat{1}_g,j_g,k_g) 
             f_3^0(\hat{\bar{1}}_g,l_g,\widetilde{(jk)}_g)\,
{A}^0_{4}(\hat{\bar{\bar{1}}}_g,i_g,\hat{2}_g,\widetilde{(jk)l}_g)\,
{J}_{2}^{(2)}(p_i,\widetilde{p_{(jk)l}}) 
  \nonumber \\ && \nonumber \\
\ph{(g1,h1)}&&+
\Bigg( F_{4}^0(\hat{1}_g,i_g,\hat{2}_g,j_g) - F_3^0(\hat{1}_g,i_g,\hat{2}_g)\,
 F_3^0(\hat{\bar{1}}_g,\hat{\bar{2}}_g,\tilde{j}_g)\nonumber\\
 &&\hspace{0.3cm}- F_3^0(\hat{2}_g,j_g,\hat{1}_g)\,
 F_3^0(\hat{\bar{1}}_g,\tilde{i}_g,\hat{\bar{2}}_g)
\Bigg) A^0_{4}(\hat{\bar{1}}_g,\hat{\bar{2}}_g,\tilde{k}_g,\tilde{l}_g)\,
{J}_{2}^{(2)}(\widetilde{p_{k}},\widetilde{p_{l}})
  \nonumber \\
\ph{(g1,h1)}&&+
\Bigg( F_{4}^0(\hat{1}_g,i_g,\hat{2}_g,l_g) - F_3^0(\hat{1}_g,i_g,\hat{2}_g)\,
 F_3^0(\hat{\bar{1}}_g,\hat{\bar{2}}_g,\tilde{l}_g)\nonumber\\
 &&\hspace{0.3cm}- F_3^0(\hat{2}_g,l_g,\hat{1}_g)\,
 F_3^0(\hat{\bar{1}}_g,\tilde{i}_g,\hat{\bar{2}}_g)
\Bigg) A^0_{4}(\hat{\bar{1}}_g,\hat{\bar{2}}_g,\tilde{k}_g,\tilde{j}_g)\,
{J}_{2}^{(2)}(\widetilde{p_{k}},\widetilde{p_{j}})
  \nonumber \\  && \nonumber\\
\ph{(w1)}&& + F_3^0(\hat{2}_g,j_g,\hat{1}_g) 
                F_3^0(\hat{\bar{1}}_g,\tilde{i}_g,\hat{\bar{2}}_g)\,
{A}^0_{4}(\hat{\bar{\bar{1}}}_g,\hat{\bar{\bar{2}}}_g,\tilde{\tilde{k}}_g,\tilde{\tilde{l}}_g)\,
{J}_{2}^{(2)}(\tilde{\tilde{p_k}},\tilde{\tilde{p_l}}) 
  \nonumber \\
\ph{(z1)}&& + F_3^0(\hat{2}_g,l_g,\hat{1}_g) 
                F_3^0(\hat{\bar{1}}_g,\tilde{i}_g,\hat{\bar{2}}_g)\,
{A}^0_{4}(\hat{\bar{\bar{1}}}_g,\hat{\bar{\bar{2}}}_g,\tilde{\tilde{k}}_g,\tilde{\tilde{j}}_g)\,
{J}_{2}^{(2)}(\tilde{\tilde{p_k}},\tilde{\tilde{p_j}}) 
 \nonumber \\
 && \nonumber\\
\ph{(g1,h1)}&&-
\Bigg( F_{4}^0(\hat{1}_g,l_g,\hat{2}_g,j_g) - F_3^0(\hat{1}_g,l_g,\hat{2}_g)\,
 F_3^0(\hat{\bar{1}}_g,\hat{\bar{2}}_g,\tilde{j}_g)\nonumber\\
 &&\hspace{0.3cm}- F_3^0(\hat{2}_g,j_g,\hat{1}_g)\,
 F_3^0(\hat{\bar{1}}_g,\tilde{l}_g,\hat{\bar{2}}_g)
\Bigg) A^0_{4}(\hat{\bar{1}}_g,\tilde{i}_g,\hat{\bar{2}}_g,\tilde{k}_g)\,
{J}_{2}^{(2)}(\widetilde{p_{i}},\widetilde{p_{k}})
  \nonumber \\
\ph{(v)}&& - \frac{1}{2}F_3^0(\hat{1}_g,l_g,\hat{2}_g) 
                F_3^0(\hat{\bar{1}}_g,\hat{\bar{2}}_g,\tilde{j}_g)\,
{A}^0_{4}(\hat{\bar{\bar{1}}}_g,\tilde{\tilde{i}}_g,\hat{\bar{\bar{2}}}_g,\tilde{\tilde{k}}_g)\,
{J}_{2}^{(2)}(\tilde{\tilde{p_i}},\tilde{\tilde{p_k}}) 
  \nonumber \\
\ph{(w)}&& - \frac{1}{2}F_3^0(\hat{2}_g,j_g,\hat{1}_g) 
                F_3^0(\hat{\bar{1}}_g,\tilde{l}_g,\hat{\bar{2}}_g)\,
{A}^0_{4}(\hat{\bar{\bar{1}}}_g,\tilde{\tilde{i}}_g,\hat{\bar{\bar{2}}}_g,\tilde{\tilde{k}}_g)\,
{J}_{2}^{(2)}(\tilde{\tilde{p_i}},\tilde{\tilde{p_k}}) \nonumber\\
&& \nonumber\\
\ph{(c1)}&& - f_3^0(\hat{2}_g,j_g,k_g) 
                F_3^0(\hat{1}_g,i_g,\hat{\bar{2}}_g)\,
{A}^0_{4}(\hat{\bar{1}}_g,\hat{\bar{\bar{2}}}_g,\widetilde{(jk)}_g,\tilde{l}_g)\,
{J}_{2}^{(2)}(\widetilde{p_{jk}},\tilde{p_l})
  \nonumber \\
\ph{(c3)}&& - f_3^0(\hat{1}_g,l_g,k_g) 
                F_3^0(\hat{\bar{1}}_g,i_g,\hat{2}_g)\,
{A}^0_{4}(\hat{\bar{\bar{1}}}_g,\hat{\bar{2}}_g,\tilde{j}_g,\widetilde{(lk)}_g)\,
{J}_{2}^{(2)}(\tilde{p_j},\widetilde{p_{lk}})
  \nonumber \\
\ph{(c5)}&& - \frac{1}{2}f_3^0(\hat{2}_g,j_g,k_g) 
                f_3^0(\hat{1}_g,l_g,\widetilde{(jk)}_g)\,
{A}^0_{4}(\hat{\bar{1}}_g,i_g,\hat{\bar{2}}_g,\widetilde{(jk)l}_g)\,
{J}_{2}^{(2)}(p_i,\widetilde{p_{(jk)l}})
  \nonumber \\
\ph{(c6)}&& - \frac{1}{2}f_3^0(\hat{1}_g,l_g,k_g) 
                f_3^0(\hat{2}_g,j_g,\widetilde{(lk)}_g)\,
{A}^0_{4}(\hat{\bar{1}}_g,i_g,\hat{\bar{2}}_g,\widetilde{(lk)j}_g)\,
{J}_{2}^{(2)}(p_i,\widetilde{p_{(lk)j}})
  \nonumber \\ && \nonumber\\
\ph{(c7)}&& - f_3^0(\hat{2}_g,l_g,k_g) 
                F_3^0(\hat{1}_g,i_g,\hat{\bar{2}}_g)\,
{A}^0_{4}(\hat{\bar{1}}_g,\hat{\bar{\bar{2}}}_g,\widetilde{(lk)}_g,\tilde{j}_g)\,
{J}_{2}^{(2)}(\widetilde{p_{lk}},\tilde{p_j})
  \nonumber \\
\ph{(c9)}&& - f_3^0(\hat{1}_g,j_g,k_g) 
                F_3^0(\hat{\bar{1}}_g,i_g,\hat{2}_g)\,
{A}^0_{4}(\hat{\bar{\bar{1}}}_g,\hat{\bar{2}}_g,\tilde{l}_g,\widetilde{(jk)}_g)\,
{J}_{2}^{(2)}(\tilde{p_l},\widetilde{p_{jk}})
  \nonumber \\\ph{(c11)}&& - \frac{1}{2}f_3^0(\hat{2}_g,l_g,k_g) 
                f_3^0(\hat{1}_g,j_g,\widetilde{(lk)}_g)\,
{A}^0_{4}(\hat{\bar{1}}_g,i_g,\hat{\bar{2}}_g,\widetilde{(lk)j}_g)\,
{J}_{2}^{(2)}(p_i,\widetilde{p_{(lk)j}})
  \nonumber \\
\ph{(c12)}&& - \frac{1}{2}f_3^0(\hat{1}_g,j_g,k_g) 
                f_3^0(\hat{2}_g,l_g,\widetilde{(jk)}_g)\,
{A}^0_{4}(\hat{\bar{1}}_g,i_g,\hat{\bar{2}}_g,\widetilde{(jk)l}_g)\,
{J}_{2}^{(2)}(p_i,\widetilde{p_{(jk)l}})
  \nonumber \\ && \nonumber\\
\ph{(d1)}&& - F_3^0(\hat{1}_g,i_g,\hat{2}_g) 
                f_3^0(\tilde{j}_g,\tilde{k}_g,\tilde{l}_g)\,
{A}^0_{4}(\hat{\bar{1}}_g,\hat{\bar{2}}_g,(\tilde{j}\tilde{k})_g,(\tilde{k}\tilde{l})_g)\,
{J}_{2}^{(2)}(p_{(\tilde{j}\tilde{k})},p_{(\tilde{k}\tilde{l})}) 
  \nonumber \\
\ph{(d2)}&& - F_3^0(\hat{1}_g,i_g,\hat{2}_g) 
                f_3^0(\tilde{l}_g,\tilde{k}_g,\tilde{j}_g)\,
{A}^0_{4}(\hat{\bar{1}}_g,\hat{\bar{2}}_g,(\tilde{l}\tilde{k})_g,(\tilde{k}\tilde{j})_g)\,
{J}_{2}^{(2)}(p_{(\tilde{l}\tilde{k})},p_{(\tilde{k}\tilde{j})})  
  \nonumber \\ && \nonumber\\
&&+\frac{1}{2}f_3^0(\hat{2}_g,j_g,k_g)F_3^0(\hat{1}_g,\hat{\bar{2}}_g,l_g)
A_4^0(\hat{\bar{1}}_g,\tilde{i}_g,\hat{\bar{\bar{2}}}_g,\widetilde{(jk)}_g)
{J}_{2}^{(2)}(\tilde{p_{i}},\widetilde{p_{jk}})\nonumber\\
&&-\frac{1}{2}F_3^0(\hat{1}_g,\hat{2}_g,l_g)f_3^0(\hat{\bar{2}}_g,\tilde{j}_g,\tilde{k}_g)
A^0_{4}(\hat{\bar{1}}_g,\tilde{i}_g,\hat{\bar{\bar{2}}}_g,(\tilde{j}\tilde{k})_g){J}_{2}^{(2)}(\tilde{p_{i}},p_{{\tilde{j}\tilde{k}}})\nonumber\\
&&+\frac{1}{2}f_3^0(\hat{2}_g,j_g,l_g)F_3^0(\hat{1}_g,\hat{\bar{2}}_g,k_g)
A_4^0(\hat{\bar{1}}_g,\tilde{i}_g,\hat{\bar{\bar{2}}}_g,\widetilde{(jl)}_g)
{J}_{2}^{(2)}(\tilde{p_{i}},\widetilde{p_{jl}})\nonumber\\
&&-\frac{1}{2}F_3^0(\hat{1}_g,\hat{2}_g,k_g)f_3^0(\hat{\bar{2}}_g,\tilde{j}_g,\tilde{l}_g)
A^0_{4}(\hat{\bar{1}}_g,\tilde{i}_g,\hat{\bar{\bar{2}}}_g,(\tilde{j}\tilde{l})_g)
{J}_{2}^{(2)}(\tilde{p_{i}},p_{{\tilde{j}\tilde{l}}})\nonumber\\ && \nonumber\\
&&+\frac{1}{2}f_3^0(\hat{1}_g,j_g,k_g)F_3^0(\hat{1}_g,\hat{\bar{2}}_g,l_g)
A_4^0(\hat{\bar{1}}_g,\tilde{i}_g,\hat{\bar{\bar{2}}}_g,\widetilde{(jk)}_g)
{J}_{2}^{(2)}(\tilde{p_{i}},\widetilde{p_{jk}})\nonumber\\
&&-\frac{1}{2}F_3^0(\hat{1}_g,\hat{2}_g,l_g)f_3^0(\hat{\bar{1}}_g,\tilde{j}_g,\tilde{k}_g)
A^0_{4}(\hat{\bar{1}}_g,\tilde{i}_g,\hat{\bar{\bar{2}}}_g,(\tilde{j}\tilde{k})_g)
{J}_{2}^{(2)}(\tilde{p_{i}},p_{{\tilde{j}\tilde{k}}})\nonumber\\
&&+\frac{1}{2}f_3^0(\hat{1}_g,j_g,l_g)F_3^0(\hat{1}_g,\hat{\bar{2}}_g,k_g)
A_4^0(\hat{\bar{1}}_g,\tilde{i}_g,\hat{\bar{\bar{2}}}_g,\widetilde{(jl)}_g)
{J}_{2}^{(2)}(\tilde{p_{i}},\widetilde{p_{jl}})\nonumber\\
&&-\frac{1}{2}F_3^0(\hat{1}_g,\hat{2}_g,k_g)f_3^0(\hat{\bar{1}}_g,\tilde{j}_g,\tilde{l}_g)
A^0_{4}(\hat{\bar{1}}_g,\tilde{i}_g,\hat{\bar{\bar{2}}}_g,(\tilde{j}\tilde{l})_g)
{J}_{2}^{(2)}(\tilde{p_{i}},p_{{\tilde{j}\tilde{l}}})
\nonumber \\ && \nonumber \\ 
\bigg\}.
\label{eq:nnlolc2}
\end{eqnarray}
As usual, ${\rm d}\hat\sigma_{NNLO}^{S,Y_6}$ introduces spurious limits from large
angle soft radiation
which are cancelled by an additional subtraction term ${\rm d}\hat\sigma_{NNLO}^{A,Y_6}$, as discussed in section~{\ref{sec:LAST}},
\begin{eqnarray}
\lefteqn{{\rm d}\hat\sigma_{NNLO}^{A,Y_6}={\cal N} \left(\frac{\alpha_s N}{2\pi}\right)^2  
{\rm d}\Phi_{4}(p_{3},\ldots,p_{6};p_{1},p_{2})
 \, \frac{2}{4!}\,\sum_{P_C(j,k,l)}\Bigg\{ }\nonumber \\
\ph{(4s1)}&&\phantom{+}\left(-S_{{\bar{\bar{2}}}\tilde{j}{\tilde{\tilde{k}}}}+S_{\bar{2}j\tilde{k}}
-S_{{\bar{\bar{1}}}\tilde{j}{\tilde{\tilde{l}}}}+S_{\bar{1}j\tilde{l}}
-S_{{\bar{2}}j{\bar{1}}}+S_{\bar{\bar{2}}\tilde{j}\bar{\bar{1}}}\right)
F_3^0(\hat{\bar{1}}_g,\tilde{i}_g,\hat{\bar{2}}_g)\nonumber\\
&&\hspace{0.3cm}\times A^0_{4}(\hat{\bar{\bar{1}}}_g,\hat{\bar{\bar{2}}}_g,\tilde{\tilde{k}}_g,\tilde{\tilde{l}}_g)
{J}_{2}^{(2)}(p_{\tilde{\tilde{k}}},p_{\tilde{\tilde{l}}})\nonumber\\
\ph{4s2}&&+\left(-S_{{\bar{\bar{2}}}\tilde{l}{\tilde{\tilde{k}}}}+S_{\bar{2}l\tilde{k}}
-S_{{\bar{\bar{1}}}\tilde{l}{\tilde{\tilde{j}}}}+S_{\bar{1}l\tilde{j}}
-S_{{\bar{2}}l{\bar{1}}}+S_{\bar{\bar{2}}\tilde{l}\bar{\bar{1}}}\right)
F_3^0(\hat{\bar{1}}_g,\tilde{i}_g,\hat{\bar{2}}_g)\nonumber\\
&&\hspace{0.3cm}\times A^0_{4}(\hat{\bar{\bar{1}}}_g,\hat{\bar{\bar{2}}}_g,\tilde{\tilde{k}}_g,\tilde{\tilde{j}}_g)
{J}_{2}^{(2)}(p_{\tilde{\tilde{k}}},p_{\tilde{\tilde{j}}})\nonumber\\
\ph{(4s3)}&&+\frac{1}{2}\left(S_{{\bar{\bar{2}}}\tilde{j}{\tilde{\tilde{k}}}}-S_{\bar{2}j\tilde{k}}
+S_{{\bar{\bar{1}}}\tilde{j}{\tilde{\tilde{k}}}}-S_{\bar{1}j\tilde{k}}
+S_{{\bar{2}}j{\bar{1}}}-S_{\bar{\bar{2}}\tilde{j}\bar{\bar{1}}}\right)
F_3^0(\hat{\bar{1}}_g,\tilde{l}_g,\hat{\bar{2}}_g)\nonumber\\
&&\hspace{0.3cm}\times A^0_{4}(\hat{\bar{\bar{1}}}_g,\tilde{\tilde{i}}_g,\hat{\bar{\bar{2}}}_g,\tilde{\tilde{k}}_g)
{J}_{2}^{(2)}(p_{\tilde{\tilde{i}}},p_{\tilde{\tilde{k}}})\nonumber\\
\ph{(4s4)}&&+\frac{1}{2}\left(S_{{\bar{\bar{2}}}\tilde{l}{\tilde{\tilde{k}}}}-S_{\bar{2}l\tilde{k}}
+S_{{\bar{\bar{1}}}\tilde{l}{\tilde{\tilde{k}}}}-S_{\bar{1}l\tilde{k}}
+S_{{\bar{2}}l{\bar{1}}}-S_{\bar{\bar{2}}\tilde{l}\bar{\bar{1}}}\right)
F_3^0(\hat{\bar{1}}_g,\tilde{j}_g,\hat{\bar{2}}_g)\nonumber\\
&&\hspace{0.3cm}\times 
A^0_{4}(\hat{\bar{\bar{1}}}_g,\tilde{\tilde{i}}_g,\hat{\bar{\bar{2}}}_g,\tilde{\tilde{k}}_g)
{J}_{2}^{(2)}(p_{\tilde{\tilde{i}}},p_{\tilde{\tilde{k}}})\nonumber\\
\ph{4s5}&&+\frac{1}{2}\left(-S_{2j((kj)l)}+S_{2j(kj)}
-S_{\bar{1}j(kj)}+S_{\bar{\bar{1}}j(l(kj))}
-S_{2j\bar{\bar{1}}}+S_{2j\bar{1}}\right)
f_3^0(\hat{\bar{1}}_g,l_g,(kj)_g)\nonumber\\
&&\hspace{0.3cm}\times A^0_{4}(\hat{\bar{\bar{1}}}_g,i_g,\hat{2}_g,\widetilde{(kj)l)_g}
{J}_{2}^{(2)}(p_{i},\widetilde{p_{(kj)l)}})\nonumber\\
\ph{4s6}&&+\frac{1}{2}\left(-S_{2l((kl)j)}+S_{2l(kl)}
-S_{\bar{1}l(kl)}+S_{\bar{\bar{1}}l(j(kl))}
-S_{2l\bar{\bar{1}}}+S_{2l\bar{1}}\right)
f_3^0(\hat{\bar{1}}_g,j_g,(kl)_g)\nonumber\\
&&\hspace{0.3cm}\times A^0_{4}(\hat{\bar{\bar{1}}}_g,i_g,\hat{2}_g,\widetilde{(kl)j}_g)
{J}_{2}^{(2)}(p_{i},\widetilde{p_{(kl)j)}})\nonumber\\
\ph{4s7}&&+\frac{1}{2}\left(-S_{1j((kj)l)}+S_{1j(kj)}
-S_{\bar{2}j(kj)}+S_{\bar{\bar{2}}j(l(kj))}
-S_{1j\bar{\bar{2}}}+S_{1j\bar{2}}\right)
f_3^0(\hat{\bar{2}}_g,l_g,(kj)_g)\nonumber\\
&&\hspace{0.3cm}\times A^0_{4}(\hat{1}_g,i_g,\hat{\bar{\bar{2}}}_g,\widetilde{(kj)l}_g)
{J}_{2}^{(2)}(p_{3},\widetilde{p_{(kj)l)}})\nonumber\\
\ph{4s8}&&+\frac{1}{2}\left(-S_{1l((kl)j)}+S_{1l(kl)}
-S_{\bar{2}l(kl)}+S_{\bar{\bar{2}}l(j(kl))}
-S_{1l\bar{\bar{2}}}+S_{1l\bar{2}}\right)
f_3^0(\hat{\bar{2}}_g,j_g,(kl)_g)\nonumber\\
&&\hspace{0.3cm}\times A^0_{4}(\hat{1}_g,i_g,\hat{\bar{\bar{2}}}_g,\widetilde{(kl)j}_g)
{J}_{2}^{(2)}(p_{i},\widetilde{p_{(kl)j)}})\nonumber\\
\bigg\}.
\label{eq:LAST2}
\end{eqnarray}

\subsection{IFFIFF topology}

The real radiation contribution to the cross section for the third topology, IFFIFF, receives contributions from twelve colour ordered amplitudes.
It is convenient to organise these amplitudes in six terms,
\begin{eqnarray}
\sum_{(i,j,k,l)\in P_C(3,4,5,6) }
A^0_{6}(\hat{1}_g,i_g,j_g,\hat{2}_g,k_g,l_g)
&=&\phantom{+}
Z^0_{6}(\hat{1}_g,3_g,4_g,\hat{2}_g,5_g,6_g) + Z^0_{6}(\hat{1}_g,3_g,4_g,\hat{2}_g,6_g,5_g) \nonumber\\
&& +Z^0_{6}(\hat{1}_g,3_g,5_g,\hat{2}_g,4_g,6_g) + Z^0_{6}(\hat{1}_g,3_g,5_g,\hat{2}_g,6_g,4_g)\nonumber\\
&& + Z^0_{6}(\hat{1}_g,3_g,6_g,\hat{2}_g,4_g,5_g) + Z^0_{6}(\hat{1}_g,3_g,6_g,\hat{2}_g,5_g,4_g),\nonumber \\
\end{eqnarray}
where each $Z_6^0$ contains 2 colour ordered amplitudes,
\begin{eqnarray}
Z_6^0(\hat{1}_g,3_g,4_g,\hat{2}_g,5_g,6_g)=A_6^0(\hat{1}_g,3_g,4_g,\hat{2}_g,5_g,6_g)+A_6^0(\hat{1}_g,4_g,3_g,\hat{2}_g,6_g,5_g),
\end{eqnarray}
so that $Z_6^0$ matches 
the symmetry of the full $F_4^0(\hat{1}_g,j_g,\hat{2}_g,k_g)$  initial-initial antenna function.

The real radiation subtraction term to be used with $Z_6^0(\hat{1}_g,i_g,j_g,\hat{2}_g,k_g,l_g)$ is,
\begin{eqnarray}
\lefteqn{{\rm d}\hat\sigma_{NNLO}^{S,Z_6}= {\cal N} \left(\frac{\alpha_s N}{2\pi}\right)^2 
{\rm d}\Phi_{4}(p_{3},\ldots,p_{6};p_{1},p_{2})
 \, \frac{2}{4!}\,\Bigg\{ }\nonumber \\
\ph{(a11)}&&\phantom{+} f^0_3(\hat{1}_g,i_g,j_g)\,
A^0_{5}(\hat{\bar{1}}_g,\widetilde{(ij)}_g,\hat{2}_g,k_g,l_g) \,
{J}_{2}^{(3)}(\widetilde{p_{ij}},p_{k},p_{l}) 
\nonumber \\
\ph{(a12)}&&+ f^0_3(i_g,j_g,\hat{2}_g)\,
{A}^0_{5}(\hat{1}_g,\widetilde{(ji)}_g,\hat{\bar{2}}_g,k_g,l_g)\,
{J}_{2}^{(3)}(\widetilde{p_{ji}},p_k,p_l) 
\nonumber \\
\ph{(a13)}&&+ f^0_3(\hat{2}_g,k_g,l_g)\,
{A}^0_{5}(\hat{1}_g,i_g,j_g,\hat{\bar{2}}_g,\widetilde{(kl)}_g)\,
{J}_{2}^{(3)}(p_i,p_j,\widetilde{p_{kl}}) 
\nonumber \\
\ph{(a14)}&&+ f^0_3(k_g,l_g,\hat{1}_g)\,
A^0_{5}(\hat{\bar{1}}_g,i_g,j_g,\hat{2}_g,\widetilde{(lk)}_g) \,
{J}_{2}^{(3)}(p_i,p_j,\widetilde{p_{lk}}) 
\nonumber \\ && \nonumber\\
\ph{(a21)}&&+ f^0_3(\hat{1}_g,j_g,i_g)\,
A^0_{5}(\hat{\bar{1}}_g,\widetilde{(ji)}_g,\hat{2}_g,l_g,k_g) \,
{J}_{2}^{(3)}(\widetilde{p_{ji}},p_{l},p_{k}) 
\nonumber \\
\ph{(a22)}&&+ f^0_3(j_g,i_g,\hat{2}_g)\,
{A}^0_{5}(\hat{1}_g,\widetilde{(ij)}_g,\hat{\bar{2}}_g,l_g,k_g)\,
{J}_{2}^{(3)}(\widetilde{p_{ij}},p_l,p_k) 
\nonumber \\
\ph{(a23)}&&+ f^0_3(\hat{2}_g,l_g,k_g)\,
{A}^0_{5}(\hat{1}_g,j_g,i_g,\hat{\bar{2}}_g,\widetilde{(lk)}_g)\,
{J}_{2}^{(3)}(p_j,p_i,\widetilde{p_{lk}}) 
\nonumber \\
\ph{(a24)}&&+ f^0_3(l_g,k_g,\hat{1}_g)\,
A^0_{5}(\hat{\bar{1}}_g,j_g,i_g,\hat{2}_g,\widetilde{(kl)}_g) \,
{J}_{2}^{(3)}(p_j,p_i,\widetilde{p_{kl}}) 
\nonumber \\ && \nonumber\\
\ph{(g1,h1)}&&+
\Bigg( F_{4}^0(\hat{1}_g,i_g,j_g,\hat{2}_g) - f_3^0(\hat{1}_g,i_g,j_g)\,
 F_3^0(\hat{\bar{1}}_g,\widetilde{(ij)}_g,\hat{2}_g)\nonumber\\
&&\hspace{0.3cm} - f_3^0(i_g,j_g,\hat{2}_g)\,
 F_3^0(\hat{1}_g,\widetilde{(ji)}_g,\hat{\bar{2}}_g)\Bigg)
A^0_{4}(\hat{\bar{1}}_g,\hat{\bar{2}}_g,\tilde{k}_g,\tilde{l}_g)\,
{J}_{2}^{(2)}(\widetilde{p_{k}},\widetilde{p_{l}})
  \nonumber \\
\ph{(g2,i2)}&&+
\Bigg( F_{4}^0(\hat{2}_g,k_g,l_g,\hat{1}_g) - f_3^0(\hat{2}_g,k_g,l_g)\,
 F_3^0(\hat{\bar{2}}_g,\widetilde{(kl)}_g,\hat{1}_g)\nonumber\\
&&\hspace{0.3cm} - f_3^0(k_g,l_g,\hat{1}_g)\,
 F_3^0(\hat{2}_g,\widetilde{(lk)}_g,\hat{\bar{1}}_g)\Bigg) 
A^0_{4}(\hat{\bar{1}}_g,\hat{\bar{2}}_g,\tilde{j}_g,\tilde{i}_g)\,
{J}_{2}^{(2)}(\widetilde{p_{j}},\widetilde{p_{i}})
  \nonumber \\
\ph{(g1,h1)}&&+
\Bigg( F_{4}^0(\hat{1}_g,j_g,i_g,\hat{2}_g) - f_3^0(\hat{1}_g,j_g,i_g)\,
 F_3^0(\hat{\bar{1}}_g,\widetilde{(ji)}_g,\hat{2}_g)\nonumber\\
&&\hspace{0.3cm} - f_3^0(j_g,i_g,\hat{2}_g)\,
 F_3^0(\hat{1}_g,\widetilde{(ij)}_g,\hat{\bar{2}}_g)\Bigg)
A^0_{4}(\hat{\bar{1}}_g,\hat{\bar{2}}_g,\tilde{l}_g,\tilde{k}_g)\,
{J}_{2}^{(2)}(\widetilde{p_{l}},\widetilde{p_{k}})
  \nonumber \\
\ph{(g2,i2)}&&+
\Bigg( F_{4}^0(\hat{2}_g,l_g,k_g,\hat{1}_g) - f_3^0(\hat{2}_g,l_g,k_g)\,
 F_3^0(\hat{\bar{2}}_g,\widetilde{(lk)}_g,\hat{1}_g)\nonumber\\
&& - f_3^0(l_g,k_g,\hat{1}_g)\,
 F_3^0(\hat{2}_g,\widetilde{(kl)}_g,\hat{\bar{1}}_g)\Bigg)
A^0_{4}(\hat{\bar{1}}_g,\hat{\bar{2}}_g,\tilde{i}_g,\tilde{j}_g)\,
{J}_{2}^{(2)}(\widetilde{p_{j}},\widetilde{p_{i}})
  \nonumber \\
\ph{(g1,h1)}&&+
\Bigg( F_{4}^0(\hat{1}_g,j_g,\hat{2}_g,k_g) - F_3^0(\hat{1}_g,j_g,\hat{2}_g)\,
 F_3^0(\hat{\bar{1}}_g,\hat{\bar{2}}_g,\tilde{k}_g)\nonumber\\
 &&\hspace{0.3cm}- F_3^0(\hat{2}_g,k_g,\hat{1}_g)\,
 F_3^0(\hat{\bar{1}}_g,\tilde{j}_g,\hat{\bar{2}}_g)
\Bigg) A^0_{4}(\hat{\bar{1}}_g,\tilde{i}_g,\hat{\bar{2}}_g,\tilde{l}_g)\,
{J}_{2}^{(2)}(\widetilde{p_{i}},\widetilde{p_{l}})
  \nonumber \\
\ph{(g1,h1)}&&+
\Bigg( F_{4}^0(\hat{1}_g,i_g,\hat{2}_g,l_g) - F_3^0(\hat{1}_g,i_g,\hat{2}_g)\,
 F_3^0(\hat{\bar{1}}_g,\hat{\bar{2}}_g,\tilde{l}_g)\nonumber\\
 &&\hspace{0.3cm}- F_3^0(\hat{2}_g,l_g,\hat{1}_g)\,
 F_3^0(\hat{\bar{1}}_g,\tilde{i}_g,\hat{\bar{2}}_g)
\Bigg) A^0_{4}(\hat{\bar{1}}_g,\tilde{j}_g,\hat{\bar{2}}_g,\tilde{k}_g)\,
{J}_{2}^{(2)}(\widetilde{p_{j}},\widetilde{p_{k}})
  \nonumber \\  && \nonumber\\
\ph{(v)}&& + \frac{1}{2}F_3^0(\hat{1}_g,j_g,\hat{2}_g) 
                F_3^0(\hat{\bar{1}}_g,\hat{\bar{2}}_g,\tilde{k}_g)\,
{A}^0_{4}(\hat{\bar{\bar{1}}}_g,\tilde{\tilde{i}}_g,\hat{\bar{\bar{2}}}_g,\tilde{\tilde{l}}_g)\,
{J}_{2}^{(2)}(\tilde{\tilde{p_i}},\tilde{\tilde{p_l}}) 
  \nonumber \\
\ph{(w)}&& + \frac{1}{2}F_3^0(\hat{2}_g,k_g,\hat{1}_g) 
                F_3^0(\hat{\bar{1}}_g,\tilde{j}_g,\hat{\bar{2}}_g)\,
{A}^0_{4}(\hat{\bar{\bar{1}}}_g,\tilde{\tilde{i}}_g,\hat{\bar{\bar{2}}}_g,\tilde{\tilde{l}}_g)\,
{J}_{2}^{(2)}(\tilde{\tilde{p_i}},\tilde{\tilde{p_l}}) 
  \nonumber \\
\ph{(x)}&& + \frac{1}{2}F_3^0(\hat{1}_g,i_g,\hat{2}_g) 
                F_3^0(\hat{\bar{1}}_g,\hat{\bar{2}}_g,\tilde{l}_g)\,
{A}^0_{4}(\hat{\bar{\bar{1}}}_g,\tilde{\tilde{j}}_g,\hat{\bar{\bar{2}}}_g,\tilde{\tilde{k}}_g)\,
{J}_{2}^{(2)}(\tilde{\tilde{p_j}},\tilde{\tilde{p_k}}) 
  \nonumber \\
\ph{(z)}&& + \frac{1}{2}F_3^0(\hat{2}_g,l_g,\hat{1}_g) 
                F_3^0(\hat{\bar{1}}_g,\tilde{i}_g,\hat{\bar{2}}_g)\,
{A}^0_{4}(\hat{\bar{\bar{1}}}_g,\tilde{\tilde{j}}_g,\hat{\bar{\bar{2}}}_g,\tilde{\tilde{k}}_g)\,
{J}_{2}^{(2)}(\tilde{\tilde{p_j}},\tilde{\tilde{p_k}}) 
  \nonumber \\ && \nonumber\\
\ph{(c1)}&& - \frac{1}{2}f_3^0(i_g,j_g,\hat{2}_g) 
                f_3^0(\hat{\bar{2}}_g,k_g,l_g)\,
{A}^0_{4}(\hat{1}_g,\widetilde{(ji)}_g,\hat{\bar{\bar{2}}}_g,\widetilde{(kl)}_g)\,
{J}_{2}^{(2)}(\widetilde{p_{ji}},\widetilde{p_{kl}})
  \nonumber \\
\ph{(c2)}&& - \frac{1}{2}f_3^0(l_g,k_g,\hat{2}_g) 
                f_3^0(\hat{\bar{2}}_g,j_g,i_g)\,
{A}^0_{4}(\hat{1}_g,\widetilde{(ji)}_g,\hat{\bar{\bar{2}}}_g,\widetilde{(kl)}_g)\,
{J}_{2}^{(2)}(\widetilde{p_{ji}},\widetilde{p_{kl}})
  \nonumber \\
\ph{(c3)}&& - \frac{1}{2}f_3^0(j_g,i_g,\hat{1}_g) 
                f_3^0(\hat{\bar{1}}_g,l_g,k_g)\,
{A}^0_{4}(\hat{\bar{\bar{1}}}_g,\widetilde{(ij)}_g,\hat{2}_g,\widetilde{(lk)}_g)\,
{J}_{2}^{(2)}(\widetilde{p_{ij}},\widetilde{p_{lk}})
  \nonumber \\
\ph{(c4)}&& - \frac{1}{2}f_3^0(k_g,l_g,\hat{1}_g) 
                f_3^0(\hat{\bar{1}}_g,i_g,j_g)\,
{A}^0_{4}(\hat{\bar{\bar{1}}}_g,\widetilde{(ij)}_g,\hat{2}_g,\widetilde{(lk)}_g)\,
{J}_{2}^{(2)}(\widetilde{p_{ij}},\widetilde{p_{lk}})
  \nonumber \\
\ph{(c5)}&& - \frac{1}{2}f_3^0(j_g,i_g,\hat{2}_g) 
                f_3^0(\hat{\bar{2}}_g,l_g,k_g)\,
{A}^0_{4}(\hat{1}_g,\widetilde{(ij)}_g,\hat{\bar{\bar{2}}}_g,\widetilde{(lk)}_g)\,
{J}_{2}^{(2)}(\widetilde{p_{ij}},\widetilde{p_{lk}})
  \nonumber \\
\ph{(c6)}&& - \frac{1}{2}f_3^0(k_g,l_g,\hat{2}_g) 
                f_3^0(\hat{\bar{2}}_g,i_g,j_g)\,
{A}^0_{4}(\hat{1}_g,\widetilde{(ij)}_g,\hat{\bar{\bar{2}}}_g,\widetilde{(lk)}_g)\,
{J}_{2}^{(2)}(\widetilde{p_{ij}},\widetilde{p_{lk}})
  \nonumber \\
\ph{(c7)}&& - \frac{1}{2}f_3^0(i_g,j_g,\hat{1}_g) 
                f_3^0(\hat{\bar{1}}_g,k_g,l_g)\,
{A}^0_{4}(\hat{\bar{\bar{1}}}_g,\widetilde{(ji)}_g,\hat{2}_g,\widetilde{(kl)}_g)\,
{J}_{2}^{(2)}(\widetilde{p_{ji}},\widetilde{p_{kl}})
  \nonumber \\
\ph{(c8)}&& - \frac{1}{2}f_3^0(l_g,k_g,\hat{1}_g) 
                f_3^0(\hat{\bar{1}}_g,j_g,i_g)\,
{A}^0_{4}(\hat{\bar{\bar{1}}}_g,\widetilde{(ji)}_g,\hat{2}_g,\widetilde{(kl)}_g)\,
{J}_{2}^{(2)}(\widetilde{p_{ji}},\widetilde{p_{kl}})
  \nonumber \\ && \nonumber\\
\ph{(d1)}&& - f_3^0(\hat{1}_g,i_g,j_g) 
                f_3^0(\hat{2}_g,k_g,l_g)\,
{A}^0_{4}(\hat{\bar{1}}_g,\widetilde{(ij)}_g,\hat{\bar{2}}_g,\widetilde{(kl)}_g)\,
{J}_{2}^{(2)}(\widetilde{p_{ij}},\widetilde{p_{kl}}) 
  \nonumber \\
\ph{(d2)}&& - f_3^0(i_g,j_g,\hat{2}_g) 
                f_3^0(k_g,l_g,\hat{1}_g)\,
{A}^0_{4}(\hat{\bar{1}}_g,\widetilde{(ji)}_g,\hat{\bar{2}}_g,\widetilde{(lk)}_g)\,
{J}_{2}^{(2)}(\widetilde{p_{ji}},\widetilde{p_{lk}}) 
  \nonumber \\
\ph{(d3)}&& - f_3^0(\hat{1}_g,j_g,i_g) 
                f_3^0(\hat{2}_g,l_g,k_g)\,
{A}^0_{4}(\hat{\bar{1}}_g,\widetilde{(ji)}_g,\hat{\bar{2}}_g,\widetilde{(lk)}_g)\,
{J}_{2}^{(2)}(\widetilde{p_{ji}},\widetilde{p_{lk}}) 
  \nonumber \\
\ph{(d4)}&& - f_3^0(j_g,i_g,\hat{2}_g) 
                f_3^0(l_g,k_g,\hat{1}_g)\,
{A}^0_{4}(\hat{\bar{1}}_g,\widetilde{(ij)}_g,\hat{\bar{2}}_g,\widetilde{(kl)}_g)\,
{J}_{2}^{(2)}(\widetilde{p_{ij}},\widetilde{p_{kl}}) 
\nonumber \\ && \nonumber\\
&&-\frac{1}{2}f_3^0(k_g,l_g,\hat{2}_g)F_3^0(\hat{1}_g,\hat{\bar{2}}_g,i_g)
A_4^0(\hat{\bar{1}}_g,\tilde{j}_g,\hat{\bar{\bar{2}}}_g,\widetilde{(lk)}_g)
{J}_{2}^{(2)}(\tilde{p_{j}},\widetilde{p_{lk}})\nonumber\\
&&+\frac{1}{2}F_3^0(\hat{1}_g,\hat{2}_g,i_g)f_3^0(\tilde{k}_g,\tilde{l}_g,\hat{\bar{2}}_g)
A^0_{4}(\hat{\bar{1}}_g,\tilde{j}_g,\hat{\bar{\bar{2}}}_g,(\tilde{l}\tilde{k})_g)
{J}_{2}^{(2)}(\tilde{p_{j}},p_{{\tilde{l}\tilde{k}}})\nonumber\\
&&-\frac{1}{2}f_3^0(l_g,k_g,\hat{2}_g)F_3^0(\hat{1}_g,\hat{\bar{2}}_g,j_g)
A^0_{4}(\hat{\bar{1}}_g,\tilde{i}_g,\hat{\bar{\bar{2}}}_g,\widetilde{(kl)}_g)
{J}_{2}^{(2)}(\tilde{p_{i}},\widetilde{p_{kl}})\nonumber\\
&&+\frac{1}{2}F_3^0(\hat{1}_g,\hat{2}_g,j_g)f_3^0(\tilde{l}_g,\tilde{k}_g,\hat{\bar{2}}_g)
A^0_{4}(\hat{\bar{1}}_g,\tilde{i}_g,\hat{\bar{\bar{2}}}_g,(\tilde{k}\tilde{l})_g)
{J}_{2}^{(2)}(\tilde{p_{i}},p_{{\tilde{k}\tilde{l}}})\nonumber\\
&&-\frac{1}{2}f_3^0(i_g,j_g,\hat{2}_g)F_3^0(\hat{1}_g,\hat{\bar{2}}_g,k_g)
A^0_{4}(\hat{\bar{1}}_g,\widetilde{(ji)}_g,\hat{\bar{\bar{2}}}_g,\tilde{l}_g)
{J}_{2}^{(2)}(\widetilde{p_{ji}},\tilde{p_{l}})\nonumber\\
&&+\frac{1}{2}F_3^0(\hat{1}_g,\hat{2}_g,k_g)f_3^0(\tilde{i}_g,\tilde{j}_g,\hat{\bar{2}}_g)
A^0_{4}(\hat{\bar{1}}_g,(\tilde{j}\tilde{i})_g,\hat{\bar{\bar{2}}}_g,\tilde{l}_g)
{J}_{2}^{(2)}(p_{{\tilde{j}\tilde{i}}},\tilde{p_{l}})\nonumber\\
&&-\frac{1}{2}f_3^0(j_g,i_g,\hat{2}_g)F_3^0(\hat{1}_g,\hat{\bar{2}}_g,l_g)
A^0_{4}(\hat{\bar{1}}_g,\widetilde{(ij)}_g,\hat{\bar{\bar{2}}}_g,\tilde{k}_g)
{J}_{2}^{(2)}(\widetilde{p_{ij}},\tilde{p_{k}})\nonumber\\
&&+\frac{1}{2}F_3^0(\hat{1}_g,\hat{2}_g,l_g)f_3^0(\tilde{j}_g,\tilde{i}_g,\hat{\bar{2}}_g)
A^0_{4}(\hat{\bar{1}}_g,(\tilde{i}\tilde{j})_g,\hat{\bar{\bar{2}}}_g,\tilde{k}_g)
{J}_{2}^{(2)}(p_{{\tilde{i}\tilde{j}}},\tilde{p_{k}})
\nonumber \\ && \nonumber\\
&&-\frac{1}{2}f_3^0(k_g,l_g,\hat{1}_g)F_3^0(\hat{\bar{1}}_g,\hat{2}_g,i_g)
A_4^0(\hat{\bar{\bar{1}}}_g,\widetilde{(lk)}_g,\hat{\bar{2}}_g,\tilde{j}_g)
{J}_{2}^{(2)}(\widetilde{p_{lk}},\tilde{p_{j}})\nonumber\\
&&+\frac{1}{2}F_3^0(\hat{1}_g,\hat{2}_g,i_g)f_3^0(\tilde{k}_g,\tilde{l}_g,\hat{\bar{1}}_g)
A^0_{4}(\hat{\bar{\bar{1}}}_g,(\tilde{l}\tilde{k})_g,\hat{\bar{2}}_g,\tilde{j}_g)
{J}_{2}^{(2)}(p_{{\tilde{l}\tilde{k}}},\tilde{p_{j}})\nonumber\\
&&-\frac{1}{2}f_3^0(l_g,k_g,\hat{1}_g)F_3^0(\hat{\bar{1}}_g,\hat{2}_g,j_g)
A^0_{4}(\hat{\bar{\bar{1}}}_g,\widetilde{(kl)}_g,\hat{\bar{2}}_g,\tilde{i}_g)
{J}_{2}^{(2)}(\widetilde{p_{kl}},\tilde{p_{i}})\nonumber\\
&&+\frac{1}{2}F_3^0(\hat{1}_g,\hat{2}_g,j_g)f_3^0(\tilde{l}_g,\tilde{k}_g,\hat{\bar{1}}_g)
A^0_{4}(\hat{\bar{\bar{1}}}_g,(\tilde{k}\tilde{l})_g,\hat{\bar{2}}_g,\tilde{i}_g)
{J}_{2}^{(2)}(p_{{\tilde{k}\tilde{l}}},\tilde{p_{i}})\nonumber\\
&&-\frac{1}{2}f_3^0(i_g,j_g,\hat{1}_g)F_3^0(\hat{\bar{1}}_g,\hat{2}_g,k_g)
A^0_{4}(\hat{\bar{\bar{1}}}_g,\tilde{l}_g,\hat{\bar{\bar{2}}}_g,\widetilde{(ji)}_g)
{J}_{2}^{(2)}(\tilde{p_{l}},\widetilde{p_{ji}})\nonumber\\
&&+\frac{1}{2}F_3^0(\hat{1}_g,\hat{2}_g,k_g)f_3^0(\tilde{i}_g,\tilde{j}_g,\hat{\bar{1}}_g)
A^0_{4}(\hat{\bar{\bar{1}}}_g,\tilde{l}_g,\hat{\bar{\bar{2}}}_g,(\tilde{j}\tilde{i})_g)
{J}_{2}^{(2)}(\tilde{p_{l}},p_{{\tilde{j}\tilde{i}}})\nonumber\\
&&-\frac{1}{2}f_3^0(j_g,i_g,\hat{1}_g)F_3^0(\hat{\bar{1}}_g,\hat{2}_g,l_g)
A^0_{4}(\hat{\bar{\bar{1}}}_g,\tilde{k}_g,\hat{\bar{2}}_g,\widetilde{(ij)}_g)
{J}_{2}^{(2)}(\tilde{p_{k}},\widetilde{p_{ij}})\nonumber\\
&&+\frac{1}{2}F_3^0(\hat{1}_g,\hat{2}_g,l_g)f_3^0(\tilde{j}_g,\tilde{i}_g,\hat{\bar{1}}_g)
A^0_{4}(\hat{\bar{\bar{1}}}_g,\tilde{k}_g,\hat{\bar{2}}_g,(\tilde{i}\tilde{j})_g)
{J}_{2}^{(2)}(\tilde{p_{k}},p_{{\tilde{i}\tilde{j}}})\bigg\}.
\label{eq:nnlolc3}
\end{eqnarray}
Once again, ${\rm d}\hat\sigma_{NNLO}^{S,Z_6}$ introduces spurious limits from large
angle soft radiation
which are cancelled by an additional subtraction term ${\rm d}\hat\sigma_{NNLO}^{A,Z_6}$, as discussed in section~{\ref{sec:LAST}},
\begin{eqnarray}
\lefteqn{{\rm d}\hat\sigma_{NNLO}^{A,Z_6}= {\cal N} \left(\frac{\alpha_s N}{2\pi}\right)^2 
{\rm d}\Phi_{4}(p_{3},\ldots,p_{6};p_{1},p_{2})
 \, \frac{2}{4!}\,\Bigg\{ }\nonumber \\
\ph{(3s)}&&\phantom{+}\frac{1}{2}\left(-S_{\bar{1}i\bar{2}}+S_{{\bar{\bar{1}}}\tilde{i}{\bar{\bar{2}}}}
+S_{\bar{2}i\tilde{j}}-S_{\bar{\bar{2}}\tilde{i}\tilde{\tilde{j}}}
+S_{\bar{1}i\tilde{j}}-S_{\bar{\bar{1}}\tilde{i}\tilde{\tilde{j}}}\right)
F_3^0(\hat{\bar{1}}_g,\hat{\bar{2}}_g,\tilde{l}_g)\nonumber\\
&&\hspace{0.3cm}\times
A^0_{4}(\hat{\bar{\bar{1}}}_g,\tilde{\tilde{j}}_g,\hat{\bar{\bar{2}}}_g,\tilde{\tilde{k}}_g)
{J}_{2}^{(2)}(p_{\tilde{\tilde{j}}},p_{\tilde{\tilde{k}}})\nonumber\\
\ph{4s}&&+\frac{1}{2}\left(-S_{\bar{1}j\bar{2}}+S_{{\bar{\bar{1}}}\tilde{j}{\bar{\bar{2}}}}
+S_{\bar{2}j\tilde{i}}-S_{\bar{\bar{2}}\tilde{j}\tilde{\tilde{i}}}
+S_{\bar{1}j\tilde{i}}-S_{\bar{\bar{1}}\tilde{j}\tilde{\tilde{i}}}\right)
F_3^0(\hat{\bar{1}}_g,\hat{\bar{2}}_g,\tilde{k}_g)\nonumber\\
&&\hspace{0.3cm}\times 
A^0_{4}(\hat{\bar{\bar{1}}}_g,\tilde{\tilde{i}}_g,\hat{\bar{\bar{2}}}_g,\tilde{\tilde{l}}_g)
{J}_{2}^{(2)}(p_{\tilde{\tilde{i}}},p_{\tilde{\tilde{l}}})\nonumber\\
\ph{5s}&&+\frac{1}{2}\left(-S_{\bar{1}k\bar{2}}+S_{{\bar{\bar{1}}}\tilde{k}{\bar{\bar{2}}}}
+S_{\bar{2}k\tilde{l}}-S_{\bar{\bar{2}}\tilde{k}\tilde{\tilde{l}}}
+S_{\bar{1}k\tilde{l}}-S_{\bar{\bar{1}}\tilde{k}\tilde{\tilde{l}}}\right)
F_3^0(\hat{\bar{1}}_g,\hat{\bar{2}}_g,\tilde{j}_g)\nonumber\\
&&\hspace{0.3cm}\times 
A^0_{4}(\hat{\bar{\bar{1}}}_g,\tilde{\tilde{i}}_g,\hat{\bar{\bar{2}}}_g,\tilde{\tilde{l}}_g)
{J}_{2}^{(2)}(p_{\tilde{\tilde{i}}},p_{\tilde{\tilde{l}}})\nonumber\\
\ph{6s}&&+\frac{1}{2}\left(-S_{\bar{1}l\bar{2}}+S_{{\bar{\bar{1}}}\tilde{l}{\bar{\bar{2}}}}
+S_{\bar{2}l\tilde{k}}-S_{\bar{\bar{2}}\tilde{l}\tilde{\tilde{k}}}
+S_{\bar{1}l\tilde{k}}-S_{\bar{\bar{1}}\tilde{l}\tilde{\tilde{k}}}\right)
F_3^0(\hat{\bar{1}}_g,\hat{\bar{2}}_g,\tilde{i}_g)\nonumber\\
&&\hspace{0.3cm}\times 
A^0_{4}(\hat{\bar{\bar{1}}}_g,\tilde{\tilde{j}}_g,\hat{\bar{\bar{2}}}_g,\tilde{\tilde{k}}_g)
{J}_{2}^{(2)}(p_{\tilde{\tilde{j}}},p_{\tilde{\tilde{k}}})\Bigg\}.\nonumber\\
\label{eq:LAST3}
\end{eqnarray}

\section{Results}
\label{sec:numerics}
In this section we will test how well the subtraction term ${\rm d}\hat\sigma^S_{NNLO}$ derived in the previous section approaches the 
double real contribution ${\rm d}\hat\sigma^R_{NNLO}$ in the single and double unresolved regions of phase space so that their difference can be integrated numerically over the unconstrained phase space in four dimensions.
 We will do this numerically by generating a series of phase space points using {\tt RAMBO} \cite{Stirling:1986ab} that approach a given double or single unresolved limit. For each generated point we compute the ratio,
\begin{equation}
R=\frac{{\rm d}\hat\sigma^R_{NNLO}}{{\rm d}\hat\sigma^S_{NNLO}}
\end{equation}
where ${\rm d}\hat\sigma^R_{NNLO}$ is the matrix element squared given in equation (\ref{eq:6gNNLO}) and 
${\rm d}\hat\sigma^S_{NNLO}$ is the subtraction term given by equations (\ref{eq:nnlotop1}), (\ref{eq:LAST1}), (\ref{eq:nnlolc2}), (\ref{eq:LAST2}),  (\ref{eq:nnlolc3}) and (\ref{eq:LAST3}) summed over all orderings (i.e. the three permutations of $X_6^0$, the four permutations of $Y_6^0$ and the six permutations of $Z_6^0$).  
The ratio of the matrix element and the subtraction term should approach unity as we get closer to any singularity.

For each unresolved configuration, we will define a variable that controls how we approach the singularity subject to the requirement that there are at least two jets in the final state with $p_T>$50 GeV. The centre-of-mass energy $\sqrt s$ is fixed to be 1000~GeV.

\subsection{Double soft limit}
\begin{figure}[ht]
\begin{minipage}[b]{0.3\linewidth}
\centering
\includegraphics[width=4cm]{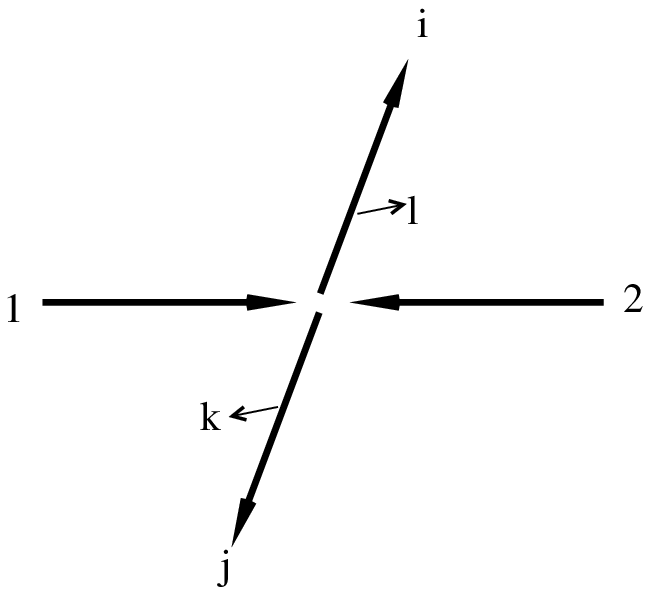}\\
\vspace{0.2cm}
(a)
\end{minipage}
\hspace{0.5cm}
\begin{minipage}[b]{0.7\linewidth}
\centering
\includegraphics[width=5cm,angle=270]{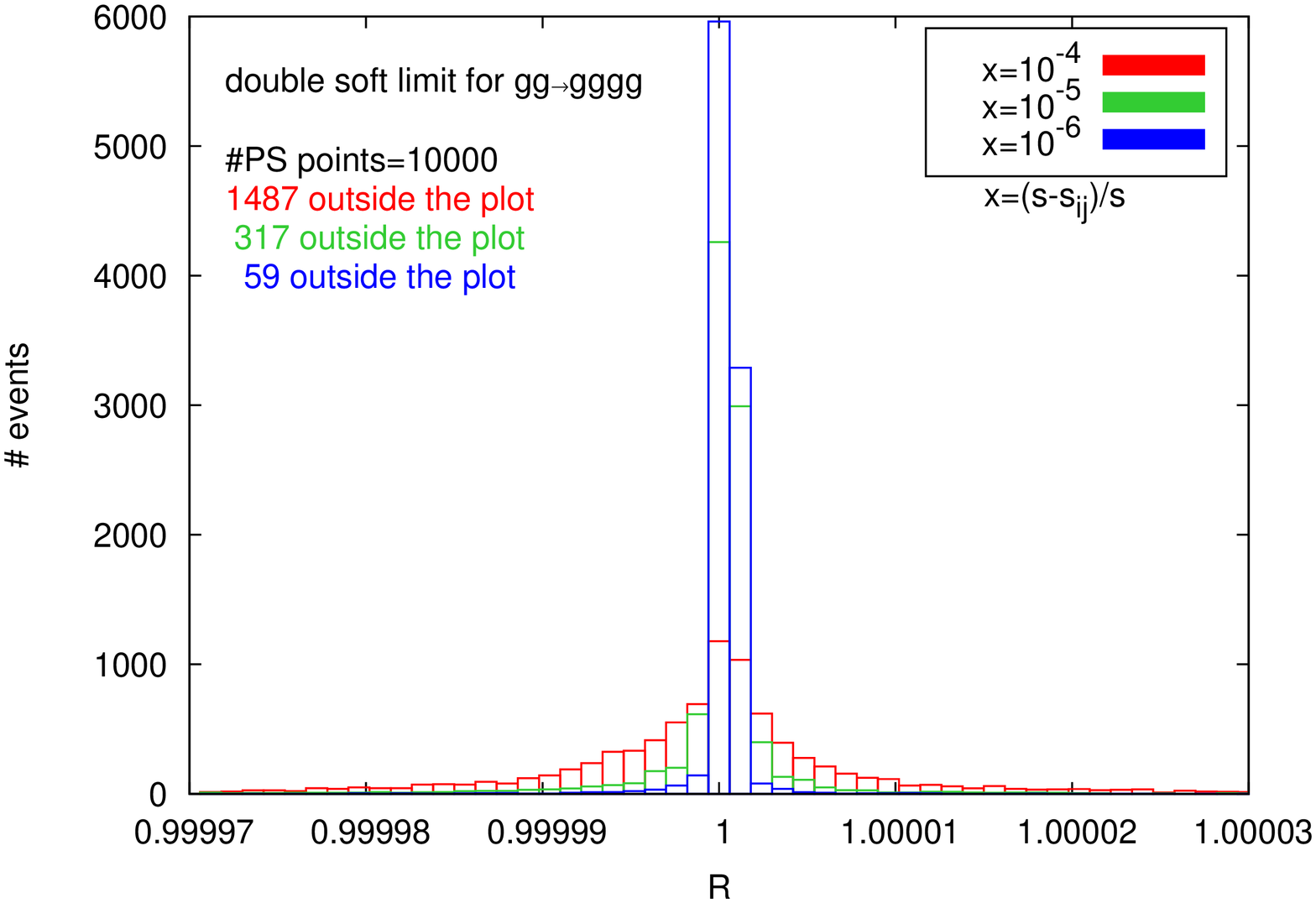}\\
\vspace{0.2cm}
(b)
\end{minipage}
\caption[Double soft limit distributions]{(a) Example configuration of a double soft event with $s_{ij}\approx s_{12}=s$.
(b) Distribution of $R$ for 10000 double soft phase space points.}
\label{fig:dsoft}
\end{figure}

A double soft configuration can be obtained by generating a four particle final state where one of the invariant masses $s_{ij}$ of two final
state particles takes nearly the full energy of the event $s$ as illustrated in fig.~\ref{fig:dsoft}(a).

For fig.~\ref{fig:dsoft}(b) we generated 10000 random double soft phase space points and show the
distribution of the ratio between the matrix element and the subtraction term. The three colours represent different values of $x=(s-s_{ij})/s$ [$x=10^{-4}$ (red), $x=10^{-5}$ (green), $x=10^{-6}$ (blue)] and we can see that for smaller values of $x$  the distribution peaks more sharply around unity. For $x=10^{-6}$ we obtained an average of $R=0.9999994$ and a standard deviation of $\sigma=4.02\times10^{-5}$. Also in the plot we give for each distribution the number of points that lie outside the range of the histogram. As expected the number of outliers systematically decreases as we approach the singular region. 

\begin{figure}[h]
\centering
\includegraphics[width=6cm,angle=270]{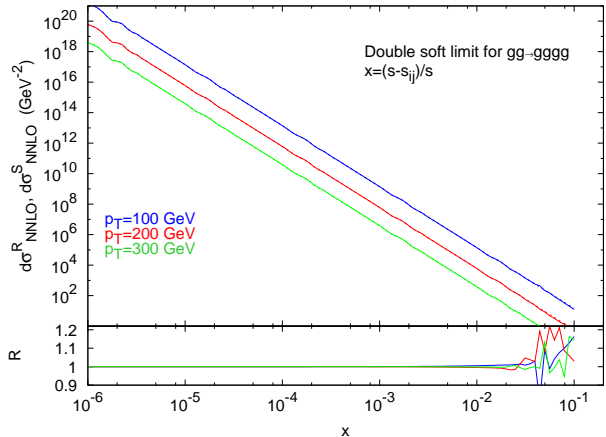}
\caption[Double soft limit subtraction]{ Matrix element squared (solid lines) and the subtraction term (dashed lines) 
    as a function of $x=(s-s_{ij})/s$ for three different values of $p_T$ for the final state.
    Also plotted is R.}
\label{fig:doublesoft}
\end{figure}
In figure \ref{fig:doublesoft} we explicitly show the behaviour of the matrix element squared and the subtraction term
as a function of $x$ and three choices of the jet transverse energy threshold, $p_T > 100$~GeV (blue), $p_T > 200$~GeV (red) and $p_T > 300$~GeV (green). It is clear that both ${\rm d}\hat\sigma^R_{NNLO}$ and ${\rm d}\hat\sigma^S_{NNLO}$ diverge in the double soft limit $x\to0$ but their ratio approaches unity.

\subsection{Triple collinear limit}
\begin{figure}[htp!]
\begin{minipage}[b]{0.3\linewidth}
\centering
\includegraphics[width=4cm]{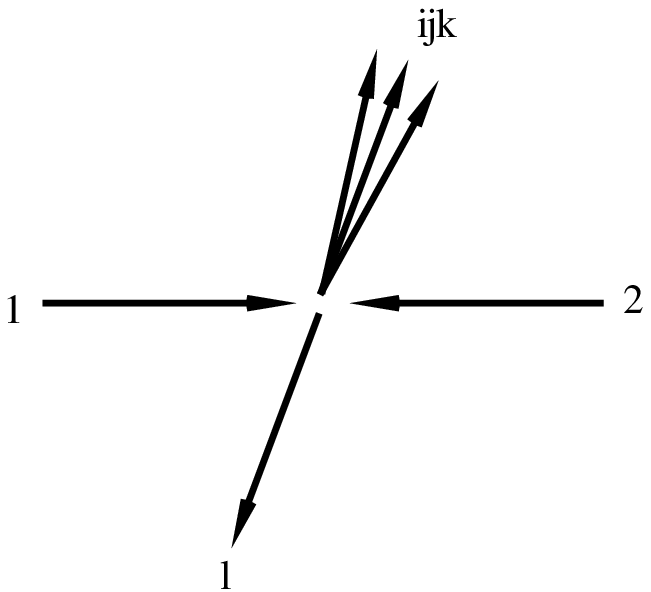}\\
\vspace{0.2cm}
(a)
\end{minipage}
\hspace{0.5cm}
\begin{minipage}[b]{0.7\linewidth}
\centering
\includegraphics[width=5cm,angle=270]{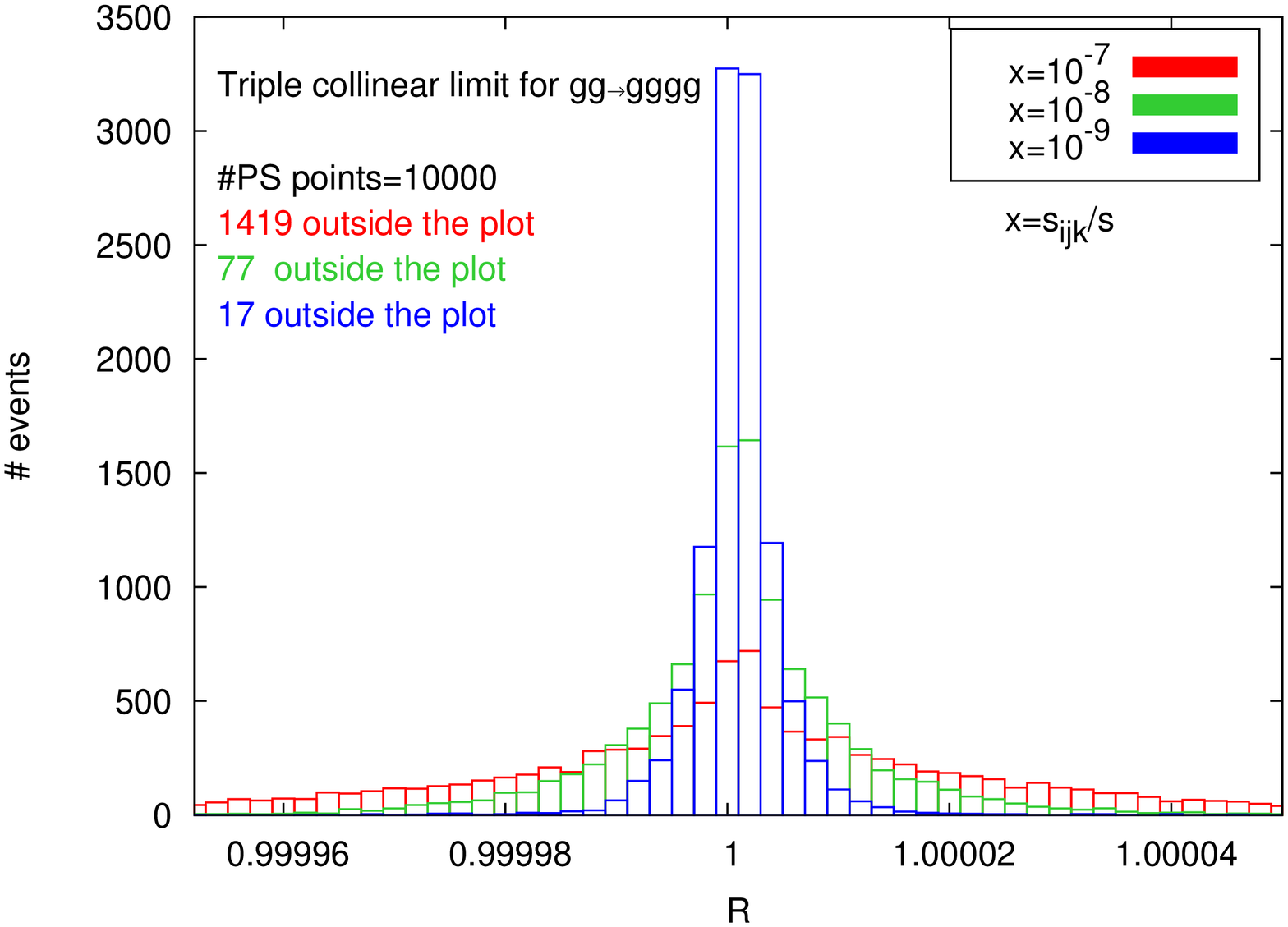}\\
\vspace{0.2cm}
(b)
\end{minipage}
\caption[Triple collinear limit final state singularity]{(a) Example configuration of a triple collinear event with $s_{ijk}\to 0$.
(b) Distribution of $R$ for 10000 triple collinear phase space points.}
\label{fig:tcollfff}
\end{figure}

\begin{figure}[htp!]
\begin{minipage}[b]{0.3\linewidth}
\centering
\includegraphics[width=4cm]{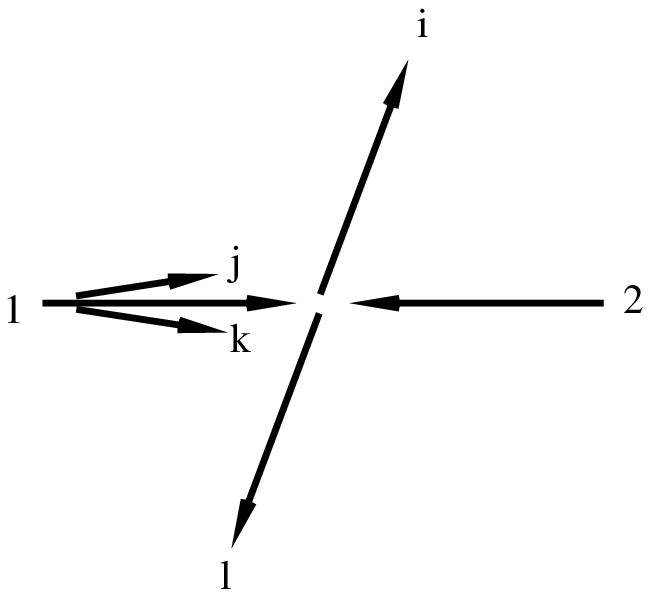}\\
\vspace{0.2cm}
(a)
\end{minipage}
\hspace{0.5cm}
\begin{minipage}[b]{0.7\linewidth}
\centering
\includegraphics[width=5cm,angle=270]{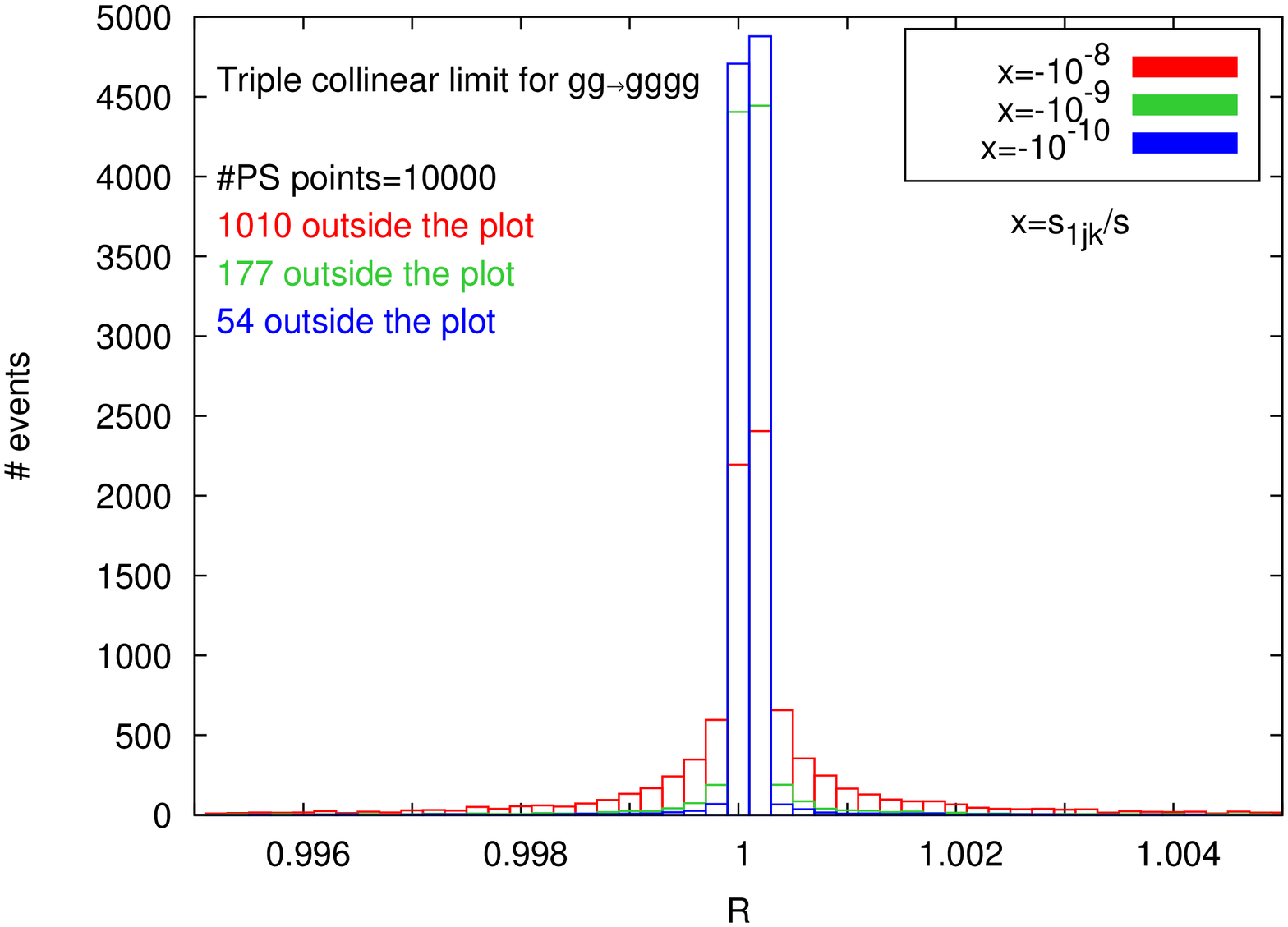}\\
\vspace{0.2cm}
(b)
\end{minipage}
\caption[Triple collinear limit initial state singularity]{(a) Example configuration of a triple collinear event with $s_{1jk}\to 0$.
(b) Distribution of $R$ for 10000 triple collinear phase space points.}
\label{fig:tcolliff}
\end{figure}

We now consider the triple collinear region of the phase space where three hard particles share a collinear direction. There are two separate configurations where the three collinear partons form either a final state (fig.~\ref{fig:tcollfff}(a)) or an initial state particle (fig.~\ref{fig:tcolliff}(a)).


For the final state case, $s_{ijk} \to 0$ and fig.~\ref{fig:tcollfff}(b) shows the distribution of $R$ for 10000 phase space points for three values of $x=s_{ijk}/s$ [$x=10^{-7}$ (red), $x=10^{-8}$ (green), $x=10^{-9}$ (blue)]. For $x=10^{-9}$ we obtained an average of $R=1.0000004$ and a standard deviation of $\sigma=4.2\times10^{-5}$. This shows that the subtraction term coincides with the matrix element squared in this limit. The number of points that lie outside the range of the histogram is also indicated on the plot and again systematically decreases as one approaches the singular region.
In principle there may be additional azimuthal correlations in the triple collinear limit.  However, we see that they do not affect the stability of the integrand.

In fig.~\ref{fig:tcolliff}(b) we perform a similar analysis for the initial state singularity. In this case $x=s_{1jk}/s$ [$x=-10^{-8}$ (red), $x=-10^{-9}$ (green), $x=-10^{-10}$ (blue)] and we have a configuration with two final state gluons collinear with the initial state gluon. The triple collinear configurations involving $p_2$ produces identical results and are not shown. For $x=-10^{-10}$ we obtained an average of $R=0.99954$ with a standard deviation of $\sigma=0.04$. Fig.~\ref{fig:tcolliff}(b) shows that this singular region is also well described by the subtraction term.

\subsection{Soft and collinear limit}
\begin{figure}[htp!]
\begin{minipage}[b]{0.3\linewidth}
\centering
\includegraphics[width=4cm]{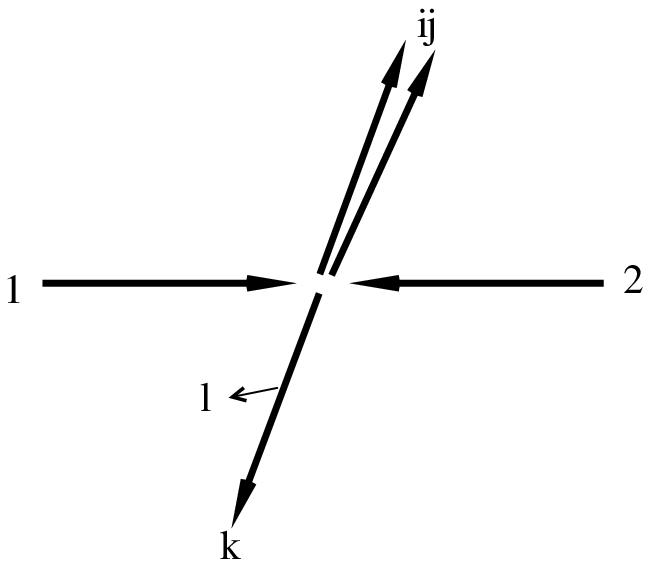}\\
\vspace{0.2cm}
(a)
\end{minipage}
\hspace{0.5cm}
\begin{minipage}[b]{0.7\linewidth}
\centering
\includegraphics[width=5cm,angle=270]{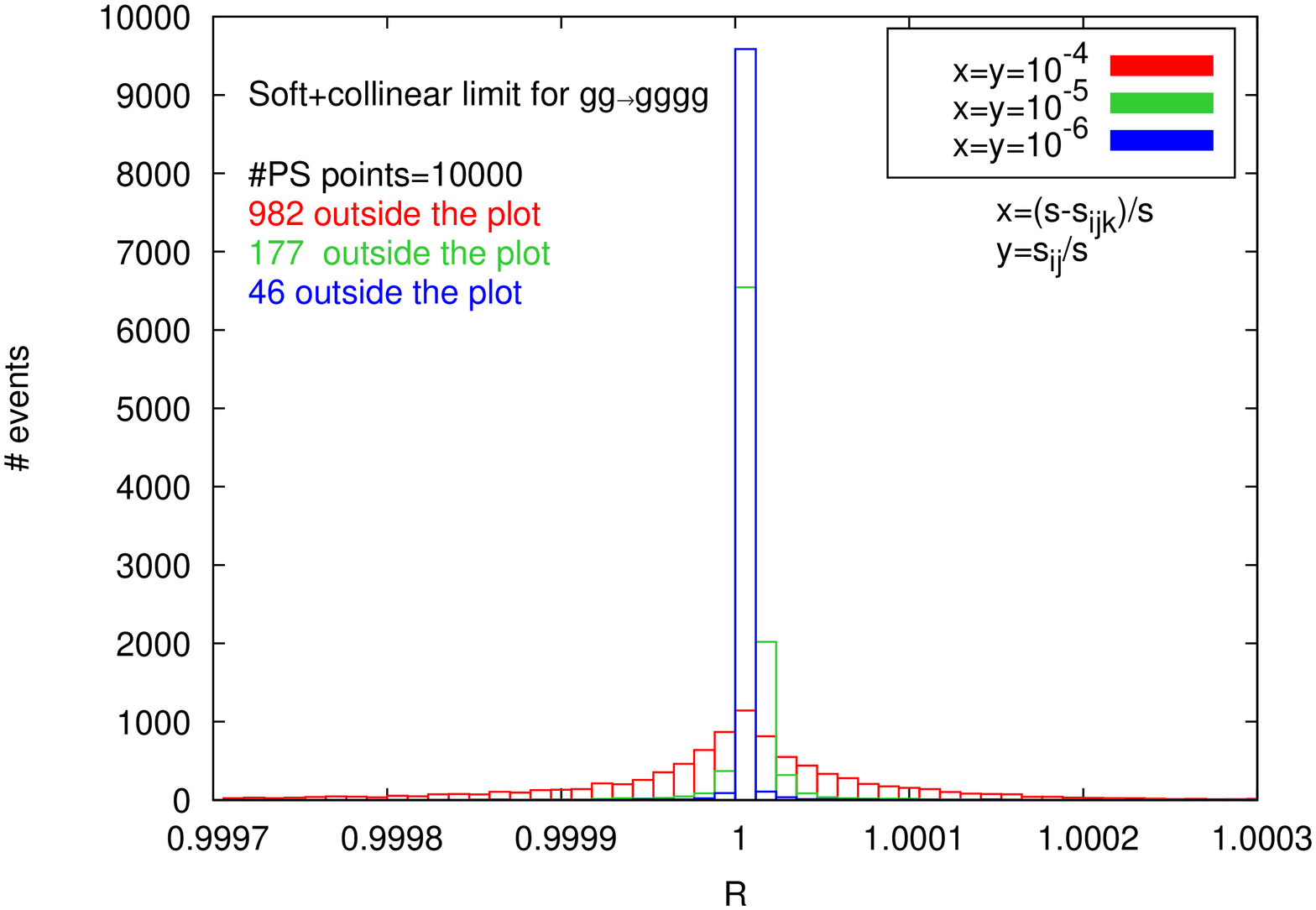}\\
\vspace{0.2cm}
(b)
\end{minipage}
\caption[Soft and collinear final state singularity]{(a) Example configuration of a soft and collinear event with $s_{ijk}\approx s_{12}=s$ and $s_{ij}\to 0$.
(b) Distribution of $R$ for 10000 soft and collinear phase space points.}
\label{fig:sftcollff}
\end{figure}
\begin{figure}[htp!]
\begin{minipage}[b]{0.3\linewidth}
\centering
\includegraphics[width=4cm]{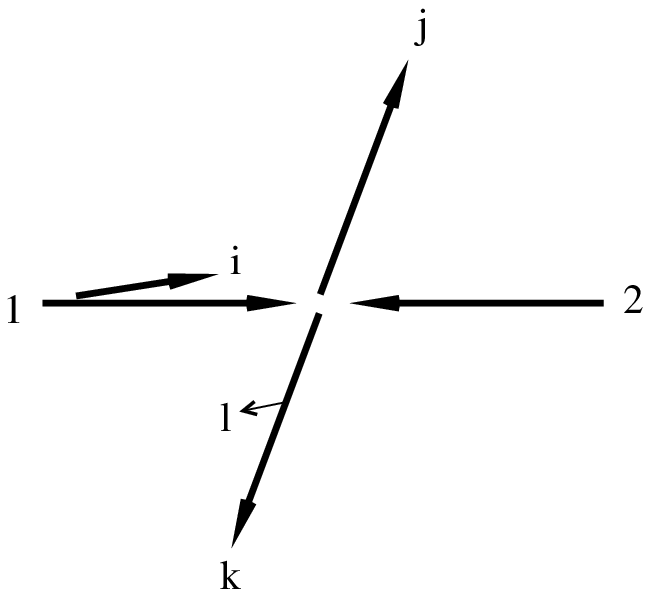}\\
\vspace{0.2cm}
(a)
\end{minipage}
\hspace{0.5cm}
\begin{minipage}[b]{0.7\linewidth}
\centering
\includegraphics[width=5cm,angle=270]{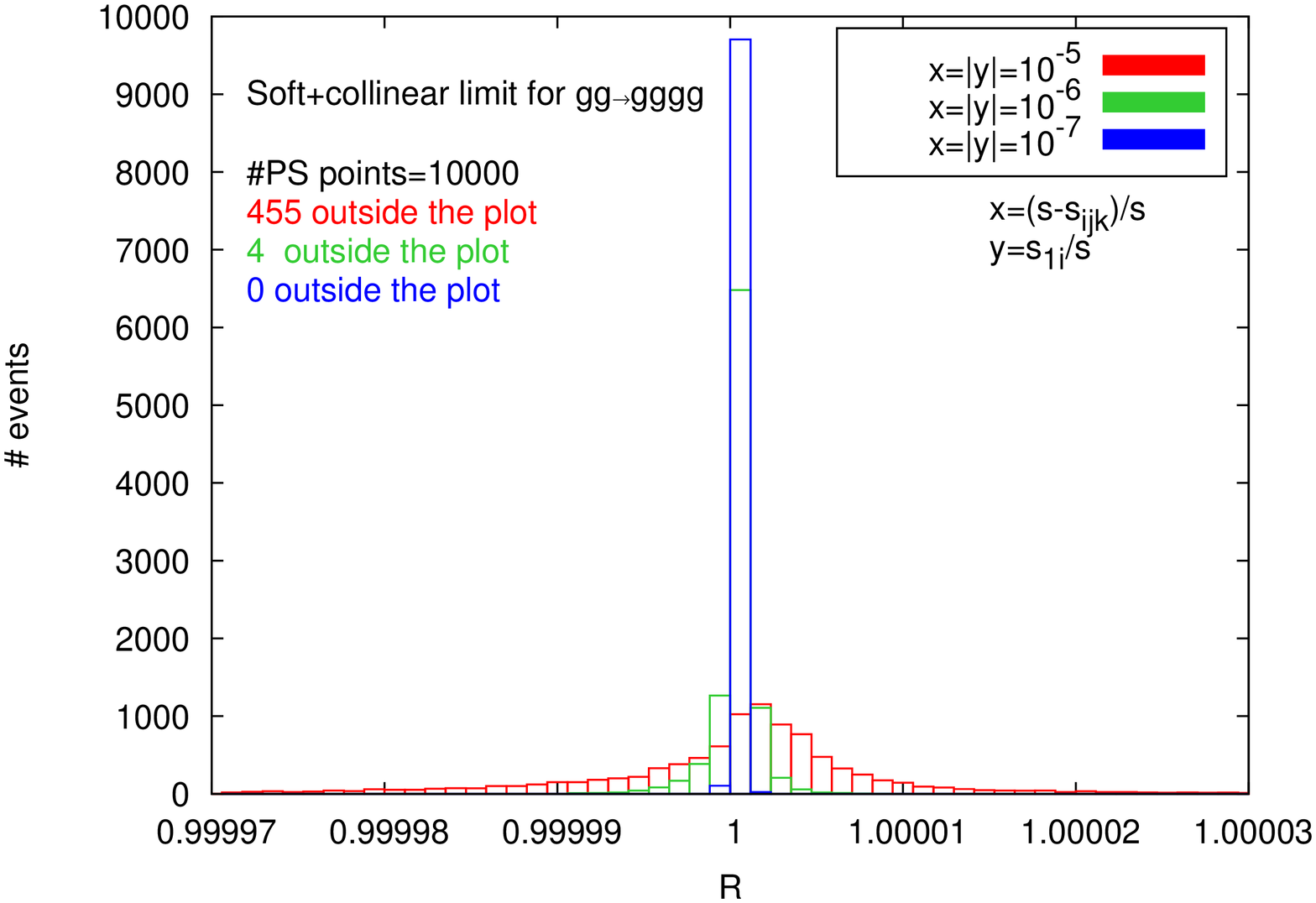}\\
\vspace{0.2cm}
(b)
\end{minipage}
\caption[Soft and collinear limit initial state singularity]{(a) Example configuration of a soft and collinear event with $s_{ijk}\approx s_{12}\equiv s$ and $s_{1i}\to 0$.
(b) Distribution of $R$ for 10000 soft and collinear phase space points.}
\label{fig:sftcollif}
\end{figure}

To probe the soft and collinear regions of the phase space, we generate an event configuration with a soft final state gluon $l$ by making a triple invariant $s_{ijk}$ close to the full center of mass energy $s_{12}$. When the two collinear gluons are in the final state, we allow the $ijk$ cluster to decay into three particles with the constraint that $s_{ij}$ is small (shown in fig.~\ref{fig:sftcollff}(a)), while for an initial state collinear singularity, we rotate the $i$, $j$ and $k$ momenta such that $s_{1i}$ is small (shown in fig.~\ref{fig:sftcollif}(a)).

In the first case we use two variables to approach this unresolved limit, $x=(s-s_{ijk})/s$ and $y=s_{ij}/s$.   
Fig.~\ref{fig:sftcollff}(b) shows three choices of $x$ and $y$, 
$x=y=10^{-4}$ (red), $x=y=10^{-5}$ (green) and $x=y=10^{-6}$ (blue).
For $x=y=10^{-6}$ we obtained an average of $R=0.99999993$ with a standard deviation of $\sigma=0.0001$. 

For the initial state singularity we define $x=(s-s_{ijk})/s$ and $y=s_{1i}/s$ and 
fig.~\ref{fig:sftcollif}(b)  shows three choices of $x$ and $y$,
$x=|y|=10^{-5}$ (red), green $x=|y|=10^{-6}$ (green)
 and $x=|y|=10^{-7}$ (blue), where we obtained an average of $R=0.99999998$ with a standard deviation of $\sigma=1.6\times10^{-7}$.

Figs.~\ref{fig:sftcollff} (b) and \ref{fig:sftcollif} (b) show that the subtraction term successfully reproduces the real radiation matrix element in the soft-collinear regions.

\subsection{Double collinear limit}
\begin{figure}[htp!]
\begin{minipage}[b]{0.3\linewidth}
\centering
\includegraphics[width=4cm]{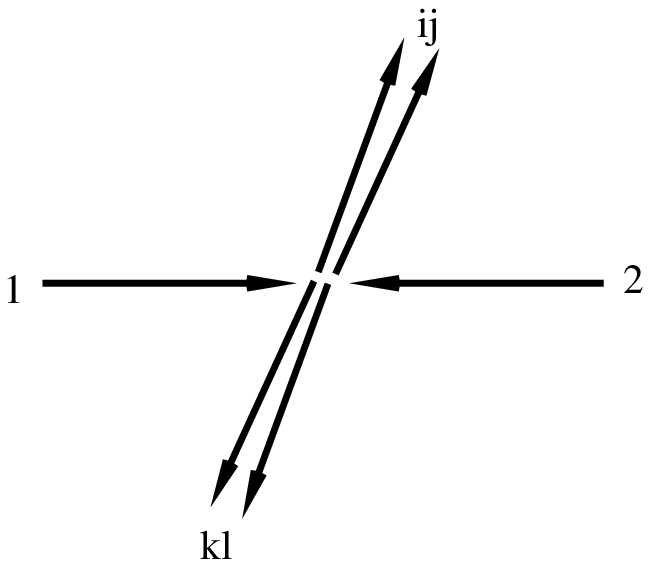}\\
\vspace{0.2cm}
(a)
\end{minipage}
\hspace{0.5cm}
\begin{minipage}[b]{0.7\linewidth}
\centering
\includegraphics[width=5cm,angle=270]{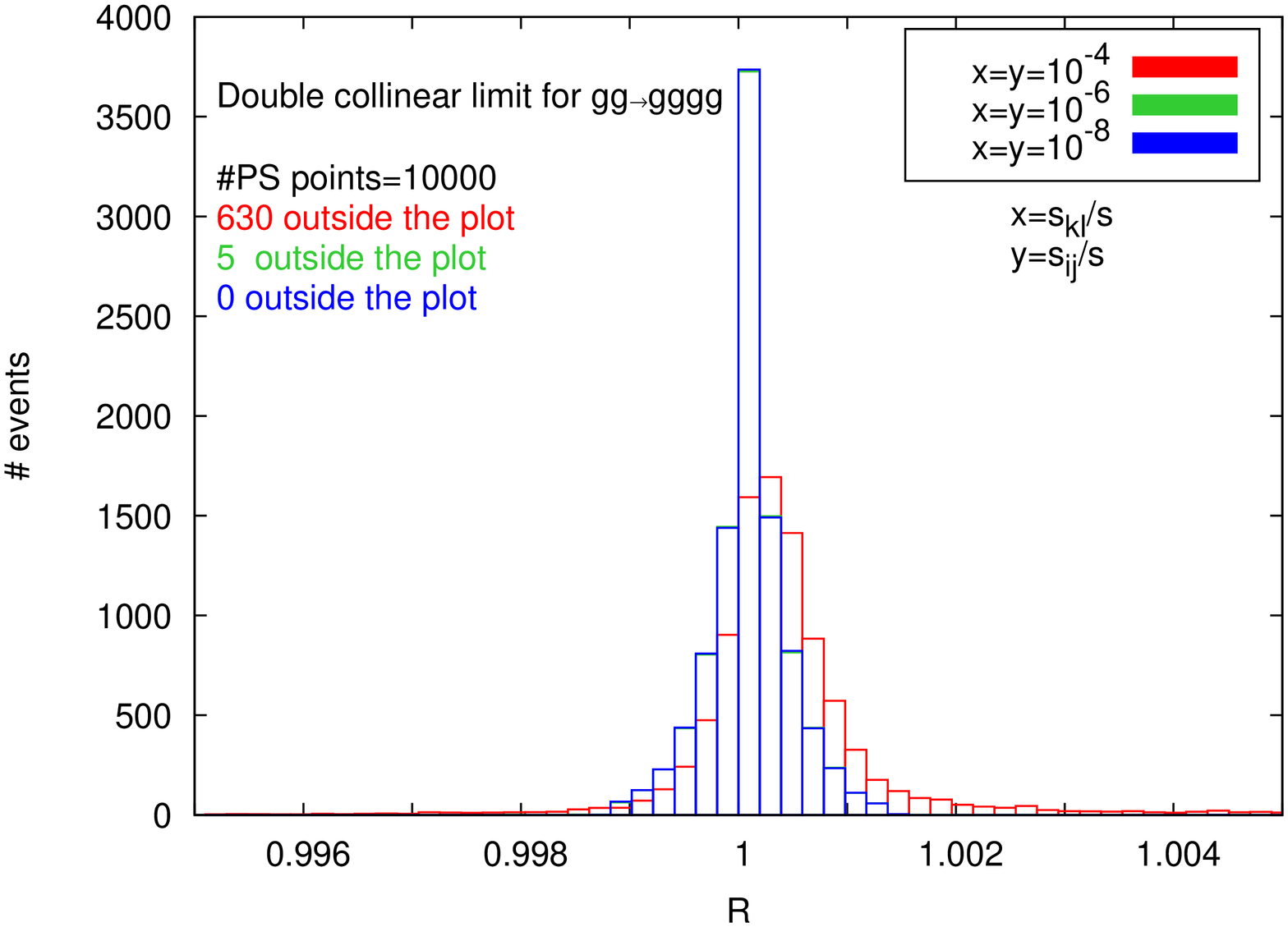}\\
\vspace{0.2cm}
(b)
\end{minipage}
\caption[Double collinear final state singularity]{(a) Example configuration of a double collinear event with $s_{ij}\to 0$ and $s_{kl}\to 0$ simultaneously.
(b) Distribution of $R$ for 10000 double collinear phase space points.}
\label{fig:dcollff}
\end{figure}

\begin{figure}[htp!]
\begin{minipage}[b]{0.3\linewidth}
\centering
\includegraphics[width=4cm]{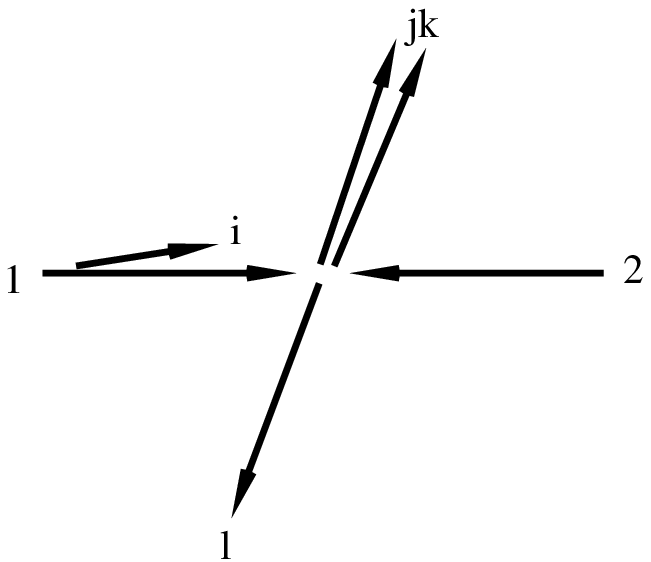}\\
\vspace{0.2cm}
(a)
\end{minipage}
\hspace{0.5cm}
\begin{minipage}[b]{0.7\linewidth}
\centering
\includegraphics[width=5cm,angle=270]{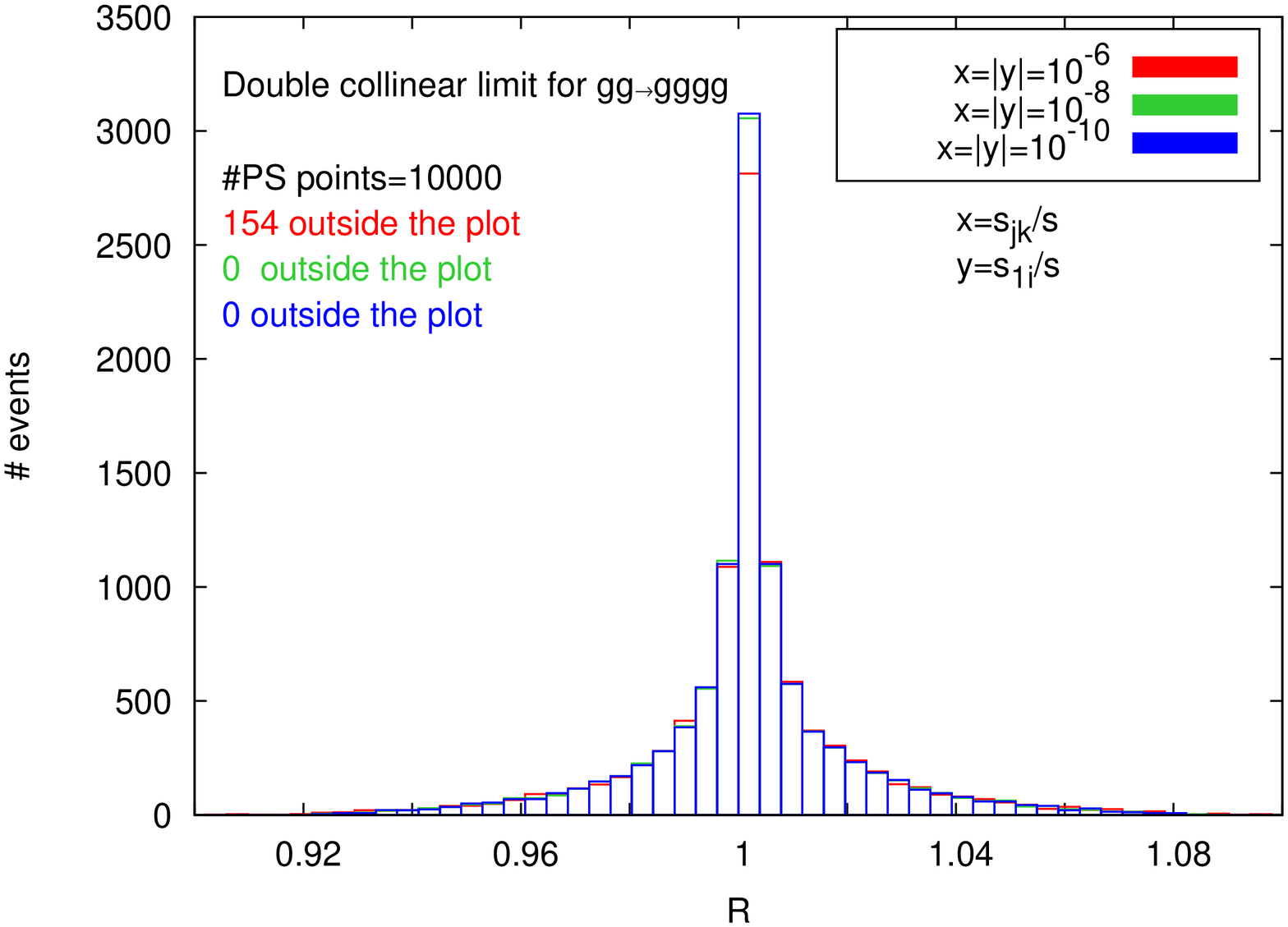}\\
\vspace{0.2cm}
(b)
\end{minipage}
\caption[Double collinear initial and final state singularity]{(a) Example configuration of a double collinear event with $s_{jk}\to 0$ and $s_{1i}\to 0$.
(b) Distribution of $R$ for 10000 double collinear phase space points.}
\label{fig:dcollif}
\end{figure}

\begin{figure}[htp!]
\begin{minipage}[b]{0.3\linewidth}
\centering
\includegraphics[width=4cm]{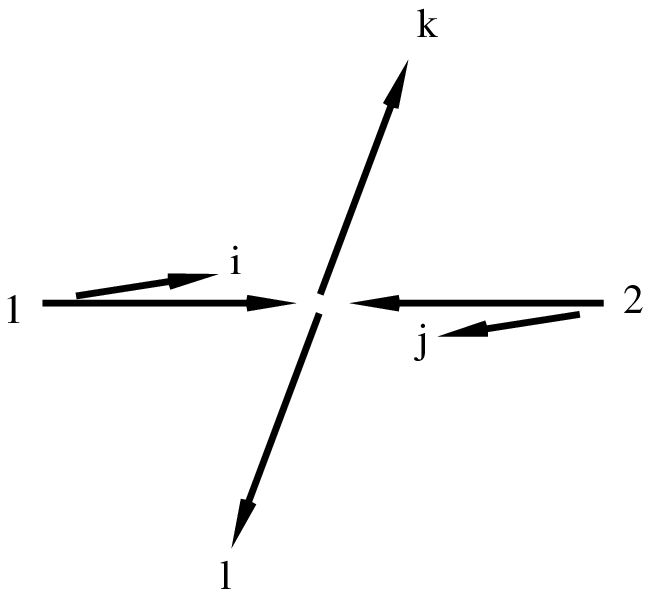}\\
\vspace{0.2cm}
(a)
\end{minipage}
\hspace{0.5cm}
\begin{minipage}[b]{0.7\linewidth}
\centering
\includegraphics[width=5cm,angle=270]{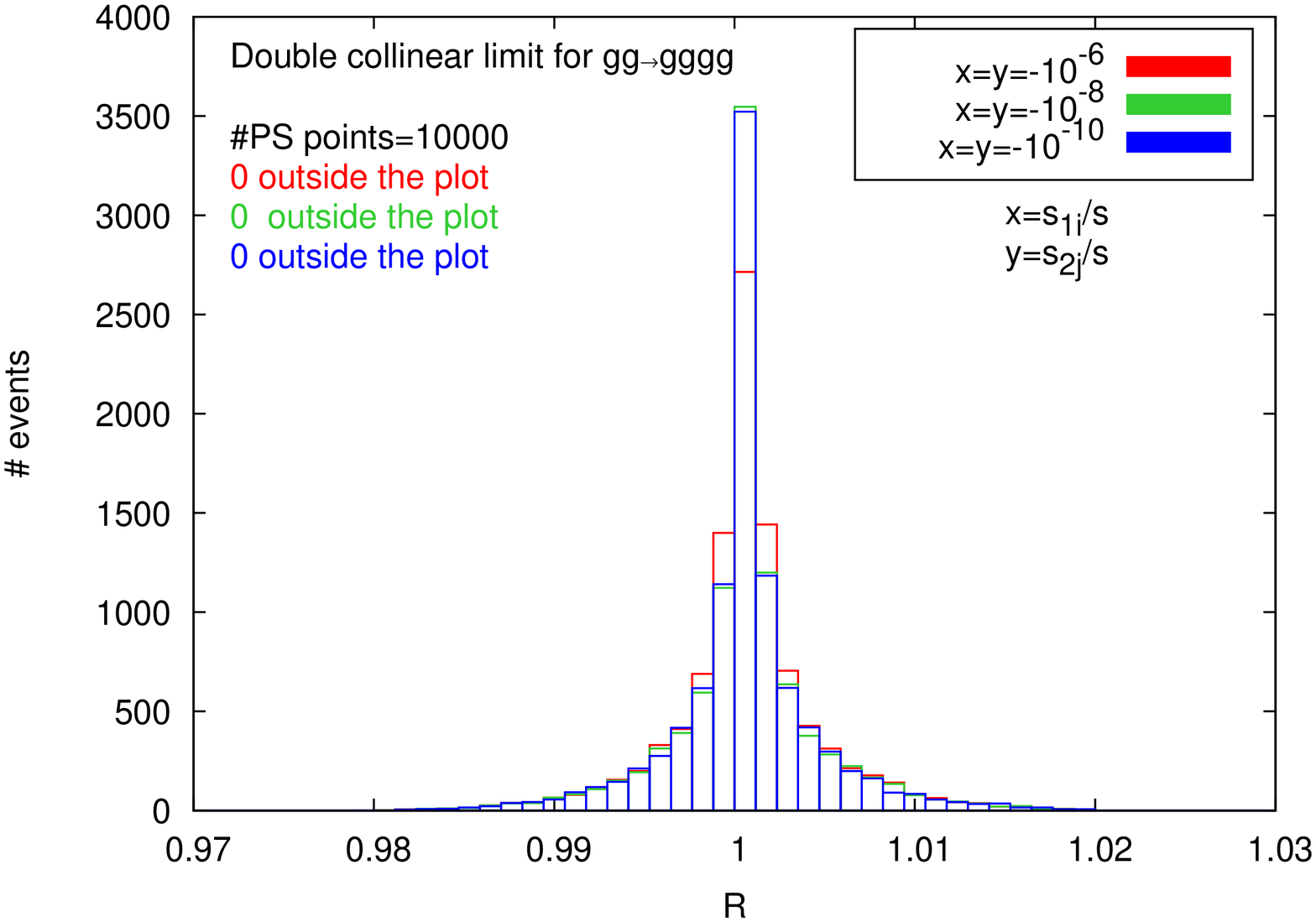}\\
\vspace{0.2cm}
(b)
\end{minipage}
\caption[Double collinear limit initial state singularity]{(a) Example configuration of a double collinear event with $s_{1i}\to 0$ and $s_{2j}\to 0$ simultaneously.
(b) Distribution of $R$ for 10000 double collinear phase space points.}
\label{fig:dcollii}
\end{figure}

There are three different topologies where two pairs of particles can become collinear separately by  demanding that two invariants vanish simultaneously. The double invariants pair may involve two final state pairs momenta (illustrated in fig.~\ref{fig:dcollff}(a)), or one final state pair and one initial state pair
(fig.~ \ref{fig:dcollif}(a)) or two initial state pairs shown in fig.~\ref{fig:dcollii}(a).

In each case we generate 10000 phase space points and plot the $R$ distribution. 
In the final-final case we set $x=s_{ij}/s$, $y=s_{kl}/s$ and show results for
$x=y=10^{-4}$ (red), $x=y=10^{-6}$ (green) and $x=y=10^{-8}$ (blue) in
in fig.~\ref{fig:dcollff}(b).
For $x=y=10^{-8}$ we obtained an average of $R=0.9999995$ with a standard deviation of $\sigma=0.00037$.

For the initial-final configuration we set $x=s_{jk}/s$, $y=s_{1i}/s$ and show
the $R$ distribution for $x=|y|=10^{-6}$ (red), $x=|y|=10^{-8}$ (green) and $x=|y|=10^{-10}$ (blue) in fig.~\ref{fig:dcollif}(b).
The average value of $R$ obtained for $x=|y|=10^{-10}$ was $R=1.00012$ with a standard deviation of $\sigma=0.018$.

In the initial-initial case we set $x=s_{1i}/s$, $y=s_{2j}/s$ and show
the $R$ distribution for $x=y=-10^{-6}$ (red), $x=y=-10^{-8}$ (green) and $x=y=-10^{-10}$ (blue) in fig.~\ref{fig:dcollii}(b)).  The average value of $R$ obtained for $x=y=-10^{-10}$ was $R=1.00001$ with a standard deviation of $\sigma=0.004$.

In all three cases we found convergence of the matrix element and the counterterm as we approach the double collinear limits. As expected the number of outliers systematically decreases as we approach the singular region.

\subsection{Subtraction of single unresolved final and initial state singularities}
In this subsection we will check that the integrand defined by the matrix element with colour ordered gluonic amplitudes and subtraction term with antenna functions is integrable over the single unresolved phase space regions. Single unresolved subtraction is well understood at NLO, but, in this case it is necessary to verify that the new NNLO subtraction term does not introduce divergences when one parton becomes unresolved. In other words it has to be correct simultaneously for both double unresolved and single unresolved configurations. In the $2\to4$ phase space these correspond to three jet configurations and, depending on the observable, these are allowed by the jet defining function through cuts on the final state momenta.

\subsubsection{Soft limit}
\begin{figure}[htp!]
\begin{minipage}[b]{0.3\linewidth}
\centering
\includegraphics[width=4cm]{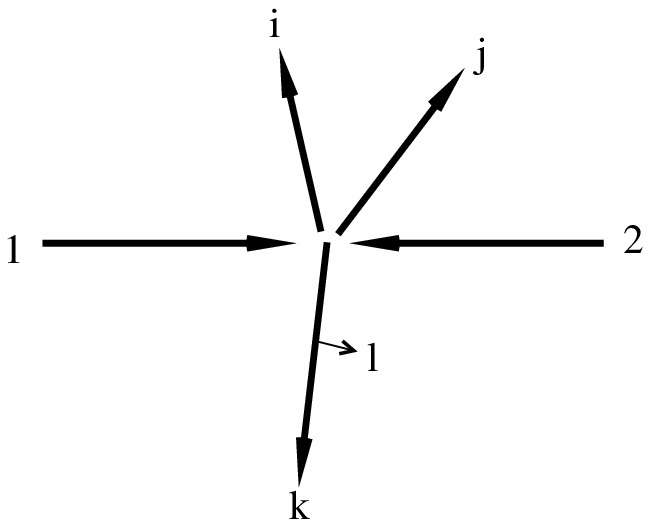}\\
\vspace{0.2cm}
(a)
\end{minipage}
\hspace{0.5cm}
\begin{minipage}[b]{0.7\linewidth}
\centering
\includegraphics[width=5cm,angle=270]{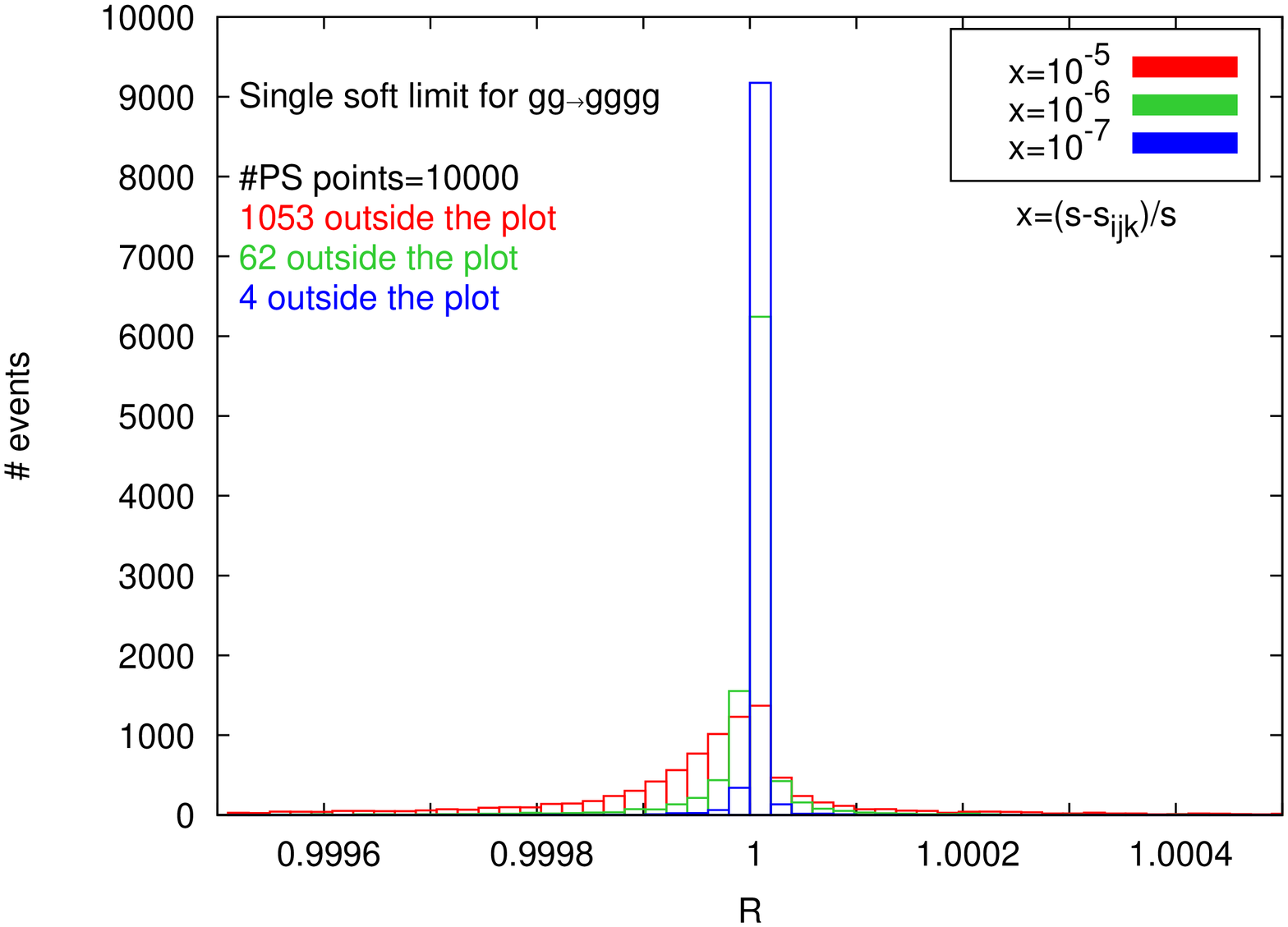}\\
\vspace{0.2cm}
(b)
\end{minipage}
\caption[Single soft singularity]{(a) Example configuration of a single soft event with $s_{ijk}\approx s_{12}=s$.
(b) Distribution of $R$ for 10000 single soft phase space points.}
\label{fig:ssoft}
\end{figure}

In figure \ref{fig:ssoft}(a) we analyse the limit where gluon $l$ is soft 
which is described by configurations where a triple invariant $s_{ijk}$ is close to the full center of mass energy $s=s_{12}$. We defined $x=(s-s_{ijk})/s$ and show the distributions for $x=10^{-5}$ (red), $x=10^{-6}$ (green) and $x=10^{-7}$ (blue)
in fig.~\ref{fig:ssoft}(b).
We see that the subtraction term converges to the matrix element as we approach the single soft limit. In this case the singularities related to soft gluons cancel and, the piece of the subtraction term described in section \ref{sec:LAST} correctly subtracts point by point the oversubtraction of large-angle soft gluon radiation. For $x=10^{-7}$ we obtained an average $R=0.999998$ and a standard deviation of $\sigma=1.9\times10^{-5}$.

\subsubsection{Collinear limit}
\label{sec:colllim}
\begin{figure}[htp!]
\begin{minipage}[b]{0.3\linewidth}
\centering
\includegraphics[width=4cm]{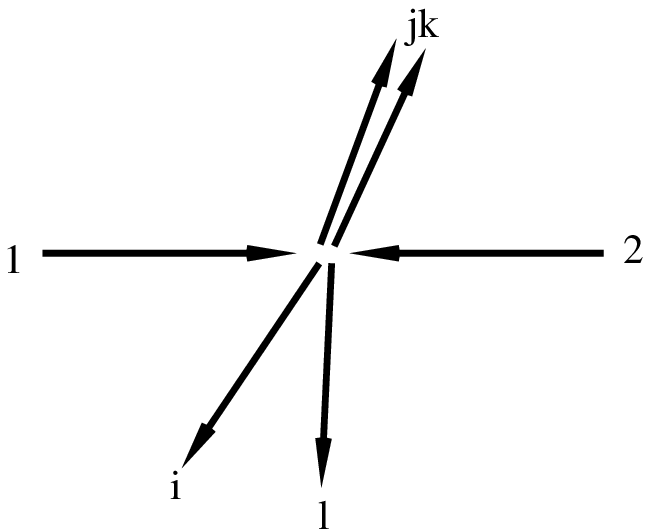}\\
\vspace{0.2cm}
(a)
\end{minipage}
\hspace{0.5cm}
\begin{minipage}[b]{0.7\linewidth}
\centering
\includegraphics[width=5cm,angle=270]{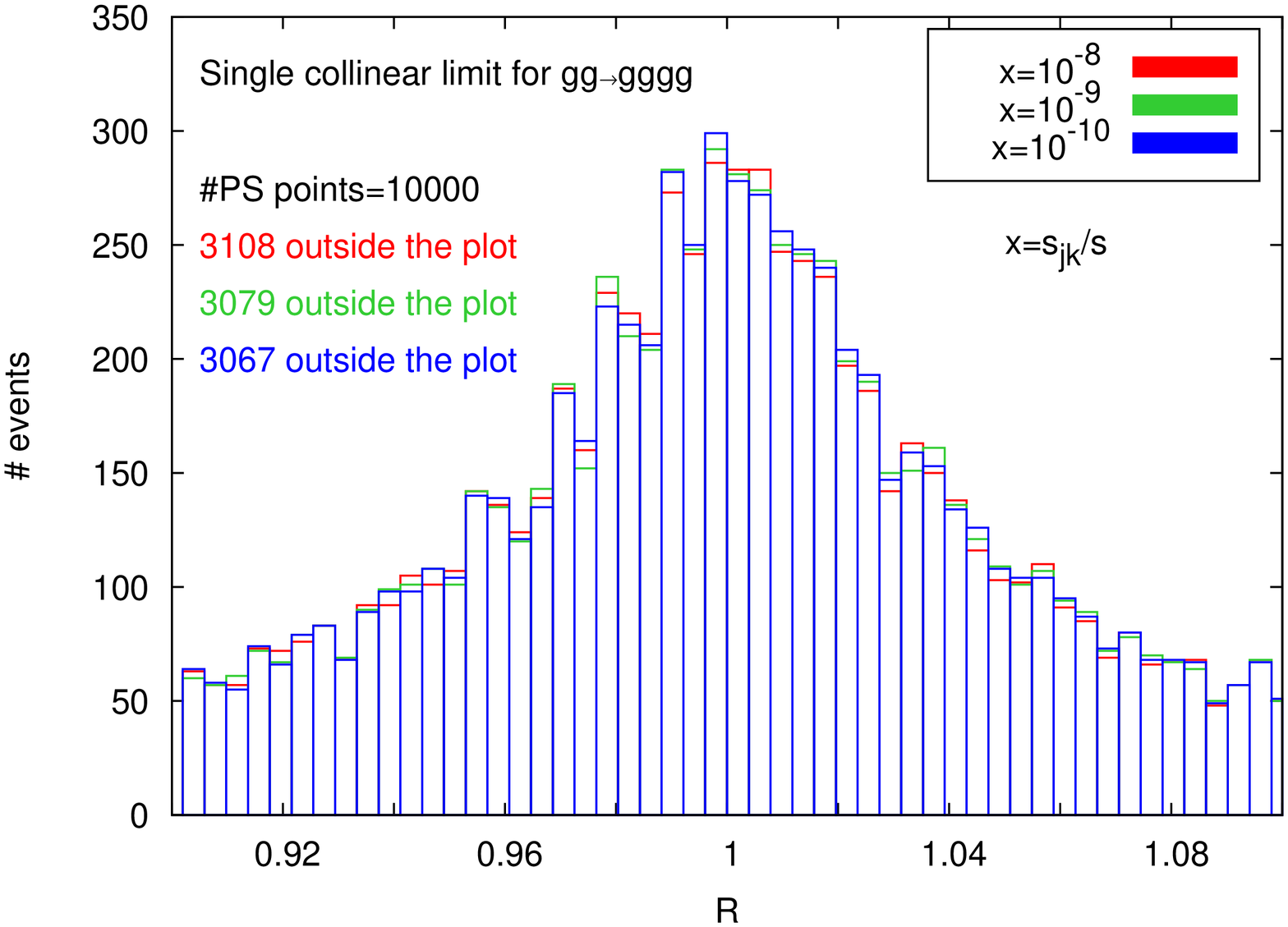}\\
\vspace{0.2cm}
(b)
\end{minipage}
\caption[Single collinear limit final state singularity]{(a) Example configuration of a single collinear event with $s_{jk}\to 0$.
(b) Distribution of $R$ for 10000 single collinear phase space points.}
\label{fig:scollff}
\end{figure}
\begin{figure}[htp!]
\begin{minipage}[b]{0.3\linewidth}
\centering
\includegraphics[width=4cm]{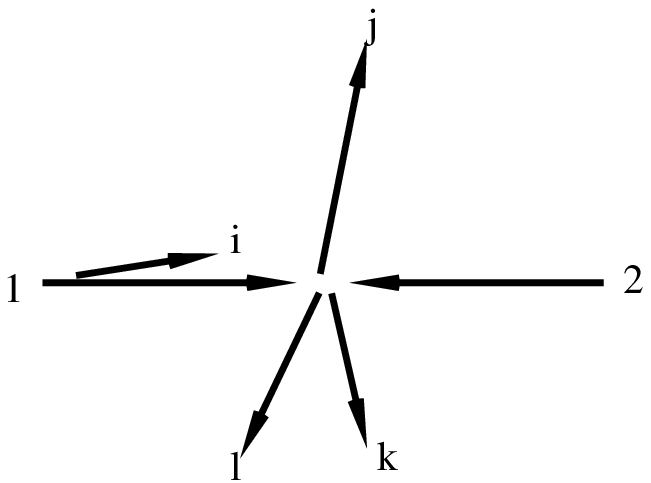}\\
\vspace{0.2cm}
(a)
\end{minipage}
\hspace{0.5cm}
\begin{minipage}[b]{0.7\linewidth}
\centering
\includegraphics[width=5cm,angle=270]{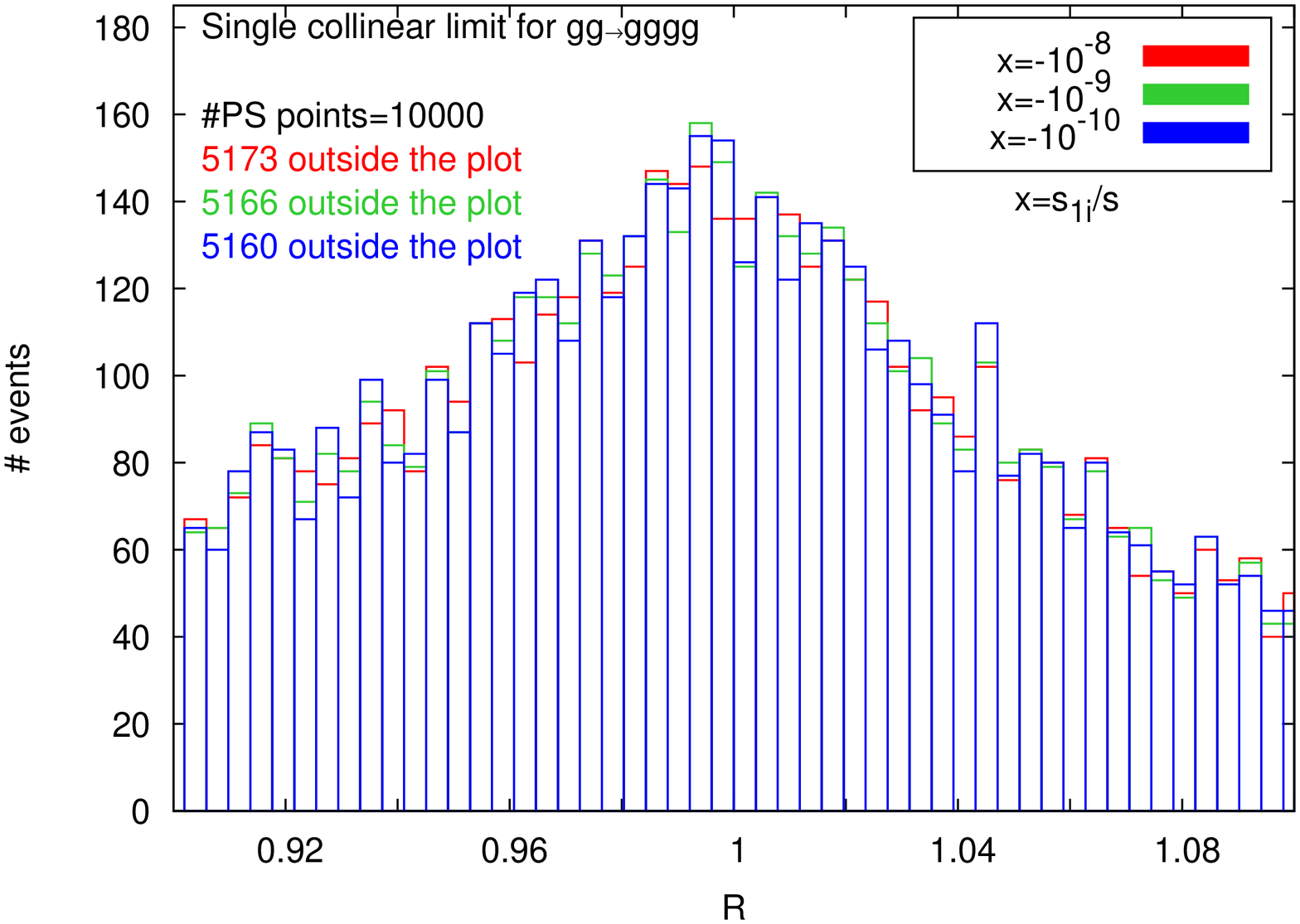}\\
\vspace{0.2cm}
(b)
\end{minipage}
\caption[Single collinear limit initial state singularity]{(a) Example configuration of a single collinear event with $s_{1i}\to 0$.
(b) Distribution of $R$ for 10000 single collinear phase space points.}
\label{fig:scollif}
\end{figure}

Finally we generate points corresponding to the final and initial state single collinear regions of the phase space. These event topologies are shown in figs.~\ref{fig:scollff}(a) and \ref{fig:scollif}(a) respectively. 

For the final-final collinear singularity, we introduce the variable $x = s_{ij}/s_{12}$ and fig.~\ref{fig:scollff}(b) shows the distribution for 
$x=10^{-8}$ (red),   $x=10^{-9}$ (green) and  $x=10^{-10}$ (blue). 
Similarly in the initial-final collinear limit, we define $x = s_{1i}/s_{12}$
and show the distributions of $R$ for the same $x$-values in fig.~\ref{fig:scollif}(b).

As we have discussed in section \ref{sec:angterms}, using scalar four-parton antennae functions the factorisation in the collinear limits where a final state gluon splits into two gluons introduces angular terms. This is the reason why the distributions in figs.~\ref{fig:scollff}(b) and \ref{fig:scollif}(b) have a much broader shape than the previous examples.  It is clear that as we approach the collinear limits $x \to 0$,
the azimuthal terms are not suppressed and the subtraction term is not, point by point, a better representation of the matrix element.

Nevertheless, the azimuthal terms coming from the single collinear limits were shown to vanish in sect.~\ref{sec:angterms}. This occurs globally after an azimuthal integration over the unresolved phase space. Here we are performing a point-by-point analysis on the integrand defined by the matrix element squared and the subtraction term. One possible strategy is to introduce the angular $\Theta_{F_3^0}(i,j,z,k_\perp)$ function defined in sect.~\ref{sec:angterms} to reconstruct the angular terms. Subtracting this additional term from the $F_4^0$ four-parton antenna functions for the  final-final  and, initial-final and initial-initial configurations (by crossing momenta to the initial state) should produce a subtraction term that is locally free of angular terms.

\begin{figure}[t]
\begin{minipage}[b]{0.3\linewidth}
\centering
\includegraphics[width=5cm,angle=270]{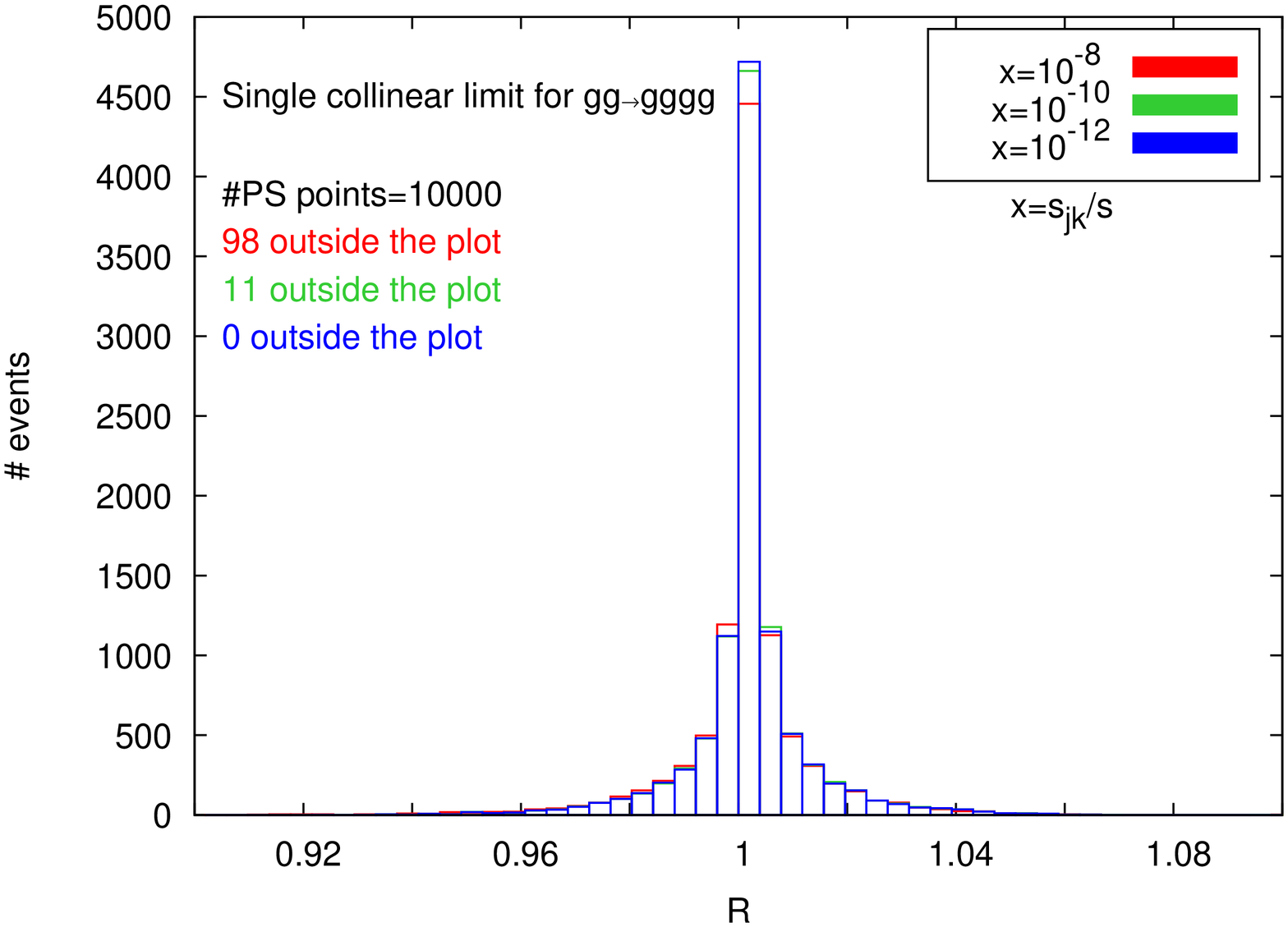}\\
\vspace{0.2cm}
(a)
\end{minipage}
\hspace{0.5cm}
\begin{minipage}[b]{0.7\linewidth}
\centering
\includegraphics[width=5cm,angle=270]{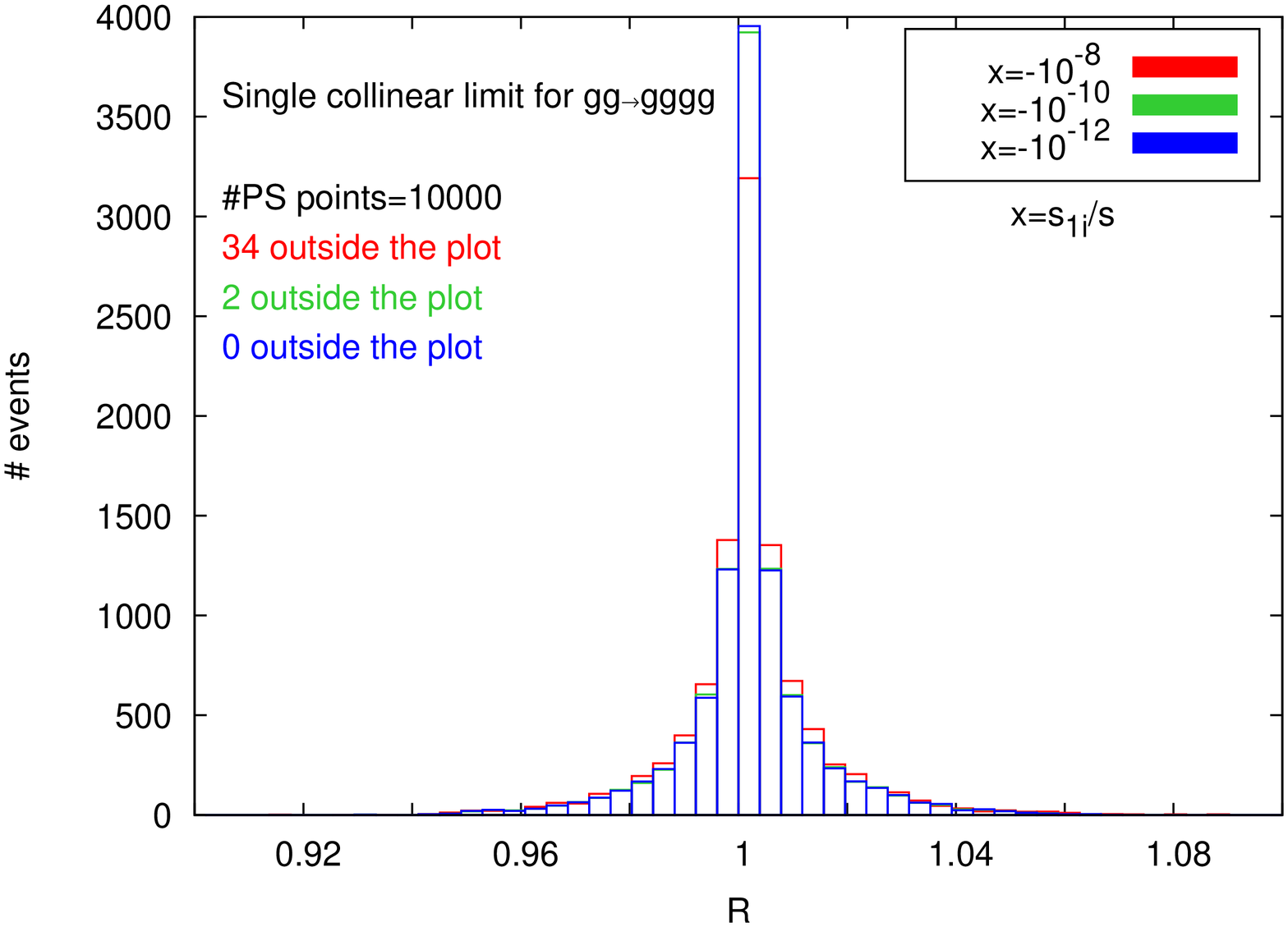}\\
\vspace{0.2cm}
(b)
\end{minipage}
\caption[Azimuthally corrected single collinear limit]{Distribution of $R$ for 10000 single collinear phase space points for the azimuthally corrected subtraction term for (a) Final state collinear singularities with $s_{ij}\to 0$ and (b) Initial state collinear singularities with $s_{1i}\to 0$.}
\label{fig:scolltheta}
\end{figure}

With this azimuthally modified subtraction term, we recompute the distributions in figs.~\ref{fig:scolltheta}. 
In fig.~\ref{fig:scolltheta}(a) we show the $R$ distribution for 10000 final state single collinear phase space points and $x=10^{-8}$ (red), $x=10^{-10}$ (green) and  $x=10^{-12}$ (blue). For $x=10^{-12}$ we obtained an average $R=0.99994$ and a standard deviation of $\sigma=0.015$.

We repeat the same analysis for the initial state collinear configuration in fig.~\ref{fig:scolltheta}(b) for $x=-10^{-8}$ (red), in green $x=-10^{-10}$ (green) and in blue $x=-10^{-12}$ (blue). For $x=-10^{-12}$ we obtained an average of $R=1.00007$ and a standard deviation of $\sigma=0.012$.

For both cases, the distributions now peaks around $R=1$ with a more pronounced peak as the limit is approached, just as in the double unresolved and single soft limits discussed earlier.  This demonstrates the convergence of the counterterm to the matrix element. 

We note that simply introducing the azimuthal correction term $\Theta_{F_3^0}(i,j,z,k_\perp)$ has the unfortunate side effect of generating new singularities in the previously analysed (double unresolved) phase space regions. For example, looking into equation (\ref{eq:angfunction}) for $\Theta_{F_3^0}(i,j,z,k_\perp)$ we see invariants in the denominator that are not compensated by a small quantity in the numerator in the double unresolved limit. The azimuthal correction term therefore introduces new divergences which are not present in the matrix element.

\begin{figure}[t]
\begin{minipage}[b]{0.3\linewidth}
\centering
\includegraphics[width=5cm,angle=270]{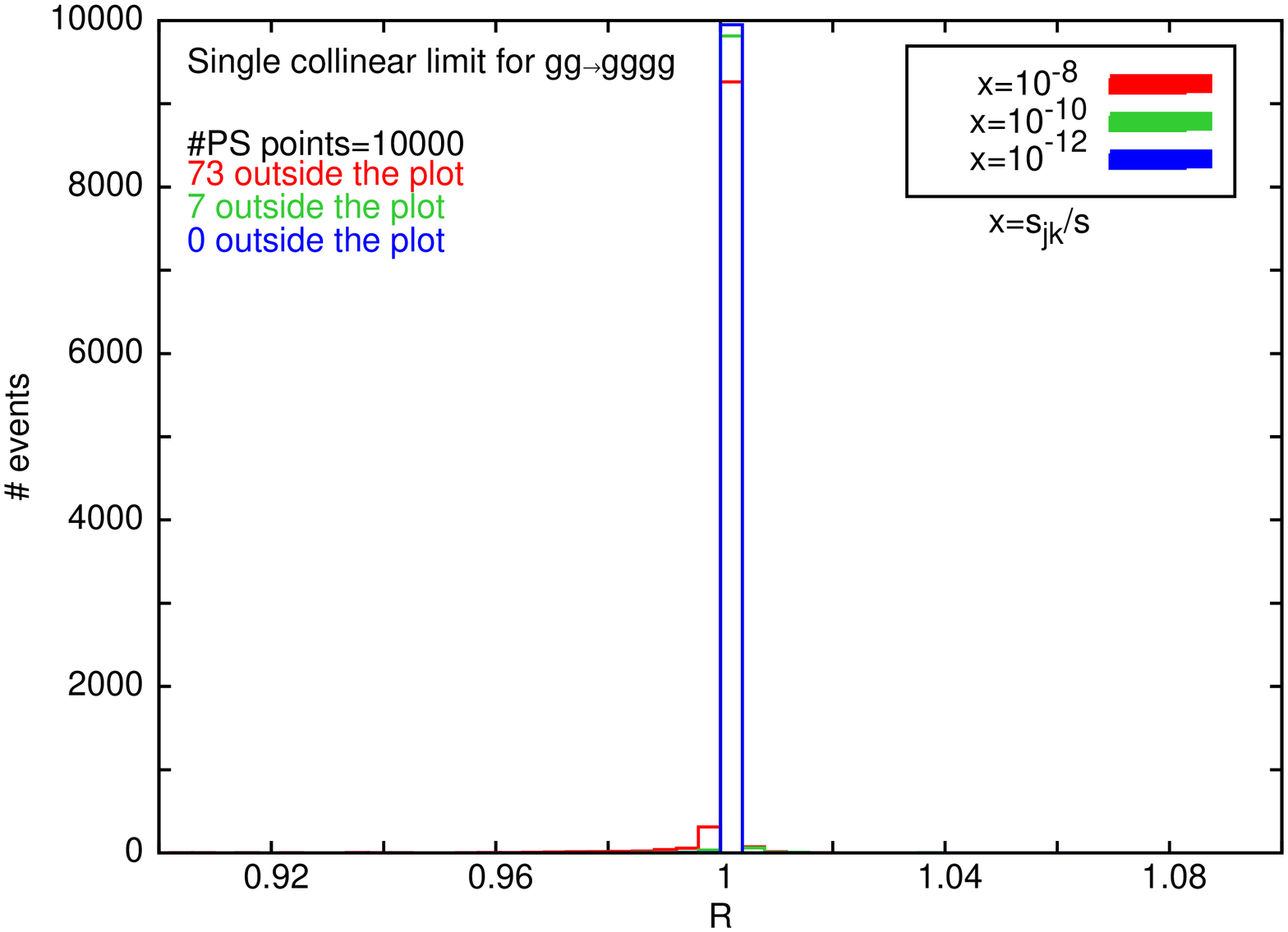}\\
\vspace{0.2cm}
(a)
\end{minipage}
\hspace{0.5cm}
\begin{minipage}[b]{0.7\linewidth}
\centering
\includegraphics[width=5cm,angle=270]{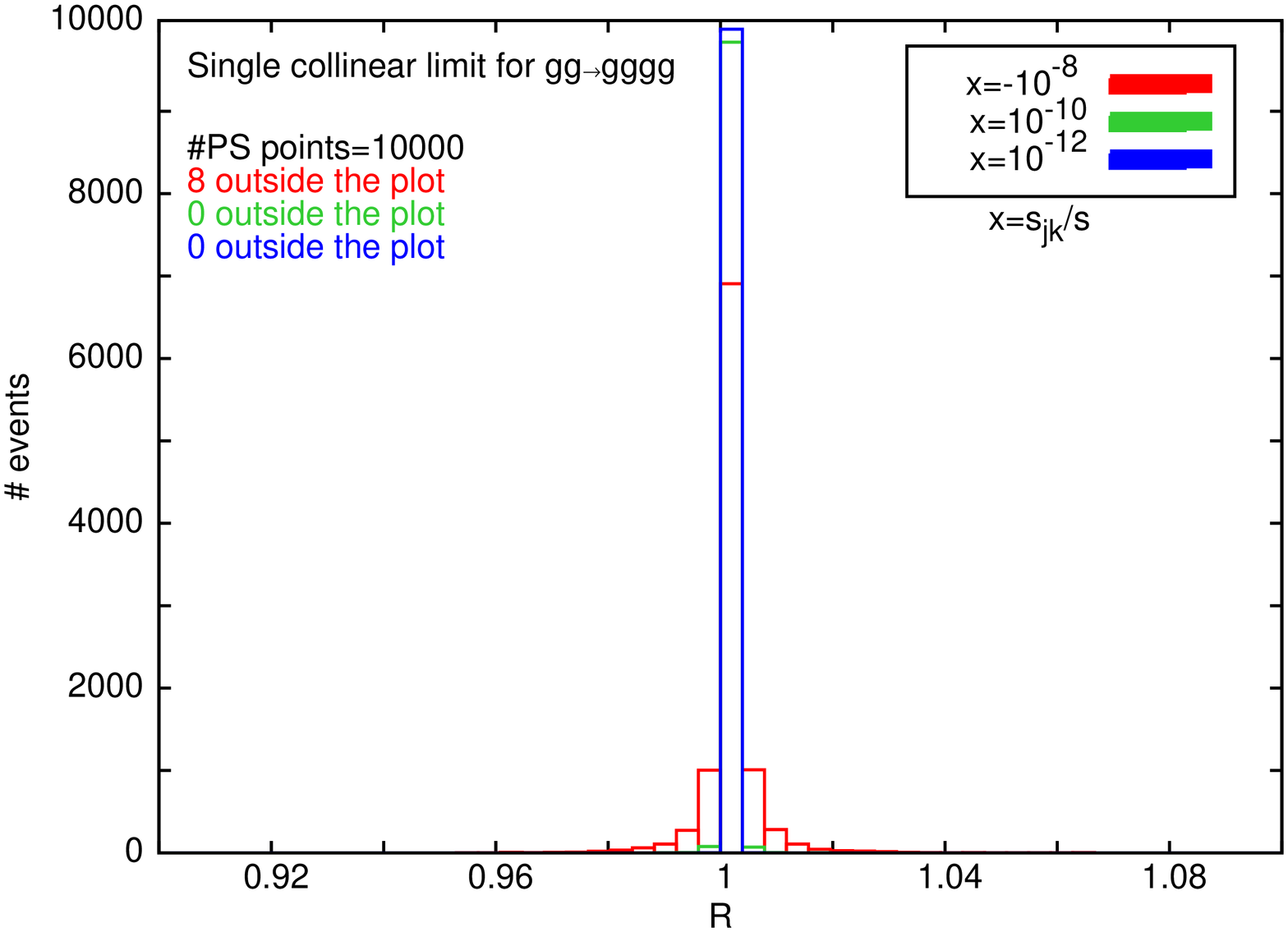}\\
\vspace{0.2cm}
(b)
\end{minipage}
\caption[Azimuthally corrected single collinear limit]{Distribution of $R$ for 10000 single collinear phase space point pairs where the pair of phase space points is related by an azimuthal rotation of $\pi/2$ about the collinear direction for (a) Final state collinear singularities with $s_{ij}\to 0$ and (b) Initial state collinear singularities with $s_{1i}\to 0$.}
\label{fig:scollrot}
\end{figure}

A second more successful approach is to try to cancel the angular terms by combining phase space points related to each other by a rotation of the system of unresolved partons \cite{Weinzierl:2006wi,GehrmannDeRidder:2007jk}.  When both collinear partons are in the final state, this can be achieved by considering pairs of phase space points which are related by rotating the collinear partons by $\pi/2$ around the resultant parton direction.
This is the direction defined by $p^\mu$ in eq.~\eqref{eq:colmap}.
Similarly, for the initial-final state collinear configurations produced when $p_i^\mu \to p^\mu+p_j^\mu$ for $i=1,2$ and with $i||j$ and $p^2=0$,  the phase space points should again be related by rotations of $\pi/2$ about the direction of $p^\mu$.  This has the consequence of rotating $p_i^\mu$ off the beam axis and therefore has to be compensated by a Lorentz boost.

The effect of combining pairs of phase space points is shown in Fig.~\ref{fig:scollrot} for the same values of small parameter: $|x|=10^{-8}$ (red),   $10^{-10}$ (green) and  $10^{-12}$ (blue) for $x = s_{ij}/s_{12}$ and $x = s_{1i}/s_{12}$ respectively. We see that the distributions for both final-final and initial-final collinear limits have a very sharp peak around $R=1$.   For the final state singularity and $x=10^{-12}$ we obtained an average of $R=0.999996$ and a standard deviation of $\sigma=0.00015$. Similarly, for the 
initial state collinear singularity and $x=-10^{-12}$ we found $R=0.9999991$ and $\sigma=0.00011$.
There is an enormous improvement compared to the raw distributions shown in Figs.~\ref{fig:scollff} and \ref{fig:scollif} and a significant improvement compared to adding an azimuthal correction term shown in Fig.~\ref{fig:scolltheta}.  The modified subtraction term clearly converges to the matrix element as we approach the single collinear limit and correctly subtracts the azimuthally enhanced terms in a point-by-point manner.

\section{Conclusions} 
\label{sec:conc}

In this paper, we have generalised the antenna subtraction method for the calculation
of higher order QCD corrections to exclusive collider observables for situations with
partons in the initial state to NNLO.  

The basic ingredients for the antenna subtraction terms, are the antenna functions,
which can be obtained from the known final-state antenna functions by simple
crossing~\cite{GehrmannDeRidder:2005cm,Daleo:2006xa}, the momentum mappings for
double unresolved configurations that have the correct behaviour in the unresolved
limits~\cite{Daleo:2006xa} and the integrated form of the antenna
functions~\cite{GehrmannDeRidder:2005cm,Daleo:2006xa,Daleo:2009yj,Boughezal:2010ty}.
Here, we derived general subtraction terms for all NNLO processes with two hadrons in the initial state.  

We focussed particular attention on the application of the antenna subtraction
formalism to construct the subtraction term relevant for the gluonic double real
radiation contribution to dijet production. The gluon scattering channel is expected
to be the dominant contribution at NNLO, but the subtraction term also involves the
four-gluon antenna function $F_4^0$ which has not been encountered in previous NNLO
calculations. The gluonic subtraction term includes a mixture of  three-parton and
four-parton antennae functions in final-final, initial-final and initial-initial
configurations. The subtraction terms for processes involving quarks will make use of
the same antenna building blocks and momentum mappings given discussed here but are
expected to be simpler.

By construction the counterterm subtracts double and single unresolved singularities
in the final and initial state. It is constructed using the analytic behaviour of the
antenna in the single and double unresolved limits. We therefore tested that our 
implementation of the antenna subtraction behaves in the expected way by comparing
the behaviour of the double radiation contribution with that of the subtraction term
numerically.  The agreement between the exact six-gluon contribution and the
approximate subtraction term represents the first sanity check on the NNLO antenna
subtraction method for gluonic processes. All of the single and double unresolved
regions of   phase space were analysed separately. The numerical results showed that
the combination of the antennae correctly describes the infrared singularity
structure of the matrix element, with the caveat of the correct handling of the
azimuthal terms associated with the single collinear limit. All of the double
unresolved and single soft singularities present in the matrix elements are cancelled
in a point-by point manner.  The single collinear terms can successfully be treated by combining phase space points related by rotations about the collinear directions. 

Together with the integrated forms of the antenna functions (see
Ref.~\cite{Daleo:2009yj} for the initial-final and  Ref.~\cite{Boughezal:2010ty} for
the initial-initial configurations), the double real subtraction terms presented here
provides a major step towards the NNLO evaluation of the dijet observables at
hadron colliders. 
Future steps include; 
\begin{itemize}
\item[(i)] completion of the analytic integration of the initial-initial antenna. 
\item[(ii)] the subtraction of infrared divergences in the gluonic mixed real-virtual correction which requires an analysis (and integration) of the one-loop three-gluon antenna function. 
\item[(iii)] analytic cancellation of infrared poles between the analytically integrated antennae present in the subtraction terms and the two-loop four-gluon and one-loop five gluon matrix elements.
\item[(iv)] full parton-level Monte Carlo implementation of the finite four-, five- and six-gluon channels.
\item[(v)] the construction of similar subtraction terms, etc., for processes
involving quarks.
\end{itemize}
The final goal is the construction of a numerical program to compute
NNLO QCD estimates of di-jet production in hadron-hadron collisions.

\acknowledgments 
We acknowledge useful discussions with James Currie, Thomas Gehrmann and Adrian Signer. This research was supported in part by the UK Science and Technology Facilities
Council and by the European Commission's Marie-Curie Research Training Network
under contract MRTN-CT-2006-035505 `Tools and Precision Calculations for Physics
Discoveries at Colliders'. EWNG gratefully acknowledges the support of the
Wolfson Foundation and the Royal Society. JP gratefully acknowledges the award
of a Funda\c{c}\~{a}o para a Ci\^encia e Tecnologia (FCT - Portugal) PhD
studentship.

\bibliographystyle{JHEP-2}


\providecommand{\href}[2]{#2}\begingroup\raggedright\endgroup

\end{document}